\RequirePackage{fix-cm}
\documentclass[smallextended]{svjour3}       
\usepackage{subfigure}
\usepackage{amsmath}
\usepackage{epstopdf}
\smartqed  
\usepackage{graphicx}
\usepackage{siunitx}
\usepackage{subfigure}
\usepackage[round]{natbib}
\usepackage{amssymb}
\begin{document}

\title{Impact of distance determinations on Galactic structure. II. Old tracers 
}

\titlerunning{Galactic structure: Old tracers}        
\authorrunning{Kunder, Valenti, Dall'Ora, Pietrukowicz, Sneden, Bono et~al.}        

\author{Andrea Kunder, Elena Valenti, Massimo Dall'Ora, Pawel Pietrukowicz, Chris Sneden, Giuseppe Bono \\ \\
\& Vittorio F. Braga, Ivan Ferraro, Giuliana Fiorentino, Giacinto Iannicola, Marcella Marconi, Clara E. Mart\'inez-V\'azquez, Matteo Monelli, Ilaria Musella, Vincenzo Ripepi, Maurizio Salaris, Peter B. Stetson
}

\institute{A. Kunder,
              Leibniz Institut f\"{u}r Astrophysik, An der Sternwarte 16, D-14467 Potsdam
 	         \email{amkunder@gmail.com}             \\ \\
	      E. Valenti, European Southern Observatory, Karl-Schwarzschild-Str. 2, D-85748,	Garching bei Munich, Germany,
		\email{evalenti@eso.org} \\  \\
	      M. Dall'Ora, INAF, Osservatorio Astronomico di Capodimonte, Via Moiarello 16, 80131 Napoli, Italy
		\email{dallora@na.astro.it} \\ \\
		P. Pietrukowicz, Warsaw University Observatory, Al. Ujazdowskie 4, 00-478 Warszawa, Poland
              \email{pietruk@astrouw.edu.pl} \\ \\
              C. Sneden, University of Texas Department of Astronomy, 2515 Speedway, Stop C1400, Austin, TX 78712
 		\email{chris@verdi.as.utexas.edu} \\ \\
		G. Bono, Universita' di Roma Tor Vergata, Via della Ricerca Scientifica 1, 00133 Roma and 
		INAF--Osservatorio Astronomico di Roma, Via Frascati 33, 00078 Monte Porzio Catone, Italy  
 		\email{bono@roma2.infn.it} \\ \\
		V.F. Braga, Instituto Milenio de Astrofisica, Santiago, Chile and
		Departamento de Fisica, Facultad de Ciencias Exactas, Universidad Andres Bello, Av. Fernandez Concha 700, Las Condes, Santiago, Chile \\ \\
		I. Ferraro, INAF--Osservatorio Astronomico di Roma, Via Frascati 33, 00078 Monte Porzio Catone, Italy  \\ \\
		G. Fiorentino, IINAF-OAS  Osservatorio di Astrofisica e Scienza dello Spazio di Bologna  \\ \\
		G. Iannicola, INAF--Osservatorio Astronomico di Roma, Via Frascati 33, 00078 Monte Porzio Catone, Italy  \\ \\
		M. Marconi , INAF, Osservatorio Astronomico di Capodimonte, Via Moiarello 16, 80131 Napoli, Italy \\ \\
		C. E. Mart\'inez-V\'azquez, Cerro Tololo Inter-American Observatory, National Optical Astronomy Observatory, Casilla 603, La Serena, Chile and IAC-Instituto de Astrofísica de Canarias, Calle Vía Lactea s/n, E-38205 La Laguna, Tenerife, Spain \\ \\
		M. Monelli, Instituto de Astrofisica de Canarias, Calle Via Lactea, 38205, La Laguna (Tenerife), Spain  \\ \\
		I. Musella, INAF, Osservatorio Astronomico di Capodimonte, Via Moiarello 16, 80131 Napoli, Italy  \\ \\
		V. Ripepi, INAF, Osservatorio Astronomico di Capodimonte, Via Moiarello 16, 80131 Napoli, Italy  \\ \\
		M. Salaris, Astrophysics Research Institute, John Moores University, IC2, Liverpool Science Park, 146 Brownlow Hill, Liverpool L3 5RF  \\ \\
		P.B. Stetson, DAO/€"HIA, NRC, 5071 West Saanich Road, Victoria, BC V9E 2E7, Canada  \\ \\
             }

\date{Accepted: June 18, 2018}

\maketitle

\begin{abstract}
Here we review the efforts of a number of recent results that use old tracers to understand the
build up of the Galaxy.  Details that lead directly to using these old
tracers to measure distances are discussed.  We concentrate on the following:  
(1) the structure and evolution of the Galactic bulge and inner Galaxy constrained from the dynamics 
of individual stars residing therein; 
(2) the spatial structure of the old Galactic bulge through photometric observations of RR Lyrae-type stars; 
(3) the three\--dimensional structure, stellar density, mass, chemical composition, and age of the Milky 
Way bulge as traced by its old stellar populations; 
(4) an overview of RR Lyrae stars known in the ultra-faint dwarfs and their relation to the Galactic halo;
and
(5) different approaches for estimating absolute and relative cluster ages. 

\end{abstract}

\vspace{15pt}
Using old stars as tracers to piece together the build-up of the Milky Way (MW) allows for a glimpse of 
the early stages of the Milky Way Galaxy.  This is because old stars have existed for
several billion years -- when the Galaxy was in its infancy -- and are therefore a foundation 
of the current ensemble of stellar populations that have since begun to also populate the MW.
The signatures of old stars that are discussed here involve primarily RR Lyrae stars (RRLs) and
Red Clump (RC) stars as probes, arguably the most widely used old stars as stellar tracers.
 
RC stars are core helium\--burning 
stars that can be considered useful distance indicators because their magnitude changes slowly and 
smoothly with age and metallicity. Such a change is predicted very accurately by stellar evolution 
models \citep[e.g.,][]{salaris+02}.  They do not univocally trace old populations, but also intermediate ages. 
However, because being a bright and distinct feature in the color\--magnitude diagram they are very numerous 
and easy to identify, they have been extensively used in the study of especially the Galactic bulge. 
It should be noted though that, according to the prescriptions of \citet{salaris+02}, when the age\--metallicity 
relation and star formation history of a given composite stellar system are unknown the derived distance 
modulus by using RC can have an error up to $\sim$0.3\,mag.  \citet{girardi16} provides an overview 
on the advantages and caveats of using RC stars for distance determination. 

RRLs, on the other hand, are unequivocally old ($\ge 10$ Gyr).  They are 
low-mass ($\sim 0.6-0.8 M_\odot$) horizontal branch (HB) stars that experience radial pulsations 
because of periodic variations of the atmosphere opacity, 
in partial ionized regions (H, He). This causes cyclic variations of luminosity and effective temperature, 
with periods ranging typically from $\sim 0.3$ to $\sim 1$ days. Their typical mean luminosity in the 
$V$-band is in the range $M_V \sim 0.5-1$ mag, making them moderately bright objects, while 
effective temperatures range from $\sim 7200$ K to $\sim 5600$ K, which transform in typical mean 
colors between $(B-V) \sim 0.2$ mag and $(B-V) \sim 0.4$ mag. Historically, RRLs have been used 
as standard candles since their mean luminosity in the $V-$band is almost constant, with some 
dependency on their metallicity and evolutionary status \citep[e.g.][]{Bono2003, Clementini2003}. 
Moreover, they are moderately bright ($\sim$ 40L$_{\odot}$), easy to detect from their light 
variations and practically ubiquitous. Also, the occurrence of the so-called Oosterhoff dichotomy 
allowed people for decades to disentangle the old component of the Galaxy in two distinct groups, 
putting strong constraints on the Galaxy formation 
mechanisms \citep[e.g.][]{Fiorentino2015, Fiorentino2016, MartinezVazquez2016}. Their role 
as standard candles and stellar population tracers has been widely investigated also from the theoretical 
point of view, on the basis of extensive and detailed nonlinear convective pulsation models
 \citep[see e.g.][and references therein]{Marconi2015}.

RRLs are much less numerous and also more time-consuming to identify than the RC. 
Both the RC and the RRLs have their own observational and theoretical advantages and disadvantages, 
however, together, they have shown to be the invaluable available tracers of old population 
for the study of the Milky Way galaxy.
In \S1, Andrea Kunder discusses how RRLs and RC giants have shaped our view of the kinematics of
the inner Galaxy.  Similarly, Pawel Pietrukowicz focuses on using these stars to understand the
spatial structure of the inner Galaxy in \S2.  The chemical composition and age of the inner Galaxy
is then described in \S3 by Elena Valenti, again utilizing RRLs and RC giants.

In \S4 and \S5, RRLs and RCs are also discussed within the context of the Milky Way satellites 
and globular clusters, to further place into context the build-up of the Milky Way.
Dwarf spheroidal (dSph) satellites of the Milky Way, and the low brightness tail of the dSph, 
the ultra-faint dwarf (UFD) galaxies, are old, metal-poor, gas poor and dark 
matter-dominated systems.  Comparing their RRL populations with that of the MW places allows
a more complete picture of galaxy formation to emerge, as shown in \S4, by
Massimo Dall'Ora, Giuseppe Bono, Giuliana Fiorentino, Marcella Marconi, 
Clara E. Martinez-Vazquez, Matteo Monelli, Ilaria Musella and Vincenzo Ripepi.

Similarly, globular clusters harbor some of the oldest stars in our Galaxy and their distances,
distribution throughout the Galaxy, and ages have long been used to as pillars to understand
the early Galaxy.  In \S5, new observations that have improved the accuracy and precision 
of stellar populations within globular clusters, as well as better stellar models, have advanced our 
ability to use these old tracers to understand the early formation of the components of Galaxy.
This section is contributed by Giuseppe Bono, Vittorio Braga,
Giuliana Fiorentino, Massimo Dall'Ora, Ivan Ferraro, Giacinto Iannicola,
Matteo Monelli, Maurizio Salaris and Peter B. Stetson.

\section{Kinematics of the Galactic bulge}
\label{intro}
The internal kinematics of the bulge using a statistical sample of stars was first analyzed by the 
Bulge Radial Velocity Assay (BRAVA) survey (Rich et al. 2007; Kunder et al. 2012).
BRAVA targeted M-giants toward the Galactic bulge in a grid covering three strips of latitude, at 
$b=-$4$^\circ, -$6$^\circ, -$8$^\circ$, that span across $-10^\circ~<~l~>~10~^\circ$.  
From a total of $\sim$10,000 stars, they showed that the bulge is in cylindrical rotation.  
Kinematic models allow at most only $\sim$10\% of the 
original model disk mass to be in the form of a ``classical'' spheroid formed by dissipational 
collapse.  Subsequent kinematic bulge surveys, probing closer to the plane and/or different
stellar populations, have confirmed this result.  However, a more complicated kinematic view
of the bulge than was first able to be disentangled by the original BRAVA results has emerged, which
is reviewed here by Andrea Kunder.

The Abundances and Radial velocity Galactic Origins Survey \citep[ARGOS;][]{freeman13} probed
$\sim$17,400 red-clump giants in the bulge -- fainter stars than probed by BRAVA, 
but having temperatures more favorable for the 
determination of metallicities.  Probing the CaT at R$\sim$11,500, the ARGOS stars could be separated into 
metallicity sub-samples, which \citet{ness13} believe
to represent different populations in the bulge.  Sample ``A" consists of stars with 
$\rm [Fe/H] \sim +0.15$~dex, which are proposed to belong to a relatively thin and
centrally concentrated part of the boxy/peanut bulge.  Sample ``B" consists of stars with 
$\rm [Fe/H] \sim -0.25$~dex, belonging to a thicker boxy/peanut bulge.  Compared to ``A",
this sample is hotter and less compact.  Sample ``C" consists of stars 
with $\rm [Fe/H]\sim-$0.7~dex and kinematically differs from component A and B in 
that it does not appear to have a latitude-independent velocity dispersion
rotation.  Sample "D" is the most metal-poor with $\rm [Fe/H]\sim-$1.0~dex.  It
is the least understood, due to the paucity of ARGOS stars with such metallicities.  
For example, there are only two stars with $\rm [Fe/H]\sim-$1.0~dex 
in the ARGOS field at ($l$, $b$) = ($-$20$^\circ$, $-$5$^\circ$), so the velocity
dispersion provided by \citet{ness13} for this field is not well constrained.  In all sub-samples cylindrical
rotation was seen, although the most metal-poor red clump giants (sample D) rotated slower
than their metal rich counterparts \citep{ness13}.  But as this signature was 
seen at latitudes 10 degrees from the plane, they associated the slower rotation 
as the consequence of contamination from the halo and metal-weak thick disk 
populations creeping into the bulge.  

The GIRAFFE Inner Bulge Survey \citep[GIBS;][]{zoccali14} is targetting red-clump 
giants closer to the plane than both BRAVA and ARGOS.  Most of the fields are at 
a resolution of R$=$6500, but a handful of fields were observed at R$=$22,500.  
Metallicities for their $\sim$5000 surveyed stars 
\citep{gonzalez15, zoccali17} were derived, as well as elemental abundances
for $\sim$400 red clump giants.  They confirmed cylindrical rotation
also at latitudes $b$=$-$2$^\circ$, and found that 
throughout  most of the bulge, a narrow metal-rich ($\rm [Fe/H]=+$0.26) population of stars 
and a broad more metal-poor ($\rm [Fe/H]=-$0.31) component appears to exist.  Both components
rotate cylindrically, although the metal-poor stars are kinematically hotter and less bar-like.

Lastly, the APOGEE survey has probed $\sim$19,000 red giant stars at positive longitudes close
to the plane \citep{zasowski16}.  The high-resolution (R$=$22,500), makes it feasible to 
obtain elemental abundances and the near-infrared wavelength regime
($\lambda$=1.51-1.70~\si{\micro\metre}) allows the plane of the bulge
to be probed, where dust and reddening is severe, but minimized by longer wavelengths.
They find that the transition from cylindrical to non-cylindrical 
rotation occurs gradually, and most notably at higher latitudes.  At a longitude of $l\sim$7$^\circ$ the
signature of cylindrical rotation fades, which is expected, as this longitude is near the end
of the boxy bulge.  Despite their large chemo-dynamical sample, they are not able to find 
distinct and separable bulge populations, although their measures of skewness is consistent
with different evolutionary histories of metal-rich ($\rm [Fe/H] =$ +0.26) and metal-poor ($\rm [Fe/H] \sim
-$0.31) bulge populations.  

All of these large surveys have shown that the bulge consists of a massive bar rotating as a solid body;
the internal kinematics of these stars are consistent with at least 90\% of the inner Galaxy 
being part of a pseudobulge and lacking a pressure supported, classical-like bulge.  
N-body barred galaxy models (boxy peanut bulge models) can explain the global kinematics, so 
our bulge appears to have formed from secular evolution of a massive early disk.

However, some finely detailed behavior of stars 
remain unexplained.  For example, it is not clear how the kinematically cooler bulge stars (which are more metal-rich)
fit together with the kinematically hotter (more metal-poor) bulge stars.  Also, some bulge
locations have shown no evidence for a metal-rich and metal-poor population, despite
being at the same latitude as other fields which do clearly separate chemically \citep[e.g.,][]{zoccali17}.

Figure~1 shows the distributions of the targets in the survey's mentioned above.  
Most survey's have focussed on the Southern bulge,
where the crowding is not as extreme.  The APOGEE survey, in contrast, using a telescope in 
the North, probes more of the Northern bulge, and has not yet been able to reach the negative 
longitudes.  With exception of the ARGOS survey, all data has been publicly released, improving the
quality and value of these surveys, and providing 
the wider scientific community the ability to productively use data for further research
with the potential to advance developments.

\begin{figure*}
  \includegraphics[height=11.4cm]{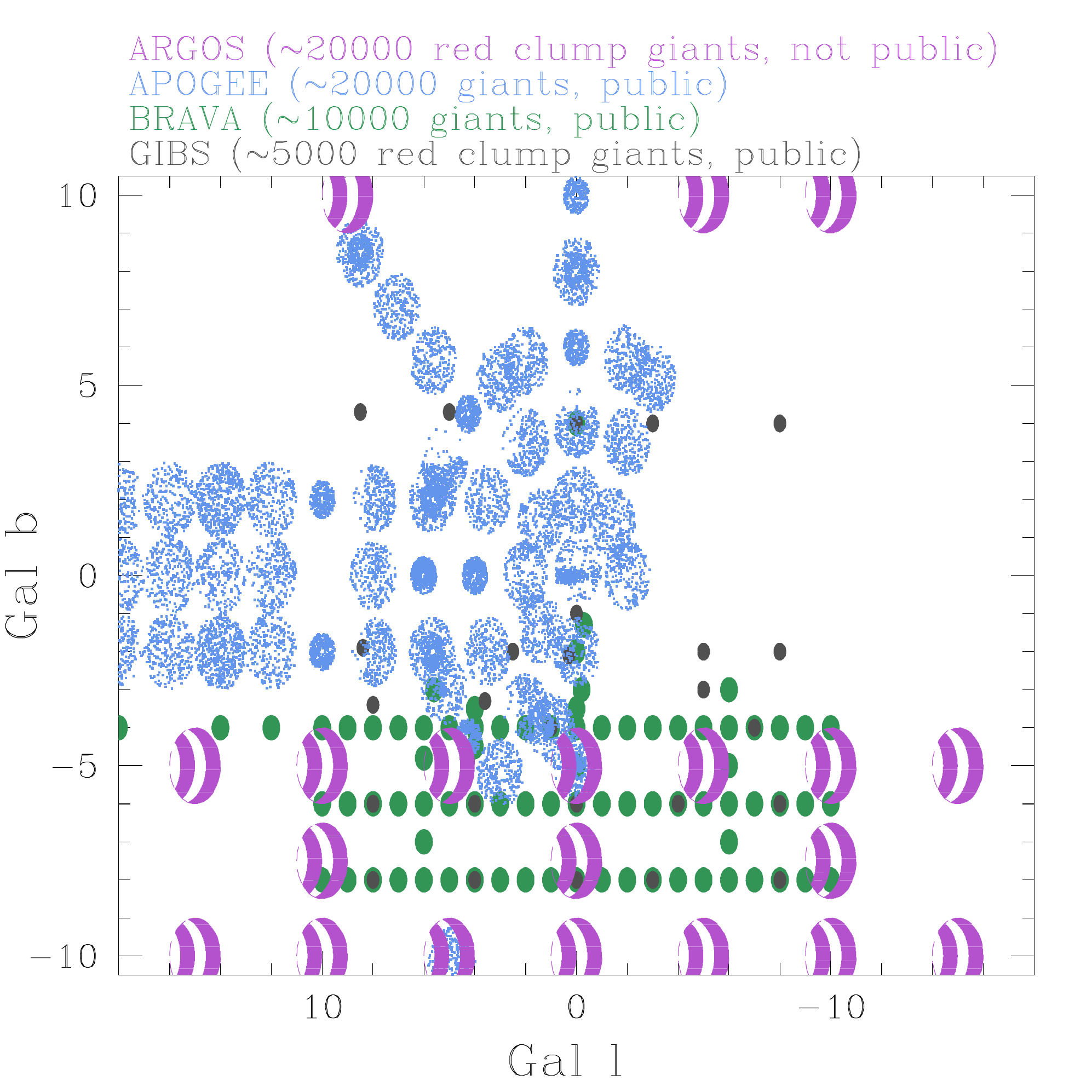}
\caption{The spatial location of the ARGOS, APOGEE, BRAVA and GIBS spectroscopic bulge surveys.}
\label{fig:1}       
\end{figure*}

\subsection{Targeted Kinematic Studies}
\label{sec:targeted}
Notable recent high resolution studies (R$\sim$20,000-30,000) of bulge field stars have been 
meticulously obtained by Johnson et al. 2011, 2012, 2013a, 2013b, 2014.  In these papers, 
along with radial velocities, numerous individual elemental abundances, for some 
stars 27 elements ranging from oxygen to erbium, are derived for a sample of $\sim$500 bulge giants.
Therefore, not only can the kinematics, $\rm [Fe/H]$ and $\rm [\alpha/Fe]$ ratios of bulge stars be compared to 
those in the thin and thick disks, but also the light odd-Z and Fe-peak (and also neutron-capture) elements 
are touched on, which also provide discriminatory power between models and other stellar 
populations.  These more detailed and targeted observations indicate that at $\rm [Fe/H] > -$0.5, 
the bulge exhibits a different chemical composition than the local thick disk in that
the bulge $\rm [\alpha/Fe]$ ratios remain enhanced to a slightly higher $\rm [Fe/H]$ than the thick disk, 
and the Fe-peak elements Co, Ni, and Cu appear enhanced compared to the disk. 

Further, these studies point to a bulge that formed rapidly ($<$1-3~Gyr), because of the 
enhanced $\rm [\alpha/Fe]$ abundances coupled with the low $\rm [La/Eu]$ ratios of the bulge 
stars \citep[see also][]{mcwilliam10}.  This confirms a very fast chemical enrichment in the bulge 
put forth by the very first detailed abundance studies of red giants in the Milky Way 
bulge \citep[e.g.,][]{mcwilliam94, zoccali06, fulbright07}.

\citet{babusiaux10,babusiaux14} compared the velocities of metal-rich and metal-poor bulge 
stars and found that higher metallicity stars in the bulge show larger vertex deviations 
of the velocity ellipsoid than more metal-poor stars.  They also found that metal-rich 
stars show an increase in their velocity dispersion with decreasing latitude (moving 
closer to the Galactic plane), while metal-poor stars show no changes in the velocity 
dispersion profiles.  They concluded that the more metal-rich stars are consistent 
with a barred population and the metal-poor stars with a spheroidal component.  
However, other high-resolution studies of bulge stars have not confirmed such 
trends and instead find consistent decrease in velocity dispersion 
with increasing $\rm [Fe/H]$ \citep[e.g.,][]{johnson14, uttenthaler12, ness13}.

Perhaps the greatest limitation in finding possible differences between a 
metal-rich and a metal-poor population in the bulge is the difficulty of finding
metal-poor stars in the bulge.  For example, within the ARGOS survey \citep{ness13}, 
0.1\% of the stars identified as lying in the bulge have $\rm [Fe/H] < -2.0$~dex.
The first metal-poor stars found close to the Galactic center
was presented by \citet{schultheis15}, who find $10$ stars with
$\rm [M/H] \sim -1.0$~dex within $\sim$200~pc from the Galactic center.  
\citet{garciaperez13} used infrared spectroscopy of 2400 bulge stars to uncover five new
metal-poor stars with $\rm -2.1 < [Fe/H] < -1.6$, and 
using optical photometry to first select metal-poor candidates, 
and \citet{schlaufman14}
uncovered three stars in the direction of the bulge with $\rm -3.0 < [Fe/H] < - 2.7$.
The Extremely Metal-poor BuLge stars with AAOmega (EMBLA) survey,
dedicated to search for metal-poor stars in the bulge, has uncovered 
$\sim$40 metal-poor stars \citep{howes14, howes15, howes16}, including a handful with
$\rm [Fe/H] < -3.0$.  
Five stars in the very metal-poor regime, at $\rm -2.7 < [Fe/H] < -2.0$
are presented in \citet{koch16}, where they find that the metal-poor stars are a broad mix,
and no single, homogeneous ``metal-poor bulge" can yet be established.

\begin{figure*}
  \includegraphics[width=0.75\textwidth]{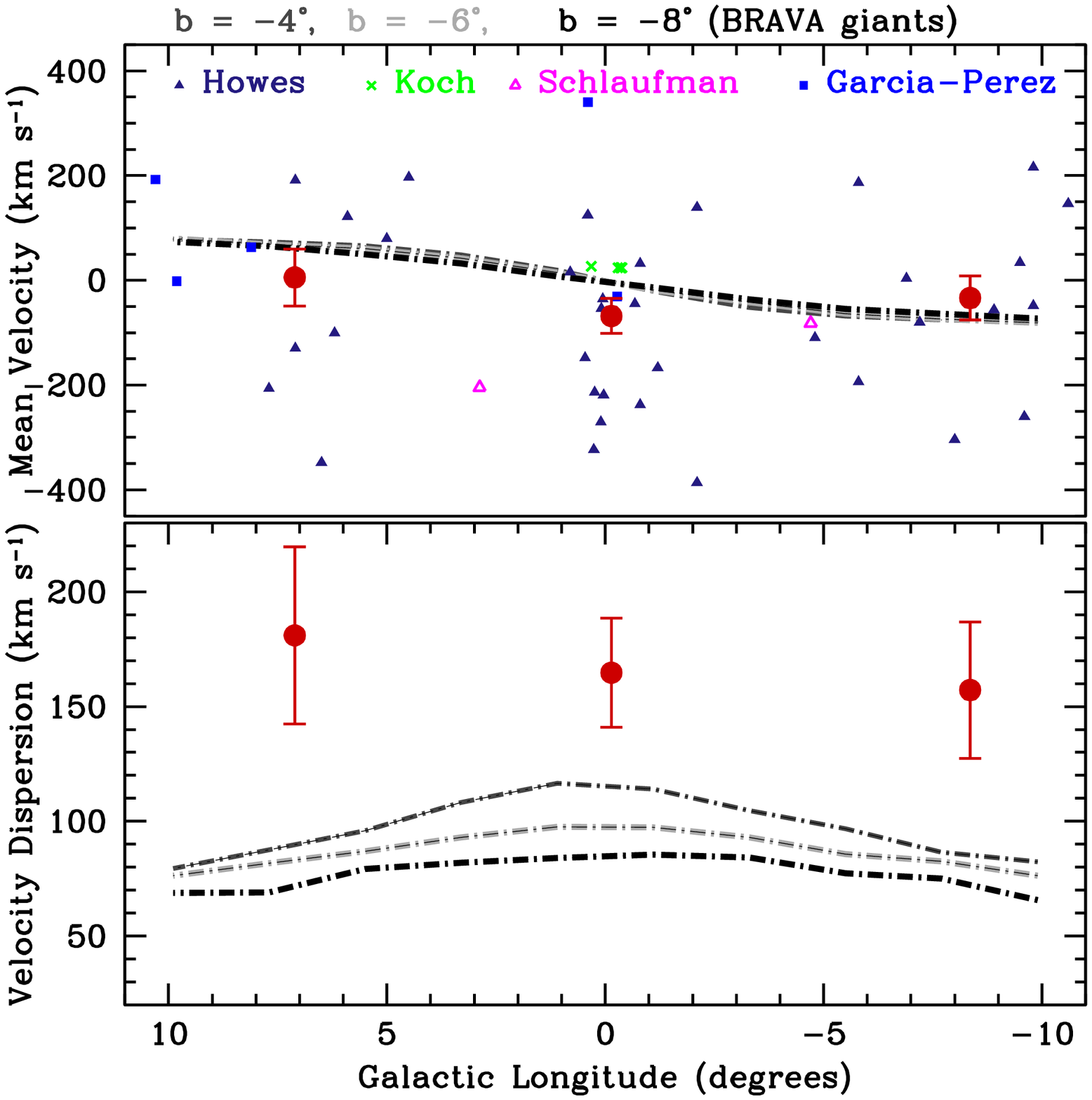}
\caption{ The velocity dispersion profile (bottom) and rotation curve (top) for the metal-poor stars 
($\rm [Fe/H] < -$2.0) observed in the bulge compared to that of the BRAVA giants at $b$ = $-$4$^\circ$, $-$6$^\circ$, 
and $-$8$^\circ$ strips (Kunder et al. 2012).  
The individual metal-poor star measurements are given in the top panel, and the 
large red circles indicate the mean Galactocentric velocity (top) and velocity dispersion (bottom).
The metal-poor stars have kinematics suggesting 
they are different from the bulge giants, although the sample size is small ($\sim$50).  It has been put
forward that these metal-poor stars are actually halo interlopers (e.g., Howes et~al. 2014,
Kunder et~al. 2015, Koch et~al. 2016).}
\label{fig:howes_rot_curve}       
\end{figure*}

Figure~\ref{fig:howes_rot_curve} shows the kinematics of these metal-poor stars compared to ``normal" bulge
giants from the BRAVA survey.  Though the number statistics are still small, their velocity 
dispersion suggests either that the metal-poor stars in the bulge 
have different kinematics than the more metal-rich stars, or that the metal-poor stars
discovered are a halo population. 

Lacking a statistical sample of metal-poor stars in the bulge, understanding their kinematics 
and placing them in context within the Galaxy is nontrivial.  Especially since the oldest and 
most metal poor stars (which may trace the dark matter) are thought 
to be found in the center of the Galaxy -- in the bulge but not sharing its kinematics 
and abundance patterns \citep{tumlinson10}, the metal-poor ``bulge" stars could provide a 
big piece of the puzzle in understanding the formation and subsequent evolution of the
Galactic bulge.  

Perhaps the easiest identifiable old, metal-poor bulge population are those horizontal 
branch stars that pulsate as RR Lyrae stars.   
Their progenitors formed long ago ($\sim$10 Gyr), so that the RRLs we see today tell us about 
conditions when the halo of the Galaxy was being formed \citep[e.g.,][]{lee92}.  
The bulge RRLs were shown to be on average $\sim$1~dex more metal-poor 
than the majority of bulge stars residing in the bar \citep{walker91}, although some of the 
bulge RRLs do appear to have metallicities that overlap in abundance with the bar population.  

The ongoing Bulge Radial Velocity Assay for RR Lyrae stars, BRAVA-RR survey \citep{kunder16}, 
aims to collect spectrographic information for RRLs located toward the inner Galaxy.
Their sample of RRLs are selected from the Optical Gravitational Lensing Experiment (OGLE),
so the periods, amplitudes and magnitudes are already precisely known.  Multi-epoch
spectroscopy (typically 3 epochs per star) is used to obtain center-of-mass radial velocity with uncertainties 
of $\sim$5-10~km~s$^{-1}$.  From radial velocities of about $sim$1000 RRLs surveyed by
BRAVA-RR in four 2-degree fields covering approximately a Galactic latitude and longitude of
$-3^\circ < b > -6^\circ$ and $-4^\circ < l > 4^\circ$, it is evident that these old and metal-poor stars are
kinematically distinct from the more metal-rich red giants in the BRAVA, GIBS, ARGOS 
and APOGEE surveys.  The RRLs show null rotation and hot (high-velocity dispersion) kinematics. 
In the ARGOS survey one also observes a slowly-rotating metal-poor population, but 
these stars are believed to be 
contamination from disk and halo stars (as it is only seen at high Galactic latitude).
In contrast, the RRLs are at low Galactic latitudes ($|b|<$7$^\circ$) and have more certain distance 
estimates, and the larger number statistics of the RRLs makes this result quantifiable.
The RR Lyrae stars trace an older, more spheroidal component in the inner Galaxy. 

\begin{figure*}
  \includegraphics[width=0.75\textwidth]{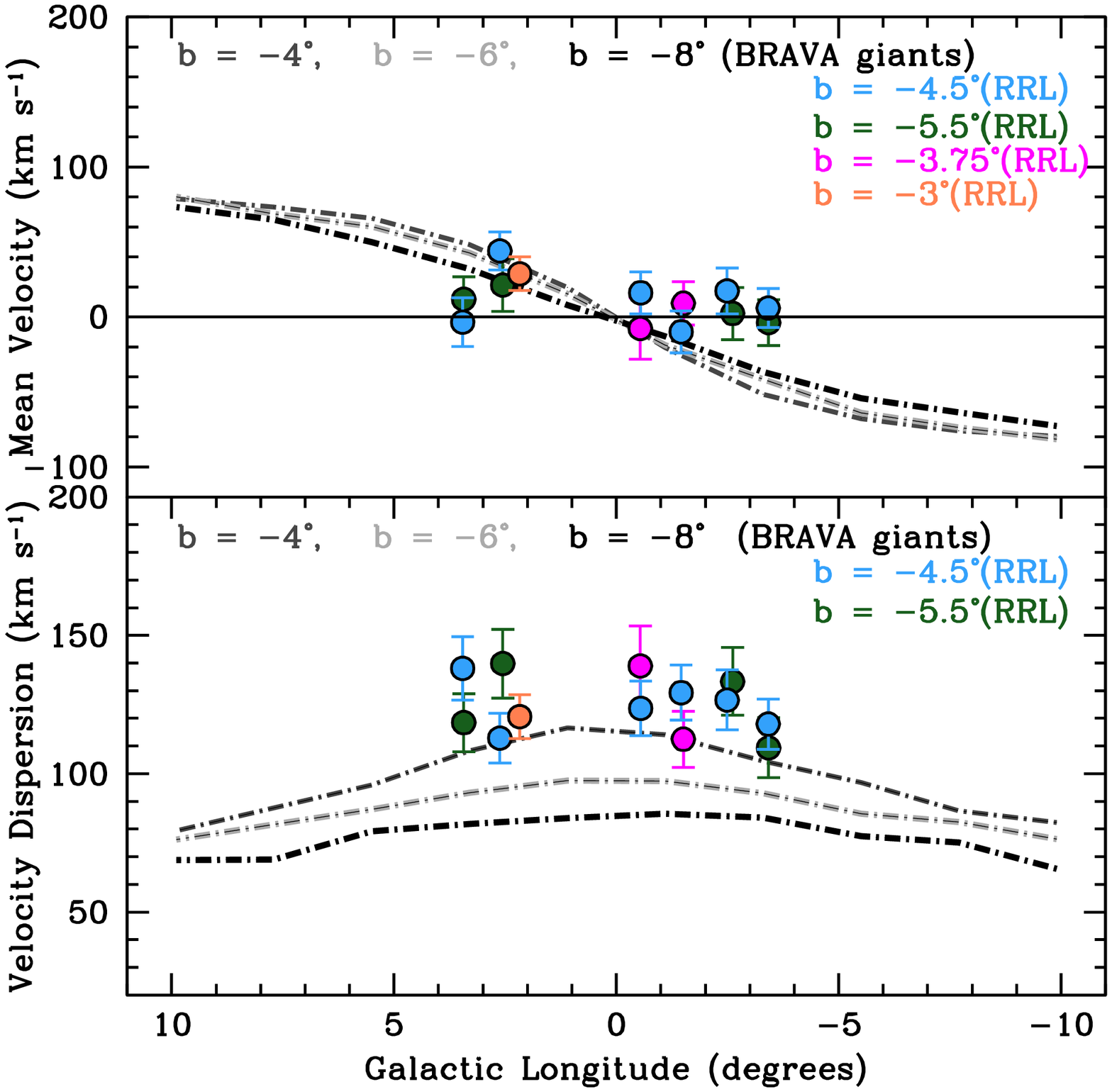}
\caption{The velocity dispersion profile (bottom) and rotation curve (top) for the $\sim$1000 RR Lyrae stars 
observed in the bulge compared to that of the BRAVA giants at $b$ = $-$4$^\circ$, $-$6$^\circ$, 
and $-$8$^\circ$ strips (Kunder et al. 2012, Kunder et al. 2016).  The bulge model showing 
these observations are consistent with a bulge
being formed from the disk is represented by the dashed lines (Shen et al. 2010).  
The RRLs have kinematics clearly distinct from the bulge giants, and are a non-rotating 
population in the inner Galaxy.}
\label{fig:2}       
\end{figure*}

The mass of this 'old' bulge is estimated to be $\sim$1\% of the total central mass,
broadly consistent with current bulge formation models, which predict that no more 
than $\sim$5\% of a merger-generated bulge \citep{shen10, ness13, dimatteo15}.
It may be that the RRL stars toward the bulge are actually an inner halo-bulge 
sample, as originally speculated in the early 1990s \citep[e.g.,][]{minniti94} and 
as at least one RRL orbit toward the Galactic bulge seems to indicate \citep{kunder15}.  

Prompted by the results from the RRLs, \citet{perezvillegas17} carried out $N$-body 
simulations for the Milky Way to investigate the kinematic and structural 
properties of the old metal-poor stellar halo in the barred inner region of the Galaxy. 
They showed that the RR Lyrae population in the Galactic bulge may
be the inward extension of the Galactic metal-poor stellar halo, and that 
especially the radial velocities of RRLs in the outer Galactic longitudes constrain 
a bulge/halo scenario.  Unfortunately, the RRLs investigated in \citet{kunder16} are confined to the
innermost 500~pc.  This is where a slow-rotating component has the smallest velocity difference 
compared to the metal-rich bulge giants ($\sim$25 km/s), and hence where population contamination 
from e.g., the halo or thick disk could more easily mask the effects of rotation.  Observations
of RRL at further longitudes in the bulge would allow us to distinguish between a bulge or halo.

\subsection{Future: Gaia}
\label{sec:gaia}
Gaia has begun collecting six-dimensional space coordinates for more than 1 billion stars in the
Milky Way.  The bulge is a difficult target for Gaia, due to the crowding and extinction,
but Gaia will still impact bulge kinematics significantly.  The Radial Velocity 
Spectrometer (RVS), which is the spectroscopic instrument for all objects 
down to $G \sim$16~mag, can cope with a crowding limit
of 35,000 stars deg$^{-2}$ \citep{Reyle05}.  In denser areas, only the brightest stars are 
observed and the completeness limit will be brighter than 16 mag.  Therefore, we can
expect the brighter giants to be surveyed throughout the bulge with RVS, but most
of the red clump stars and RRLs will be lacking Gaia radial velocities.

Never-the-less, the 
astrometric instrument has been designed to cope with object densities up to 750,000 
stars per square degree and down to $G \sim$20~mag.  Therefore, for a large area of the
bulge, at least some of the horizontal branch will be reachable for useful proper-motions, although no 
useful (5 $\sigma$) parallaxes are expected for these stars.  At end of mission,
Gaia will have $\sim$54 Gaia transits covering the bulge \citep{Clementini2016}.

\begin{figure*}
  \includegraphics[width=0.75\textwidth]{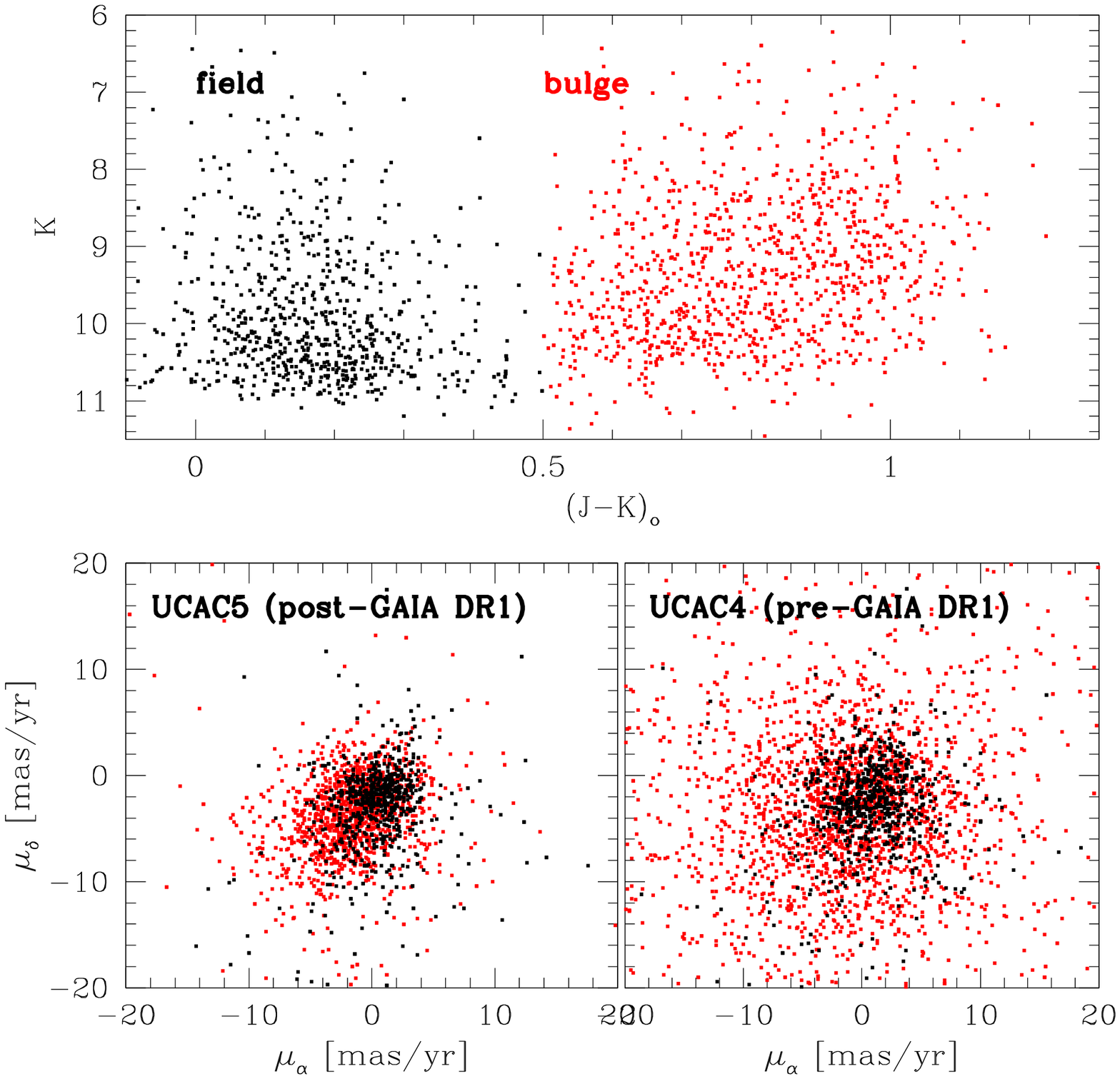}
\caption{{\it Top:} The color magnitude diagram of $\sim$2500 APOGEE giants that have 
UCAC5 proper motions with uncertainties smaller than 2 mas/yr.  {\it Bottom:}  The UCAC5 (left)
and UCAC4 (right) proper motions of the APOGEE giants with proper motion uncertainties 
smaller than 2 mas/yr.  Already with Gaia DR1, a sharper kinematic view of the bulge than previously feasible,
is possible.}
\label{fig:pms_bul}       
\end{figure*}

Figure~\ref{fig:pms_bul} shows the proper motions of the giants in the direction of the bulge surveyed in 
APOGEE DR13 post- and pre-Gaia DR1.  Before Gaia DR1 was available, the proper motions 
were not of the quality that allowed one to easily distinguish between a field and bulge
stellar population.  Color-magnitude diagrams can help in differentiating bulge stars from field
stars, since bulge stars tend to be redder, but the large and variable extinction and the sheer number of
field stars along the line of sight toward the bulge, makes color cuts not always reliable.

With the precise positions of stars measured in Gaia DR1, significant improvements 
in the astrometric solutions were able to be obtained, leading to the release of UCAC5.
For over 107 million stars, now proper motions with typical accuracies of 1 to 2 mas/yr ($R$= 11 to 15 mag), 
and about 5 mas/yr at 16th mag, exist.  With Gaia, we are approaching the possibility of using proper motion
information to separate the high degree of disk contamination from the bulge.

\section{Spatial structure of the RR Lyrae star population toward the Galactic bulge}

RR Lyrae-type variable stars can be found everywhere in the Milky Way,
but they are particularly numerous in the Galactic bulge.
Historically, \cite{vanGent1932,vanGent1933}
was the first who noticed that RRLs observed close
to the central regions of the Milky Way concentrate toward the Galaxy
center. More than a decade later, \cite{Baade1946}
in a relatively unobscured area, today called ``Baade's Window" in his honor,
found a strong predominance of RRLs
indicating a presence of Population II stars in the central area
of the Milky Way. He assumed that the center of this population
coincides with the Galactic center and assessed the distance to the 
center of the Galaxy using RRLs, obtaining a distance
of 8.7~kpc \citep{Baade1951}.

Until the early 1990s about one thousand RRLs
inhomogeneously distributed over the Galactic bulge were known.
Following the advent of massive photometric surveys, particularly
focused on searches for microlensing events, the number
of new RRLs toward the bulge has increased.
215 such objects were discovered during the first phase
of the OGLE
\citep{udalski+92}, conducted on the 1.0-m Swope telescope
at Las Campanas Observatory, Chile, in years 1992--1995.
About 1800 RR Lyrae pulsators were detected by the MACHO
microlensing survey \citep{Alcock_etal1995} which used the
1.27-m Great Melbourne Telescope at the Mount Stromlo Observatory,
Australia, in years 1992--1999. \cite{Alcock_etal1998} examined
mean magnitudes and colors of the new pulsators and found that
the bulk of the population is not barred. Only stars located in the
inner fields closer to the Galactic center ($l<4^{\circ}$, $b>-4^{\circ}$)
seem to follow the barred distribution observed for intermediate-age
red clump giants \citep{Stanek_etal1994}. \cite{Minniti_etal1998}
used this sample to show that between about 0.3 kpc and 3 kpc from
the Galactic center, the spatial density distribution of RRLs
can be represented by a power law of an index of $-3.0$.

Analysis of the data from the second phase of the OGLE project (OGLE-II),
conducted on the dedicated 1.3-m Warsaw Telescope at Las Campanas
Observatory in years 1997--2000, brought a much larger set of 
2713 RRLs \citep{Mizerski2003}. Based on the sample of 1888 fundamental-mode
RRLs from OGLE-II, \cite{Collinge_etal2006} robustly detected
the signature of a barred structure in the old population within the inner
$\pm3^{\circ}$ of Galactic longitude. Later, about 3000 fundamental-mode
RRLs from the MACHO database were used to investigate
the metallicity distribution of these stars \citep{KunderChaboyer2008} and
interstellar extinction toward the Galactic bulge \citep{Kunder_etal2008}.
It was determined that bulge variables have the average metallicity
[Fe/H]$=-1.25$ dex, with a broad range from [Fe/H]$=-2.26$ to $-0.15$ dex,
on the \cite{ZinnWest1984} metallicity scale. \cite{KunderChaboyer2008}
searched for the evidence of the Galactic bar and found a marginal
signature of a bar at Galactic latitudes $|b|<3.5^{\circ}$. The absence
of a strong bar in the RR Lyrae population clearly indicated that they
represent a different population than the metal-rich bulge.
However, the shape of this population was far from being fully known.
More data covering preferably the whole bulge area were needed.

In 2001, the OGLE project started its third phase with a new mosaic
eight-CCD camera attached to the Warsaw Telescope.
One of the OGLE-III results was the release of a collection
of 16,836 RRLs found in an area of 69 deg$^2$
mostly south of the Galactic equator \citep{Soszynski_etal2011}.
The collection composed of 11,756 fundamental mode (RRab) stars,
4989 overtone pulsators (RRc), and 91 double-mode (RRd) stars.
OGLE provided time-series photometry in two standard filters: $V$ and $I$.
This sample was promptly analyzed by \cite{Pietrukowicz_etal2012}
who demonstrated that the bulge RRLs form
a metal-uniform population, slightly elongated in its inner part.
The authors found that the photometrically derived metallicity
distribution for RRab stars is sharply peaked at
[Fe/H]$=-1.02\pm0.18$ dex with a dispersion of 0.25 dex,
on the \cite{Jurcsik1995} metallicity scale. This result agreed
very well with the one from \cite{KunderChaboyer2008},
since the \cite{Jurcsik1995} scale is shifted roughly by $+0.24$ dex
with respect to the \cite{ZinnWest1984} scale. 

\cite{Pietrukowicz_etal2012} also estimated the distance
to the Milky Way center based on the bulge RRLs 
to be $R_0=8.54+/-0.42$ kpc. Here, the theoretical
period-luminosity-metallicity (PLZ) relations in $V$ and $I$ bands
published by \cite{Catelan_etal2004} were used; the zero points of these
relations were calibrated to the data obtained for the well-studied
representative globular cluster M3 \citep{Catelan2004}.
In their analysis, \cite{Pietrukowicz_etal2012} made a simple
assumption on a linear relation between $I$-band extinction
$A_I$ and reddening $E(V-I)$. At that time it was the only reasonable
way to unredden mean magnitudes of the RRLs.
They showed that, for RRab stars as well as for RRc stars, in the
inner regions ($|l|<3^{\circ}$, $|b|<4^{\circ}$) the old population indeed
tends to follow the barred distribution of the bulge red clump giants.

A year later \cite{dekany+13} combined optical and near-infrared
data for the OGLE-III bulge RRLs to study
the RRL spatial distribution. The authors used mean $I$-band magnitudes
from OGLE and $K_s$-band magnitudes from the near-infrared VISTA
Variables in the V\'ia L\'actea (VVV) survey \citep{minniti_vvv}.
VVV was one of the ESO (European Southern Observatory) public surveys
carried out on the 4.1-m Visible and Infrared Survey Telescope for
Astronomy (VISTA) in years 2010--2015. Observations were taken in
$ZYJHK_s$ filters and included the Milky Way bulge and an adjacent
section of the Galactic plane, covering a total area of about 562 deg$^2$
\citep{Saito_etal2012}. The monitoring campaign was conducted only in
the $K_s$ band. 

The approach applied by \cite{dekany+13} is expected to bring
more precise results than the ones based on optical data alone. 
That is because PLZ relations have decreasing
metallicity dependence toward longer wavelengths and measurements
in near-infrared wavebands are much less sensitive to interstellar
reddening than optical ones. \cite{dekany+13} concluded
that the population of RRLs does not trace a strong bar,
but have a more spheroidal, centrally concentrated distribution
with only a mild elongation in its very center at an angle
$i=12.5^{\circ}\pm0.5^{\circ}$ with respect to the line of sight
from the Sun to the Galactic center.

The fourth phase of the OGLE project (OGLE-IV) \citep{Udalski_etal2015},
which was launch in 2010, covers practically the whole bulge in $V$
and $I$ passbands. With the installation of a 32-CCD mosaic camera of
a total field of view of 1.4 deg$^2$ OGLE became a truly wide-field
variability survey. The OGLE-IV collection of RRLs toward
the Galactic bulge was released by \cite{Soszynski_etal2014}.
This collection contains data on 38,257 variables detected over
182 deg$^2$: 27,258 RRab, 10,825 RRc, and 174 RRd stars.
The survey also includes the central part of the Sagittarius Dwarf
Spheroidal Galaxy with the globular cluster M54 in its core.

Analysis of this set of data was undertaken also by the OGLE team
and presented in \cite{pietrukowicz+15}. Due to some practical
reasons the analysis was based only on RRab-type pulsators, and the
part closest to the Galactic plane ($|b|$$<$3) was avoided.
RRab stars are more numerous. On average, they are intrinsically
brighter in the $I$-band and have higher amplitudes than RRc stars.
What is extremely important, RRab variables with their characteristic
saw-tooth-shaped light curves, in comparison to nearly sinusoidal
light curves of RRc stars, are harder to overlook. This makes
the searches for these type of variables more likely to yield better
completeness ratios.
Another very practical property of RRab stars is that based on the
pulsation period and shape of the light curve one can assess
metallicity of the star \citep{Jurcsik1995,JurcsikKovacs1996,Smolec2005}.

\begin{figure*}
\begin{center}
\includegraphics[width=0.46\textwidth]{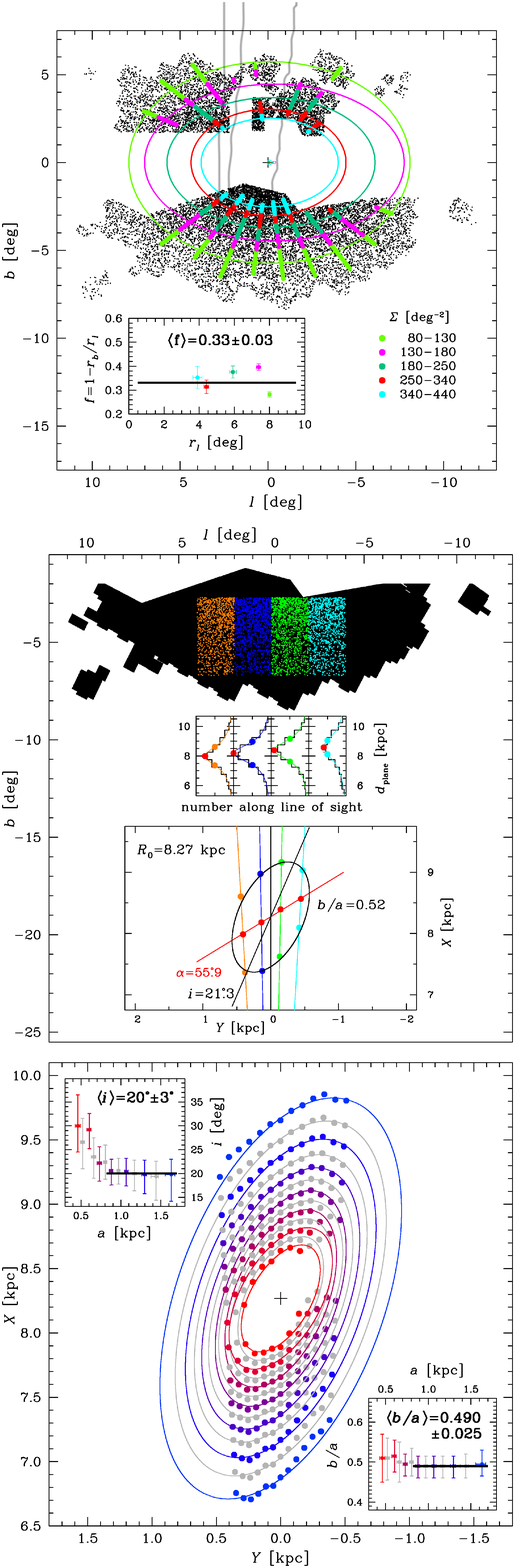}
\caption{Results of the analysis of density distribution of OGLE-IV
bulge RRab stars \citep{pietrukowicz+15}.
Upper panel: Constant surface density lines in the sky are well
represented by ellipses with a mean flattening $f=0.33\pm0.03$.
Middle panel: The maxima of the density distributions along four
selected lines of sight clearly get closer to us
with the increasing Galactic longitude. This strongly indicates
the presence of a tilted axis in the plane.
Lower panel: In the projection onto the Galactic plane, points of
the same density level form inclined ellipses. Conclusion:
the old bulge population has the shape of a triaxial ellipsoid with
the major axis inclined to us at an angle $i=20^{\circ}\pm3^{\circ}$.}
\label{bulgeRRstructure}
\end{center}
\end{figure*}

\cite{pietrukowicz+15} found again that the spatial
distribution of the inner bulge RRLs trace closely
the barred structure formed of intermediate-age red clump giants.
According to the most recent models of the Galactic bar, it is close
to being a prolate ellipsoid. Based on OGLE-III data \cite{cao+13}
found the following axis ratios and inclination of the major axis:
1.0:0.43:0.40, $i=29.4^{\circ}$. A similar result was obtained by
\cite{wegg+13} using VVV red clump data: 1.0:0.63:0.26,
$i=27^{\circ}\pm2^{\circ}$. This time in their analysis,
\cite{pietrukowicz+15} dereddened mean $I$-band magnitudes of
RRLs using a new relation derived by \cite{Nataf_etal2013}.
The relation was based on optical measurements from OGLE-III and
near-infrared measurements from 2MASS and VVV for bulge red clump giants.
After this correction the obtained distance distribution to the bulge
RRLs turned out to be smoother in comparison with
the previously used simple linear relation. They found the maximum
of the distribution or distance to the Galactic center
at $R_0=8.27\pm0.01({\rm stat})\pm0.40({\rm sys})$~kpc, which is
in very good agreement with estimates from other measuring methods.
\cite{pietrukowicz+15} showed that the spatial distribution
of the bulge RRLs has the shape of a triaxial ellipsoid
with proportions 1:0.49:0.39 and the major axis located in the
Galactic plane and inclined at an angle $i=20^{\circ}\pm3^{\circ}$
to the Sun-(Galactic center) line of sight (see Figure \ref{bulgeRRstructure}).

Differences between the \citet{dekany+13} results and those from OGLE-IV
are the following:
(1)  \citet{dekany+13} used a smaller sample of about 7700 RRab
stars from the previous phase of the OGLE survey (OGLE-III),
while the OGLE-IV collection \citep{Soszynski_etal2014} contain
nearly 27,300 pulsators of this type. 
(2)  The \citet{dekany+13} results are based on a lower number of
collected $K_s$-band measurements per light curve (about 40 by 2013),
which could affect the average infrared brightness of the variables
and could slightly smear the observed structure.  However, the amplitude
variation in the $K_s$-band is a factor of $\sim$3 smaller than that in
the $I$- or $V$-band.
(3)  Because the \citet{dekany+13} results uses also infrared magnitudes,
the de-reddening process of the RRLs differers from that adopted in
\cite{pietrukowicz+15}.
Resolution of the discrepancy between the OGLE-IV RRLs presented in
\cite{pietrukowicz+15} and the VVV RRLs presented in  \citet{dekany+13} is
ongoing.

The obtained sharp ellipsoidal shape 
does not depend on the real final distance to the studied objects.
The true inclination angle as well as the axis ratios may be
slightly different than reported. These values may also change with
the galactocentric distance. This will be known once almost all
RRLs (or at least RRab-type variables) are detected from
the inner bulge to the outer Galactic halo. Unfortunately,
searches for RRLs in obscured Galactic plane regions
are very difficult. That is because near-infrared light curves
of these pulsating stars have often symmetric, nearly sinusoidal shape.
If the number of data points per light curve is small and the time
coverage too short, RRLs can be easily confused with
other variables, particularly with contact eclipsing binaries
and spotted variables. However, the first detection of RRLs
in the vicinity of the Galactic center, in the so-called nuclear
bulge has been made \citep{minniti+16, dong17}.

A clear result from the analysis of the OGLE-IV bulge RRab
variables is that their spatial density distribution in the
galactocentric distance range from about 0.2 kpc to 2.8 kpc
can be described as a single power law with an index of $-2.96\pm0.03$.
\cite{pietrukowicz+15} also was not able to see an
X-shaped structure in the RRLs as it is observed in the case
of bulge red clump giants \citep{nataf+10,McWilliamZoccali2010}.
This is expected, as it was found that only metal-rich bulge populations have this
feature \citep{ness+12}.

Another discovery by \cite{pietrukowicz+15} is that RRab stars
form two (or even more) very close sequences in the period--amplitude
(or Bailey) diagram. This is interpreted as the presence of multiple old
populations being likely the result of mergers in the early history of the
Milky Way. So far, there are no hints that the two major old populations
have different structure. They seem to be well mixed together.
\cite{LeeJang2016} suggest that the observed period shift between
the sequences can be explained by a small difference in the helium abundance.

Recently, \cite{perezvillegas17} used N-body simulations
to investigate the structural and kinematic properties
of the old population in the barred inner region of the Galaxy.
They showed that the RR Lyrae population in the bulge is consistent with
being the inward extension of the Galactic metal-poor stellar halo, as
suggested by \cite{Minniti_etal1998} and \cite{Pietrukowicz2016}.
\cite{perezvillegas17} followed the evolution of the metal-poor
population through the formation and evolution of the more massive bar
and boxy/peanut bulge and found the density distribution to change from
oblate to triaxial. They found that at the final time of the simulations
(after 5 Gyr), the axis ratios of the triaxial old stellar halo
in its inner part reached $b/a\sim0.6$ and $c/a\sim0.5$.
The ratios increase with the distance from the center to roughly
1.0 and 0.7, respectively, at a distance of 5 kpc. These results are
very consistent with the observations of the bulge RRLs from
OGLE-IV and also spectroscopic observations of old stars in the Milky Way
halo by the SDSS survey. According to the latter studies, the Galactic halo
has an oblate shape \citep{Juric_etal2008, carollo08, kinman12}. Ongoing 
photometric surveys, such as Gaia, VVV eXtended (VVVX) and OGLE-IV with 
extended coverage of the Galactic bulge
and disk will continue in completing the picture of the old bulge population
drawn by RRLs.


\section{The Galactic bulge: 3D structure, chemical composition, and age traced by its old stellar population}
\label{sec:intro}

The formation and evolution of galaxies is still a heavily debated question of modern astrophysics. 
As one of the major stellar components of the Milky Way, the bulge provides critical and unique insights 
on the formation and evolution of the Galaxy, as well as of the external galaxies. Indeed with the current 
observational facilities, it is the only bulge in which we are able to resolve stars down to the old main sequence 
turnoff, hence providing accurate studies of the stellar populations in almost every evolutionary stage. 
The physical, kinematics and chemical properties of the stellar populations in the bulge allow us to discriminate 
among various theoretical models for the formation and evolution of bulges at large, setting tight constraints 
on the role that different processes (i.e. dynamical instabilities, hierarchical mergers, gravitational collapse) may 
have taken place.

However, this advantage comes with the need of covering a large area of the sky ($\rm \sim 500\,deg^2$). 
Therefore, to understand the global properties of the bulge one should look for reliable tracers of distance 
and age that can be easily observed in any region of the sky in the bulge direction. In this framework, 
Red Clump stars and RRLs play a crucial role, and indeed over the decades studies of
these stars have been essential to build our current knowledge of the Galactic bulge. The 
present section put together by Elena Valenti is not intended to be a complete review of the 
Galactic bulge properties but an overview of  the bulge structure, chemical composition, and age based on 
the observational results on RC stars and RRLs.  

The age\--metallicity relation and star formation history of the RCs in the bulge is not well-known, 
so an error of up to $\sim$0.3\,mag can be introduced when trying to use these stars to derive 
a distance.  At the distance of the bulge this translates 
to $\sim$1 kpc.  However, considering that overall the bulge is metal rich and old ($\gtrsim$10\,Gyr), one 
could use the corresponding population effect correction and therefore reduce the error on the distance.
RRLs represent, on the other hand, much more accurate distance candles.  Unlike the RC stars, they trace 
univocally the oldest stellar population of any given complex system (see \S2 above by P. Pietrukowicz). 

In \S\,\ref{sec:3D}, a global view of the three\--dimensional structure of the bulge addressing 
the observational evidences is presented, that leads to determination of the bar properties, the 
stellar density and the mass. The chemical composition and age of the bulge stellar populations are 
reviewed in \S\,\ref{sec:chemical}, and \ref{sec:age}, respectively.  Finally, \S\,\ref{sec:end} summarizes the 
main stellar properties that combined define our current knowledge of the Milky Way bulge, 
and the pieces of evidence that are still lacking in order to improve and possibly complete the global picture 
are highlighted.  Note that occasionally up to date versions of relevant figures obtained using state 
of the art observational data are presented. 
 
\subsection{The 3D structure}
\label{sec:3D}

\subsubsection{The bar boxy/peanut/X\--shaped structure}
Today it is well known that the Milky Way is a barred galaxy.  Although the first observational evidence 
of the presence of a bar in the innermost region of the Galaxy was presented by \citet{blitz+91} by using 
the stellar density profile at $2.4\mu $m by \citet{matsumoto+82}, its existence was hypothesized nearly 
20 years earlier. Indeed, to explain the departures from circular motions seen in the HI line profile 
at 21 cm, \citet{deVaucouleurs64} suggested for the first time that the Milky Way could host a bar 
in its inner regions.  After then, over the decades, many different tracers \-- e.g. gas 
kinematics \citep{binney+91}, stellar surface profile \citep{weiland+94,dwek+95,binney+97}, 
microlensing experiments \citep{udalski+00,alcock+00}, and OH/IR and SiO maser 
kinematics \citep{habing+06} \-- have been used to confirm the presence of the bar and to constrain 
its properties. 

Still, the strongest observational evidence for the presence of the bar comes from the use of RC stars 
as standard candles to deproject the stellar density distribution in the Galaxy inner region. By using the 
color\--magnitude diagram (CMD) derived from the OGLE \citep{udalski+92} photometry in  
Baade's Window $(l=-1^{\circ}, b=-3.9^{\circ})$ and in two additional fields at $(\pm-5^{\circ}, -3.5^{\circ})$, 
\citet{stanek+94} found that the mean RC magnitude at positive longitudes was brighter than the one 
observed at negative longitude. Under the assumption that there is no continuous metallicity and age 
gradient along the longitude, the observed change in mean RC magnitude across the field was interpreted 
in terms of distance. Stars at positive longitudes are brighter, hence closer, than those at negative longitudes. 
By using a triaxial model for the bulge, \citet{stanek+94} derived a bar pivot angle of $\Theta = 45^{\circ}$. 

Following this pioneering work, many studies \citep[see][]{stanek+97,bissantz+02,babusiaux+05,
benjamin+05,nishiyama+05,rattenbury+07,lopez-corredoira+07,cabrera-lavers+08,cao+13,wegg+13} 
have used RC stars to constrain triaxial bar models. Table\,\ref{tab:bar} lists the axis scale lengths 
and orientation angles of the bar as derived by \citet{rattenbury+07,cao+13,wegg+13}, which 
among all similar studies, are those considering the largest bulge area, therefore possibly 
providing more accurate estimates of the bar physical properties. 
\begin{table}
\caption{Axis scale lengths and orientation angle of the bulge main bar based on RC stars distribution.}
\label{tab:bar}       
\begin{tabular}{ccccccr}
\hline\noalign{\smallskip}
{\it x}$_0$ & {\it y}$_0$ & {\it z}$_0$ & $\Theta$  & Data & Area & Reference \\
(kpc) & (kpc) & (kpc) & (deg) & & (deg$^2$) &  \\
\noalign{\smallskip}\hline\noalign{\smallskip}
1.2 & 0.40 & 0.30 &24\--27 & OGLE-II  & $\sim$11 & \citet{rattenbury+07}\\
1.0 & 0.41 & 0.38 &29\--32 & OGLE-III & $\sim$90 &\citet{cao+13}\\
0.7 & 0.44 & 0.18 & 27 & VVV & $\sim$300 & \citet{wegg+13} \\
\noalign{\smallskip}\hline
\end{tabular}
\end{table}
\begin{figure*}
  \includegraphics[width=1\textwidth]{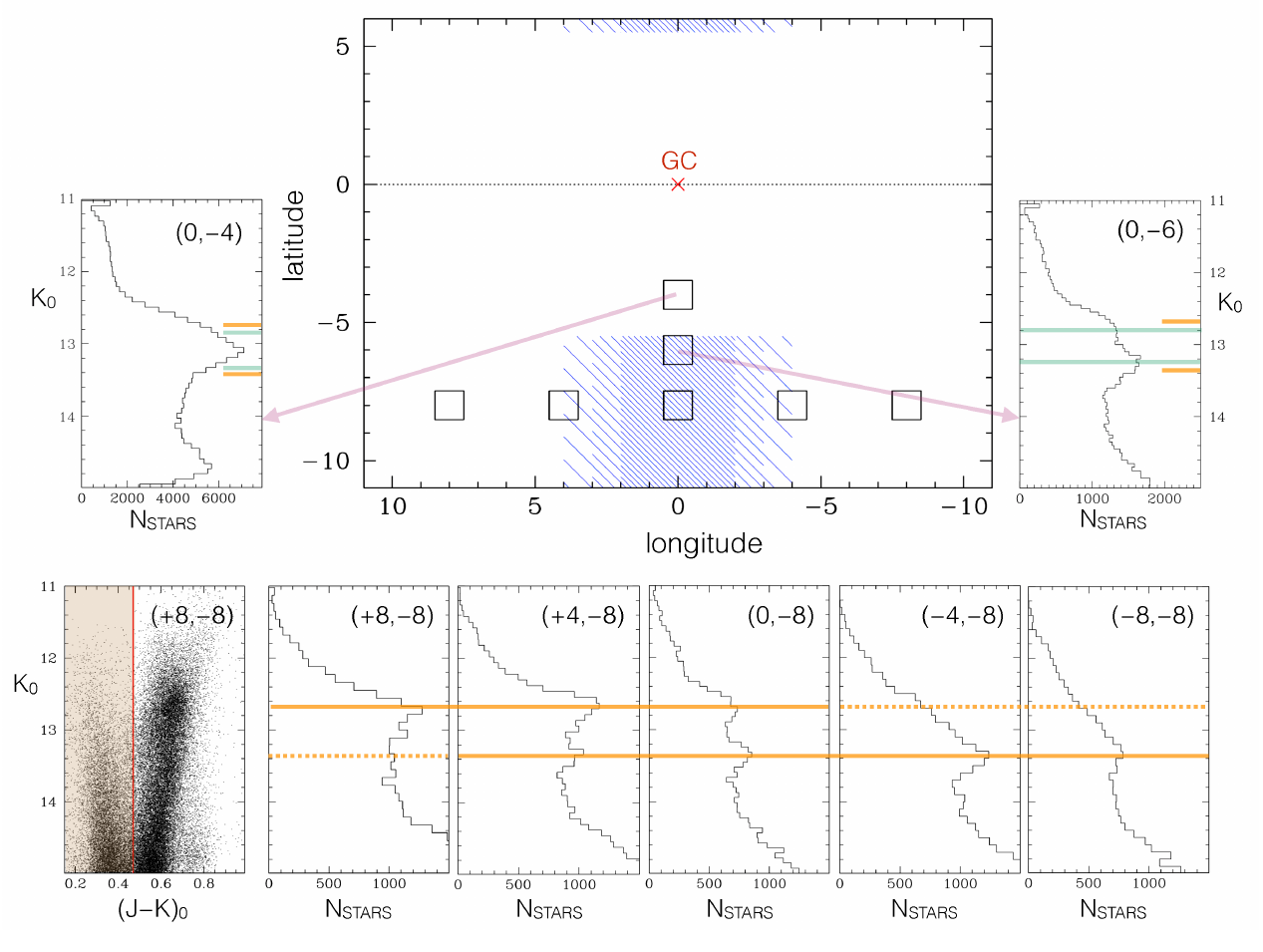}
\caption{Luminosity function of the bulge RGB and RC, in several fields. The large box (middle) marks 
the bulge area covered by the VVV survey, with the seven fields for which the RGB luminosity function 
is probed. The shaded blue rectangle refers to the region where the double RC is 
visible ($|b|>5^{\circ}, |l|<4^{\circ}$). The shading is finer where the double RC is clearly evident and 
becomes wider when one of the two RCs becomes significantly weaker than the other. The bottom 
left corner panel shows the CMD of the field at ($-8^{\circ}, +8^{\circ}$) as an example of the colour selection 
applied to exclude the main sequence of the foreground disc. Orange horizontal lines mark the 
magnitude of the two RCs in the five fields at $b=-8^{\circ}$ shown in the bottom panels. Green horizontal 
lines mark the magnitude of the two RCs at $b=-6^{\circ}$, shown at the top right. Only one RC is visible 
at $b=-4^{\circ}$ (Baade's Window), shown at the top left. 
Figure reproduced from \citet[][their Figure\,1]{gonzalez15}. }
\label{fig:Xshape}       
\end{figure*}
 
Owing to the RC stars distribution across the bulge, we now know that the bar has a 
boxy/peanut/X\--shape structure in its outer regions: a typical characteristic of all barred 
galaxies when seen edge\--on \citep[][for a recent review]{laurikainen+14,laurikainen+16}. 
\citet{mcwilliam10} and \citet{zoc10-rio} were the first who noticed that the distribution of 
the RC stars in some fields in the outer regions ($|b|>5^{\circ}$) along the bulge minor 
axis ($l=0^{\circ}$) was bimodal, suggesting the presence of a double RC. The observed 
split in the RC mean magnitude was then confirmed by \citet{manuXshape} and \citet{nataf+10} 
by using 2MASS and OGLE photometry, respectively. The authors explained the split in 
the RC as signature of two southern arms of an X\--shape structure crossing the line of 
sight.  An alternative explanation is that the double RC phenomenon is a 
manifestation of multiple populations observed in globular clusters (GCs) in the 
metal-rich regime \citep{lee15, joo17}. 

Shortly after, the 2MASS\--based 3D map of the RC distribution over a bulge area 
of $170\,deg^2$ by \citet{saito+11} confirmed the presence of the X\--shape and showed 
that the two RC over-densities were only visible in the outer bulge (i.e. $b<-5^{\circ}$, 
and $b>5^{\circ}$) along the deprojected minor axis ($|l|\leq5^{\circ}$, see also 
Figure\,\ref{fig:Xshape}).  
Thanks to the superior quality, in terms of photometric depth 
and spatial resolution, of the near\--IR Vista Variable in the Via 
Lactea \citep[VVV,][]{minniti_vvv,Saito_etal2012}  \citet{wegg+13} modelled the observed 
RC distribution across the whole bulge area ($\sim 300\,deg^2$) thus providing the first 
complete map of its X\--shape structure.  
Although not specifically obtained through the 
study of RC stars, it is worth mentioning in this context the latest work by \citet{ness+16} 
based on WISE images, in which the X\--shape nature of the Milky Way bulge revealed 
itself with unquestionable doubts (see their Figure\,2).

There is now a general consensus that the majority of the observed Milky Way bulge structure is a 
natural consequence of the evolution of the bar. The bar heats the disk in the vertical 
direction giving rise to the typical boxy/peanut shape. Dynamical instabilities cause bending 
and buckling of the elongating stellar orbits within the bar, resulting in an X\--shape when 
seen edge\--on \citep{raha+91,merritt+94,patsis+02,athanassoula+05,bureau+05,debattista+06}. 
However, the possible presence of a metal\--poor spheroid {\it embedded} in the boxy/peanut bulge 
seems to be suggested by a number of fairly recent studies investigating the correlation between 
RC stars chemical and kinematics properties, as well as  the spatial distribution of other stellar 
tracers, such as RRL, Mira and Type\,II Cepheids (T2C) variables. While the readers are 
referred to \S1 above for an overview on the bulge kinematics, 
here it is however worth mentioning that metal\--poor ([Fe/H]$\lesssim0$) stars in the Baade's 
window show negligible vertex deviation ($l_{\nu}\sim 0$) consistent with those of a spheroid. 
Conversely, metal\--rich ([Fe/H]$\gtrsim0$) stars exhibit significant  ($l_{\nu}\sim 40$) elongated 
motions typical of galactic bars \citep{babusiaux10}. In addition, based on the spectroscopic data 
provided by the ARGOS survey \citep[i.e. $\sim 14,000$ RC stars;][]{freeman13}, \citet{ness+12} 
demonstrated that only the distribution of metal\--rich stars shows the split in RC, a univocal 
signature of the bar X\--shape. On the other hand, metal\--poor stars show only a single RC peak.
The scenario in which the metal\--poor bulge stars, and therefore possibly the oldest population, 
do not trace the bar structure is also supported by the observed distribution of 
Mira \citep{catchpole+16}, RRLs \citep{dekany+13}, and T2C \citep{bhardwaj+17} variables. 
In particular, by using a combination of near\--IR and optical data from VVV and OGLE-III, \citet{bhardwaj+17} found 
that their sample of T2C population in the Galactic bulge shows a centrally concentrated spatial distribution, 
similar to OGLE-IV counterparts of metal\--poor RRLs from the VVV. 
Mira stars are also consistent with belonging to a boxy bulge\citep{lopezcorredoira17}.
It should be noted, however, that the spatial distribution of RRLs in the bulge is somehow still debated 
given that \citet{pietrukowicz+15} by using OGLE-IV data did confirmed the presence of a bar in their spatial distribution, 
although the extension, pivot angle and ellipticity of the structure traced by the RRLs are 
significantly smaller than that traced by RC stars (see \S2 above).

\subsubsection{The stellar density map}
\label{sec:density}

\begin{figure*}
  \includegraphics[width=1\textwidth]{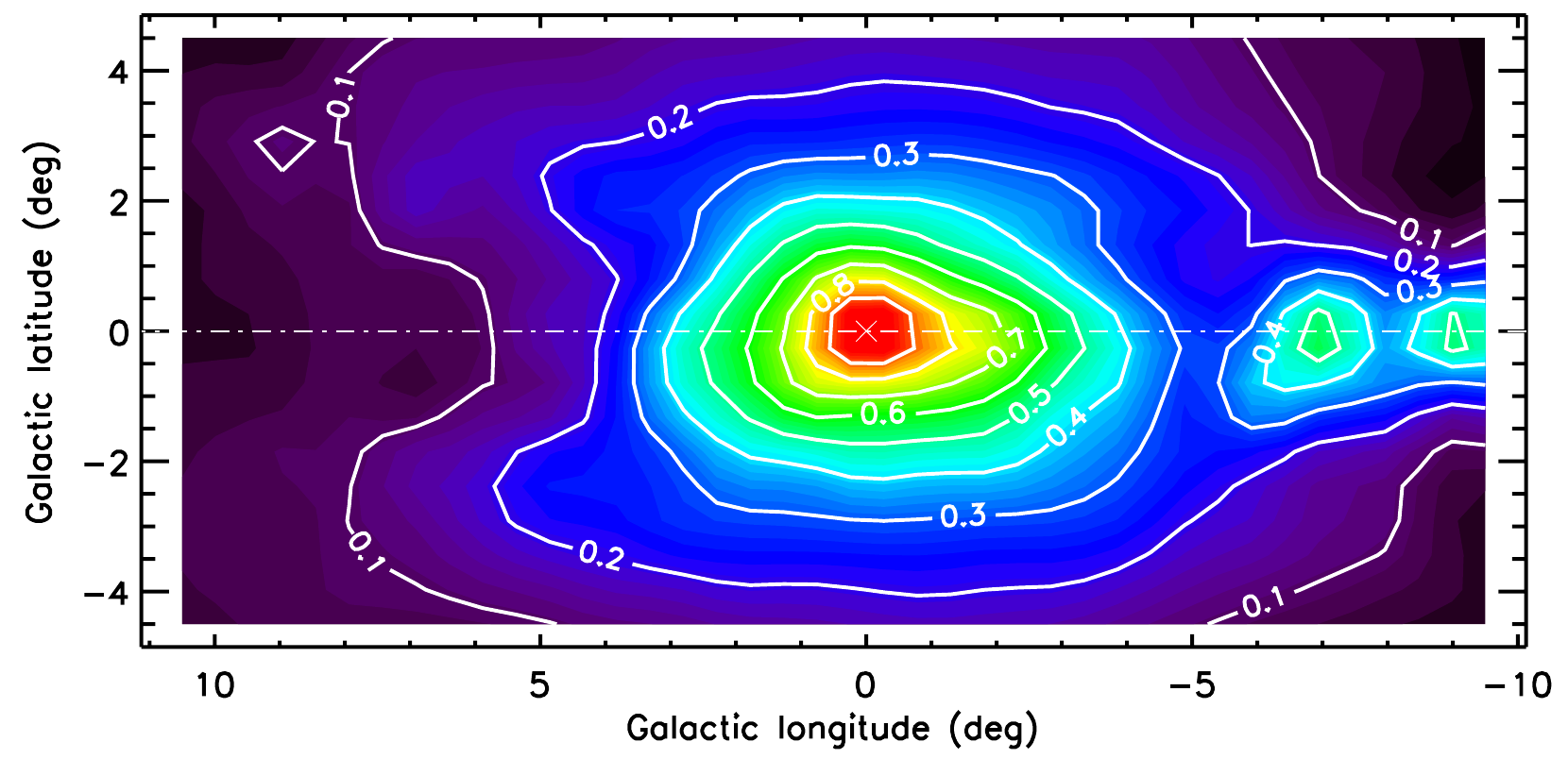}
\caption{Density map in the longitude\--latitude plane based on VVV RC star counts from \citet{valenti+16}. 
Star counts are normalized to the maximum ({\it Max}). Solid contours are isodensity curves, linearly spaced 
by 0.1\,$\times$\,{\it Max}\,deg$^{-2}$. Figure reproduced from \citet[][{\it The 3D structure of the Galactic bulge}, 
their Figure\,4]{zoccalivalenti16}. }
\label{fig:DensMap}       
\end{figure*}

More than two decades ago, \citet{weiland+94} presented the first low angular resolution map 
at 1.25, 2.2, 3.5 and 4.9\,$\mu$m of the whole Milky Way bulge based on the COBE/DIRBE data. 
After correction for extinction and subtraction of an empirical model for the Galactic disk, the derived 
surface brightness profile of the bulge was then used to study its global morphology and structure. 
However, more detailed investigations of the bulge innermost region (i.e. $|b|\leq5^{\circ}, |l|\leq10^{\circ}$) 
have been possible only recently thanks to the VVV survey.  \citet{valenti+16} presented the first stellar 
density profile of the bulge (see Figure\,\ref{fig:DensMap}) reaching latitude $b=0^{\circ}$. 
Specifically, by counting RC stars within the CMD as obtained from accurate PSF\--fitting photometry 
of VVV data and previously corrected for extinction by using the reddening map of \citet{gonzalez+11b}, 
they derived a new stellar density map that allowed to investigate the morphology of the innermost regions 
with unprecedented accuracy. As seen from Figure\,\ref{fig:DensMap}, the vertical extent of the 
isodensity contours is larger at $l>0^\circ$. This is an expected consequence of a bar whose closest 
side points towards positive longitude. The high stellar density peak in the innermost region 
(i.e. $|l|\leq1^\circ$ and $|b|\leq1^\circ$) spatially matches the $\sigma$\--peak found by GIBS, 
the kinematics survey of RC  \citep{zoccali14} and by Valenti et al. (2018, {\it in prep}). The 
stellar density maximum is found in the region $|l|\leq1^\circ$ and $|b|\leq0.5^\circ$, and slightly 
asymmetric with respect to the bulge minor axis.


\begin{figure}
\begin{center}
  \includegraphics[width=1\textwidth]{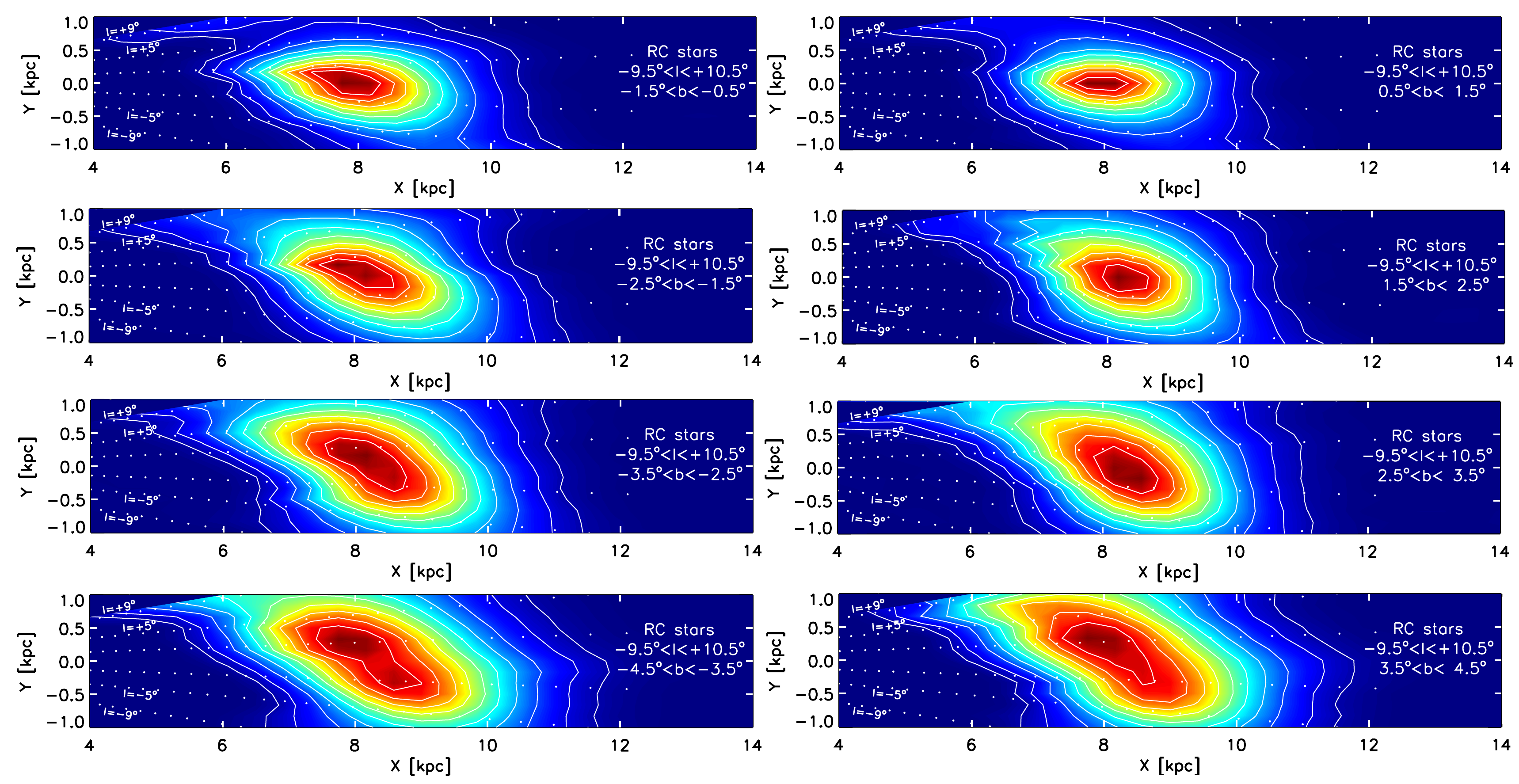}
\caption{ Deprojection of the RC density shown in Figure\,\ref{fig:DensMap} at different latitudes.}
\label{fig:RC_XY}   
\end{center}    
\end{figure}

The observed overall elongation of the density contours towards the negative longitudes is, nevertheless, found to be progressively less pronounced when moving closer to the Galactic plane. Indeed, as shown in Figure\,\ref{fig:RC_XY} the deprojected density maps become more and more spherically concentrated when RC stars at lower latitudes are considered, hence suggesting the presence of a quasi\--axisymmetric structure in the innermost region. This evidence supports the claim by \citet{gerhard+12} that: the variation in the RC slope at $b=\pm1^\circ$ and $|l|\leq10^\circ$ observed by \citet[][by using OGLE data]{nishiyama+05} and by \citet[][based on VVV photometry]{gonzalez11c}, and interpreted by these authors as evidence for the presence of a nuclear bar in the inner bulge, is instead caused by a variation of the stellar density distribution along the line of sight.

\begin{figure*}
 \includegraphics[width=1\textwidth]{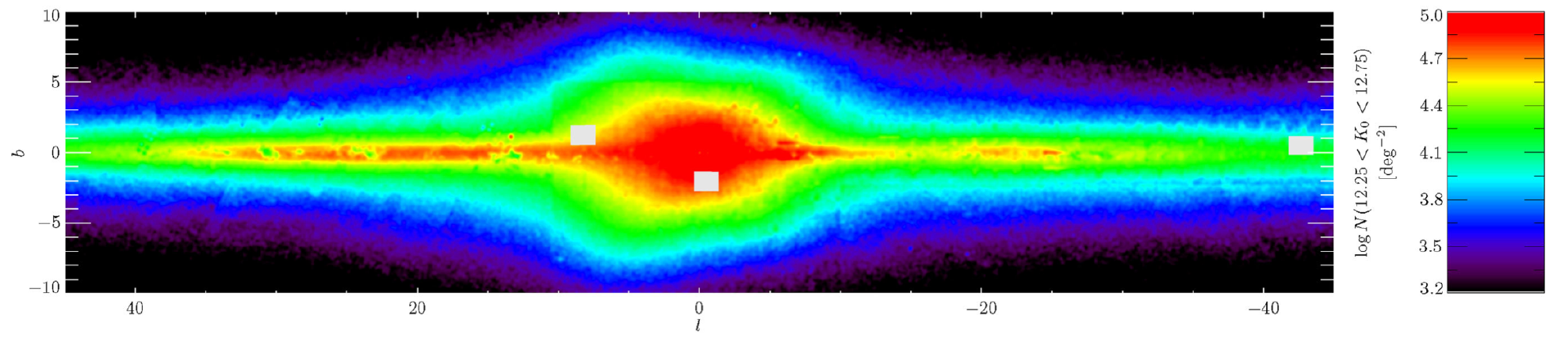}
\caption{Surface density map of RC stars based on VVV, UKIDSS and 2MASS $K_s$\--band photometry. Figure reproduced from \citet[][{\it Structure of the Milky Way's bar outside the bulge}, their Figure\,1]{wegg+15}.}
\label{fig:LongBar}       
\end{figure*}

By using a combination of UKIDSS, VVV, 2MASS and GLIMPSE data, \citet{wegg+15} presented the so far largest  (i.e. $\sim\,1900\,deg^2$) density map of the Milky Way and long bar based on RC stars (see Figure\,\ref{fig:LongBar}). A particularly interesting result of this study is that the orientation angle of the long bar constrained by the best\--fit model to the observed density map is consistent with that of the triaxial bulge (i.e. the main bar, $28^\circ \-- 33^\circ$). In other words, unlike several previous studies suggesting the presence of a long bar tilted by $\sim 45^\circ$ with respect to the Sun\--Galactic centre line \citep[i.e. in addition to the main bar;][]{benjamin+05,lopez-corredoira+07,cabrera-lavers+07,cabrera-lavers+08,vallenari+08,churchwell+09,amores+13}, the long bar with a semimajor axis of $\sim\,4.6$\,kpc in length as modelled by \citet{wegg+15} appears to be the natural extension of the bulge main bar at higher longitude. This result nicely fits the scenario proposed by \citet{martinez-valpuesta+11} and \citet{romero-gomez+11} based on N\--body simulations, where long and main bar are parts of the same structure. According to this, the boxy/peanut shape bulge would then {\it simply} be the central vertical extension of a longer and flatter single bar.

\subsubsection{The mass}
\label{sec:mass}
One of the fundamental question of Galactic astronomy is the determination of the mass distribution in the Milky Way because in general the mass of a given system is more likely the key element driving its evolution \citep[see][for a detailed review]{courteau+14}.
In this context, over the past three decades many studies addressed this specific question, deriving the dynamical mass of the Galaxy bulge either by matching the galactic rotation curve inside $\sim$1\,kpc, or by measuring the kinematics (i.e. velocity and velocity dispersion) of a variety of different tracers (i.e. stellar and gas). Then by using an observed luminosity profile (generally in K-band) one can derive the M/L ratio, which ultimately leads to the mass of the bulge \citep{sellwood+88}. 
Historically, the M/L ratio derived from fitting the rotation curve has been often found to be $\sim 2 \-- 3$, while the M/L ratio derived from stellar kinematics $\sim 1$.
Such discrepancy has been often explained using the argument that the mass derived from the rotation curve is overestimated because of the presence of large non\--circular motions that distort the rotation curve \citep{sofue96,yoshino+08}.
Indeed, the accuracy on the measurement of the rotation curve strongly depends on the accuracy on the distance to the Galactic centre and the solar circular velocity. In addition, the rotation curves as derived from H$\alpha$ and, in general from other gas tracers (i.e. HI, HII) are often influenced by non\--circular components (i.e. inflow, outflow, streaming motions), rather than an ordered (regular) circular motion. This inevitably led to very different results for the mass of the bulge.  \citet{chemin+15} recently reviewed the uncertainties and bias affecting the determination of the Milky Way rotation curves and the consequent effects on the derived mass distribution. Table\,\ref{tab:mass} lists a number of studies, together with the adopted observables/diagnostics, that over the years tackled the problem of deriving the bulge mass. Although the reader should refrain from considering Table\,\ref{tab:mass} a complete compilation, it is evident that the large spread in the listed values make the mass of the bulge still poorly constrained. Most estimates cluster to $1.5\times10^{10}M_\odot$, however a few authors found values as large as $3\times10^{10}M_\odot$ \citep{sellwood+88} or as small as $0.6\times10^{10}M_\odot$ \citep{robin+12}.
\begin{table}
\caption{Estimates of the bulge mass over the last three decades.}
\label{tab:mass}       
\begin{tabular}{ccr}
\hline\noalign{\smallskip}
Mass & Diagnostics & Reference \\
($\times10^{10}M_\odot$) &  &  \\
\noalign{\smallskip}\hline\noalign{\smallskip}
3.0 & Rotation curve \& M/L$\sim3$ & \citet{sellwood+88} \\
1.2 & 2.4\,micron map IRT Spacelab 2 mission & \citet{kent92}\\
2.0 & Stellar kinematics \& COBE brightness profile & \citet{zhao+94}\\
1.3 & COBE brightness profile & \citet{dwek+95} \\
1.6 & microlensing depth \& COBE brightness profile & \citet{han+95} \\
2.8 & Stellar kinematics \& COBE brightness profile & \citet{blum95} \\
2.4 & Star counts DENIS near\--IR (Besan\c{c}on model) & \citet{picaud+04} \\
1.8 & Gas rotation curve & \citet{sofue+09} \\
0.6 & star counts, kinematics and metallicity  (Besan\c{c}on model) & \citet{robin+12}\\
0.8 & Unified rotation curve, CS \& CO & \citet{sofue13} \\
1.8 & OGLE RC star counts & \citet{cao+13} \\
1.8 & M2M dynamical model \& VVV RC density & \citet{portail+15} \\
2.0 & VVV RC star counts \& observed IMF & \citet{valenti+16}\\
\noalign{\smallskip}\hline
\end{tabular}
\end{table}

In this context, a special mention is deserved for the two most recent works by \citet{portail+15} and \citet{valenti+16} which, although following different methodologies, they both use the distribution of the RC stars as derived by the VVV photometry.
\citet{portail+15} used made\--to\--measure dynamical model of the bulge, with different dark matter halo to match the stellar kinematics from BRAVA \citep{rich07,kunder12} and the 3D surface brightness profile derived by \citet{wegg+13}. Their best\--fit model is consistent with the bulge having a dynamical mass of $1.8\pm0.07\times10^{10}M_\odot$, with a dark matter content that varies with the adopted IMF. When the observed IMF of \citet{zoccali+00} is considered about $0.7\times10^{10}M_\odot$ (i.e. 40\%) of dark matter is required in the bulge region. In addition, they estimated that the total stellar mass involved in the peanut shape accounts for $\sim$20\% of the total stellar bulge mass.

\begin{figure}[h]
\begin{center}
  \includegraphics[width=0.5\textwidth]{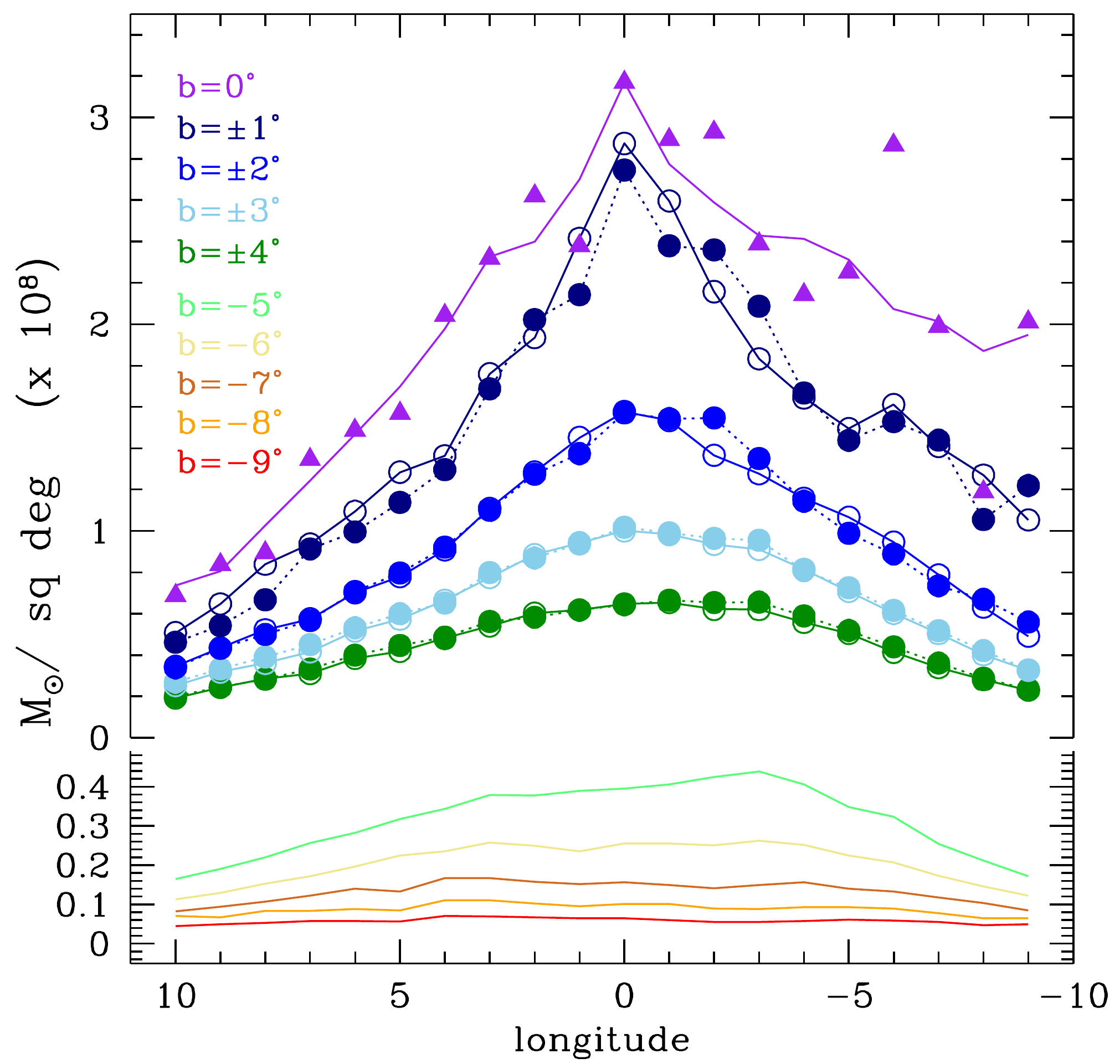}
\caption{ Stellar mass profile across latitude in each VVV field for fixed longitudes. Filled symbols with dotted lines refer to $b<0$, empty symbols with solid lines to $b>0$ fields. The lower part of the plot has an expanded y\--axis for a better display of the high negative longitudes, owing to the higher extinction, hence higher incompleteness in those fields. Figure reproduced from \citet[][{\it Stellar density profile and mass of the Milky Way bulge from VVV data}, their Figure\,5.]{valenti+16}.}
\label{fig:mass}       
\end{center}
\end{figure}

On the other hands, by scaling the observed VVV RC stellar density map (see Figure\,\ref{fig:DensMap}) with the observed bulge luminosity function from \citet{zoccali+00}, and \citet{zoccali+03}, \citet{valenti+16} provided the first empirical, hence no model\--dependent,  estimate of the bulge stellar mass. From the observed stellar mass profile shown in Figure\,\ref{fig:mass}, the authors estimated that the mass in stars and remnants of the Milky Way bulge in the region $|b|<9.5^\circ$ and $|l|<10^\circ$ is $2.0\pm0.3\times10^{10}M_\odot$.  

These two latest estimates are found compatible within the quoted errors, and they might be even more close when considering that the empirical estimate by \citet{valenti+16} refers to a larger volume that is not limited along the line of sight. 

\subsection{The chemical composition}
\label{sec:chemical}
Because the chemical content of any given stellar system retains crucial information to unveil its origin, formation and evolution \citep{mcwilliam16}, after the pioneering works of \citet{frogel+84} and \citet{rich88} several studies over the decades focussed on the determination of the bulge stars metallicity and abundances distribution to understand how the bulge formed.
What follows is not meant to be a comprehensive compilation of all such studies for which one should dedicated a entire single review, but rather a summary of our current knowledge of the chemical composition of the bulge based on the latest results from RC and RRLs.
\subsubsection{The metallicity distribution}
\begin{figure}[h]
\begin{center}
 \includegraphics[width=0.5\textwidth]{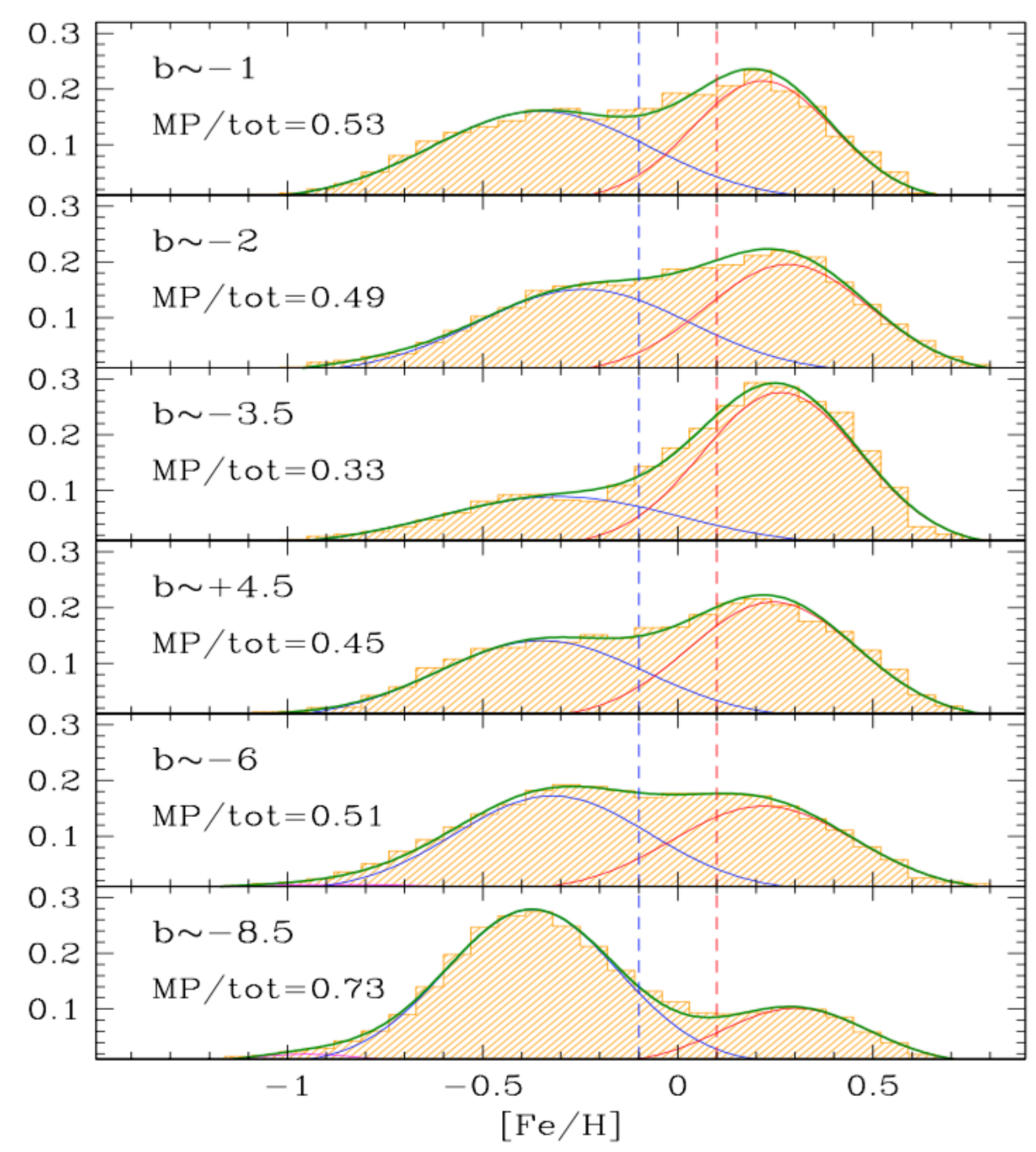}
 \caption{MDF of RC stars at constant latitudes. The fraction of metal\--poor stars, compared to the total, is given in each panel. Vertical dashed lines mark the limits of the metal\--poor and metal\--rich populations. Figure adapted from \citet[][{\it GIBS-III: MDFs and kinematics for 26 fields}, their Figure\,7]{zoccali17}.}
 \label{fig:mdf}
\end{center}
\end{figure}

\begin{figure*}
 \includegraphics[width=1\textwidth]{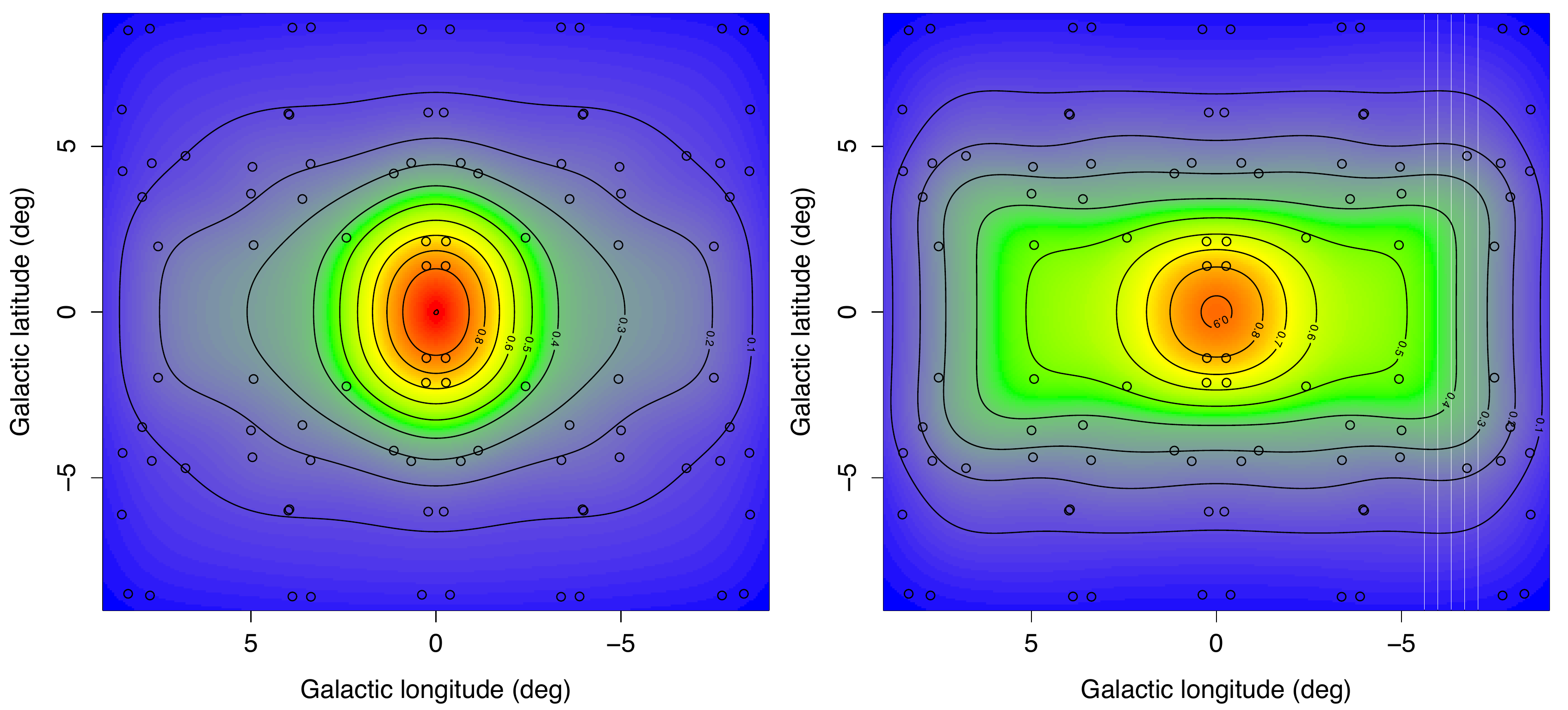}
 \caption{Density map of metal\--poor (left) and metal\--rich (right) RC stars obtained using the MDF of GIBS fields and the total number of RC stars from \citet{valenti+16}. Figure adapted from \citet[][{\it GIBS-III: MDFs and kinematics for 26 fields}, their Figure\,9]{zoccali17}.}
 \label{fig:MPRdens}
\end{figure*}

As emphasized by \citet{matteucci+99,ferreras+03}, the peak and shape of the metallicity distribution functions (MDF) provide important constraints on the IMF, star formation efficiency, as well as to the possible gas infall timescale.
However, until less than a decade ago, accurate MDF based on high\--resolution spectroscopy were available only for a handful number of sparse fields mainly located along the bulge minor axis  
\citep[see i.e.][, and reference therein]{mcwilliam94,fulbright07,rich07,johnson11,gonzalez+11d,hill11,rich+12}. The derived MDFs were consistent across various studies, which all agreed in finding the bulge population to be on average metal\--rich, although spanning a fairly broad metallicity range (e.g. $-1.5\lesssim$[Fe/H]$\lesssim+0.5$).
\begin{table}[h]
\caption{For each spectroscopic survey, the total number of stars, the total number of targeted fields and the region within the bulge covered by the observations are given.}
\label{tab:spec}       
\begin{tabular}{cccc}
\hline\noalign{\smallskip}
Survey & Total RC stars & Number of fields & Bulge region  \\
\noalign{\smallskip}\hline\noalign{\smallskip}
ARGOS & 14,000 & 27 & $-10^\circ < b < -5^\circ$, $|l|\lesssim+30^\circ$ \\
GIBS &5,500 & 26 & $-8^\circ < b < -1^\circ$, $b=+4, |l|\lesssim+30^\circ$ \\
Gaia-ESO & 1,200 & 5 & $-10^\circ < b < -4^\circ$, $-10^\circ < l < 7^\circ$ \\
\noalign{\smallskip}\hline\noalign{\smallskip}
\noalign{\smallskip}\hline
\end{tabular}
\end{table}
Our comprehension of the MDF of the bulge has improved tremendously thanks to three spectroscopic surveys, namely ARGOS \citep{freeman13}, GIBS \citep{zoccali14}, and ESO\--Gaia \citep{rojas-arriagada+14}, that all together have provided spectra for more than 20,000 RC stars across most of the inner and outer bulge regions (see Table\,\ref{tab:spec} for further details).
The MDF derived by these surveys confirmed previous results although extending them on a much larger area. The mean bulge population across all fields is metal\--rich with a small fraction of stars with [Fe/H]$>+0.5$\,dex and [Fe/H]$<-1.5$\,dex. Only in the outermost fields ($b>-7^\circ$, $|l|>10^\circ$) observed by ARGOS the MDF reaches metallicity as low as $\sim$-2.5\,dex. In addition, a mild vertical gradient is found when considering the mean metallicity of each field, such as the metallicity increases moving inwards along the bulge minor axis, hence confirming what suggested previously by \citet{minniti+95}, and \citet{zoccali+08}. However, thanks to statistically robust target samples a detailed study of the MDF shape has been possible for the first time, revealing the presence of multiple components. The observed overall metallicity gradient is therefore explained as a consequence of  the presence of two \citep[see][]{zoccali17,rojas-arriagada+14} or more \citep[see][]{ness13b} components with different mean metallicity. As evident from Figure\,\ref{fig:mdf}, the variations of the relative contribution of these components  across the fields (i.e. metal\--rich stars component becoming progressively less prominent towards the outer region) mimic the observed gradient. However, \citet{zoccali17} found also that at latitudes smaller than $|b|=3^\circ$ the metal\--poor component becomes important again (see first 2 top panels of Figure\,\ref{fig:mdf}), its relative fraction increases again close to the plane. To further investigate the spatial distribution of the two components they mapped their distribution by coupling the relative fractions derived by GIBS with the bulge stellar density from \citet{valenti+16}. The result, shown in Figure\,\ref{fig:MPRdens},  demonstrates that the metal\--poor component has a spheroid\--like spatial distribution, versus a boxy distribution of the metal\--rich component. In addition, the metal\--poor component shows a steeper radial density gradient.

Although as mentioned before the bulge kinematics is the subject of another paper in this 
volume (see \S1 above by A. Kunder), here I will only briefly mention that the two components were found to have also different kinematics. Indeed, as already found by the BRAVA \citep{kunder12} and ARGOS survey \citep{ness13}, in the outer bulge ($|b|>4^\circ$) the metal\--poor component has a higher radial velocity dispersion compared to the metal\--rich one, at all longitudes. However, \citet{zoccali17} showed that such behavior is reversed in the inner bulge. Specifically, the velocity dispersion of the metal\--poor stars at $b=-3.5^\circ, -2^\circ$ becomes similar to that of the metal\--rich counterpart, and progressively becomes smaller at $b=-1^\circ$.\\
While the chemical abundances of the RC stars in the bulge is one of the topics that received more attention in the recent years, the number of studies addressing the chemical content of the oldest bulge population, such the RRLs, is still very limited. Perhaps mostly due to the observational challenges that spectroscopic observations of RRL face, as of today there is no high\--medium resolution spectroscopic measurements of a sizeable sample of RRLs in the bulge. In K\--band RRLs are in general about 0.5 mag fainter than RC stars, hence their brightness makes them suitable targets at high resolution only with 4\,m\--class telescope or above, depending on the bulge region. In addition, because they are much less numerous than RC stars, and so more sparsely distributed RRLs are not even suitable targets for the vast majority of the current multiplexing spectrograph facilities. An additional complication is the fact that the metallicity derived from the line equivalent width measurements strongly depends upon the pulsation phase at which the star was observed. This necessarily implies a good knowledge of the variables.  All of these factors make their observations very telescope time consuming.\\
As of today the only spectroscopic study of a sizeable sample of bulge RRLs has been presented by \citet{walker91}, who derived the MDF of 59 RRLs in the Baade's window. The individual star metallicities were derived through the $\Delta$S method \citep[i.e. low\--resolution, see][for a detailed description of the $\Delta$S method]{suntzeff+91} and their distribution is found to cluster around $\rm [Fe/H] = -$1\,dex. Although the MDF is relatively broad, spanning a range of about 1\,dex, $-1.7\lesssim$[Fe/H]$\lesssim-0.5$, its very sharp peak accounts for $\approx$80\% of the entire sample.  Based on the derived MDF, the authors concluded that the RRL are being produced by the metal\--poor tail of K giants distribution (see also Figure~\ref{fig:mdf}).\\
Recently, \citet{pietrukowicz+15} provided a photometric MDF based on more than 27,000 RRLs from the OGLE-IV catalogs and located in the bulge region between $|l|\lesssim10^\circ$ and $-8^\circ\lesssim b\lesssim-2^\circ$, $+2^\circ\lesssim b\lesssim+5^\circ$. The photometric MDF is much broader than the spectroscopic one, as it spans mostly the range $-2.5\lesssim$[Fe/H]$\lesssim+0.5$, although the peak is found at the same metallicity, [Fe/H]=--1\,dex. The authors showed that there is no correlation between the distance and the shape of the MDF, however they find a very mild, but statistically significant,  radial metallicity gradient (i.e. the metal\--rich population increases towards the centre). Based on the analysis of the Bailey diagram (i.e. period\--amplitude diagram) the authors argue for the existence of two different population of RRLs with likely different metallicity, similar to the bulge RC counterparts. However, it should be mentioned that unlike what is observed in the MDF of RC stars, these 2 populations of RRLs with different metallicity do not probably change in relative fraction given that the global MDF conserves its shape throughout the total covered bulge area. Moreover, because of the lack of RRLs spectroscopic measurements in the metal\--rich regime, $[Fe/H]>-0.5$\,dex \citep[see][]{walker91}, one should refrain from drawing any firm conclusion from the available RRL MDF.
\subsubsection{The $\alpha$\--elements abundances}
\begin{figure}[h]
\begin{center}
\includegraphics[width=1\textwidth]{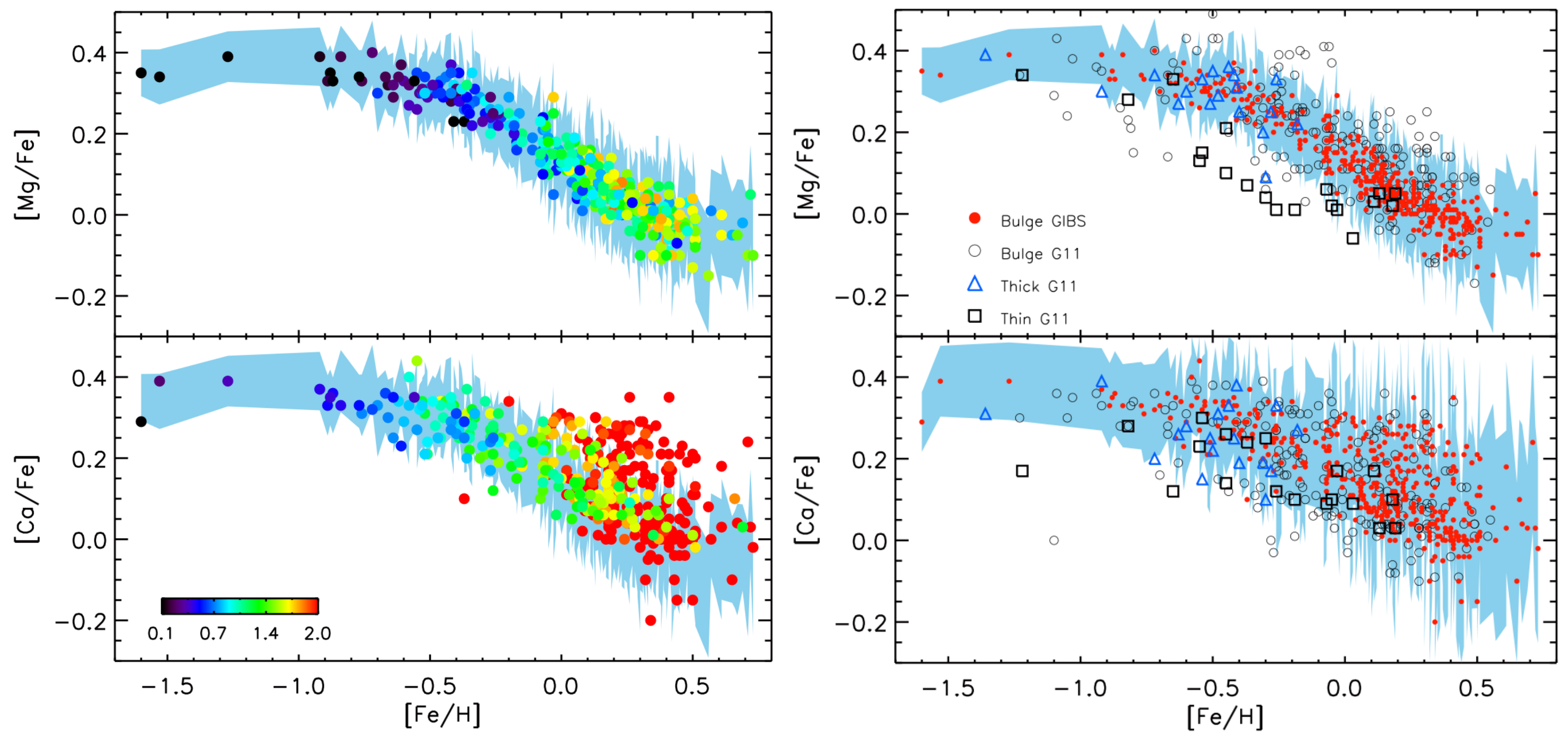}
 \caption{Left: Distribution of [Mg/Fe] ({\it upper panel}) and [Ca/Fe] ({\it lower panel}) as a function of [Fe/H]. The abundances of [Mg/Fe] and [Ca/Fe] are color\--coded according to their estimated uncertainty. The uncertainty contour of [Mg/Fe] is also shown in the background of both panels. \--- Right: Distribution of [Mg/Fe] ({\it upper panel}) and [Ca/Fe] ({\it lower panel}) as a function of [Fe/H] used as diagnostic of the formation timescale of the bulge. Red circles refers to GIBS abundances, whereas black symbols mark the abundances from \citet{gonzalez+11d}. Figure adapted from \citet[][{\it The GIRAFFE Inner Bulge Survey (GIBS). II. Metallicity distributions and alpha element abundances at fixed Galactic latitude}, their Figures\,8 and 9]{gonzalez15}.}
 \label{fig:alphas}
 \end{center}
\end{figure}

The detailed study of the chemical abundances and abundance patterns in bulge stars provides a unique tool to understand the chemical evolution enrichment of the bulge, and therefore to set tight constraints on its formation scenario.
The elemental abundance distributions, and the abundance ratio of certain critical elements such as Fe-peak, CNO, and $\alpha$-elements (i.e. those synthesized from $\alpha$ particles as O, Ne, Mg, Si, Ti, Ca and S) are particularly suitable for this purpose. Indeed, these elements are synthesized in stars of different masses, hence released into the interstellar medium on different timescales. \\
Because most of the chemical information on RC stars comes from the analysis of the $\alpha$\--elements, what follows is a summary of the picture built upon those measurements. It is not meant to be a comprehensive review of the global chemical composition of the bulge, for which the readers are instead encouraged to refer to \citet{mcwilliam16}.\\
As mentioned above a particularly useful abundance ratio is [$\alpha$/Fe]. Due to the time delay in the bulk of Fe and iron\--peak elements production \citep[mostly due to SNe\,Ia, see][]{nomoto+84} relative to $\alpha$-elements \citep[due to SNe\,II, see][]{woosley+95}, the [$\alpha$/Fe] abundance ratio can be efficiently used as a {\it cosmic clock} \citep[see e.g.][and references therein]{mcwilliam97,wyse00}.  For this reason many studies in the past addressed this question providing [$\alpha$/Fe] ratios for relatively small sample of K and M giants in few bulge regions \citep[see][and references therein]{mcwilliam94,rich05,cunha+06,fulbright07,lecureur+07,rich07,melendez+08,alves-brito+10,johnson11,gonzalez+11d,hill11,rich+12,bensby+13,johnson14,bensby+17}. As it has been the case for the MDF, the advent of the recent spectroscopic surveys ARGOS \citep{ness13b} and GIBS \citep{gonzalez15} provided $\alpha$\--element abundances for thousands of RC stars over a large area, hence allowing to study the  [$\alpha$/Fe]  trends as a function of the position in the bulge.\\
All previous and very recent studies agree on finding the bulge to be $\alpha$\--elements enhanced with respect to the solar value, thus suggesting a fast bulge formation scenario. As shown in Figure\,\ref{fig:alphas}, the $\alpha$\--element abundances of bulge stars with [Fe/H]$<-0.3$ are enhanced over iron by $\sim 0.3$\,dex, whereas metal\--rich stars show a decrease in [$\alpha$/Fe] reaching 0 for metallicity above the solar values. However, as discussed in \citet{gonzalez+11d} the direct translation of this trend to absolute timescales is not easy because the SNe\,Ia delay time can depend on different production channels. This is the reason why a relative approach through the comparison of  [$\alpha$/Fe]  trends observed in different Galactic components turns to be more reliable. From the comparison between bulge RC and giants in the thin and thick disk (see Figure\,\ref{fig:alphas}, right panels), the $\alpha$\--elements enhancement of the bulge with respect to the thin disk is evident across most of the entire metallicity regime. At solar metallicity, the bulge and thin disk are both $\alpha$\--poor. On the other hands, the thick disk giants are found to be as $\alpha$\--enhanced as the bulge, although they never reach the high metallicity tail of bulge stars. A possible interpretation of this relative trends is that the metal\--poor bulge population experienced  a fast formation scenario similar to the thick disk, whereas the metal\--rich bulge population underwent a more extended (i.e. longer) star formation, on a timescale similar to that of the thin disk.
\subsection{The age}
\label{sec:age}
An accurate dating of the bulge stellar component allows one to gauge at which lookback (i.e., at which redshift) one should look for possible analogs of the Milky Way, when their bulge formation processes were about to start, well on their way, or even already concluded. Indeed, with an age of  $\sim$10\,Gyr or older, it is at z$\gtrsim$2 that such analogs can be searched, or at lower redshift if significant fraction of the stellar component is found to be several Gyr younger \citep[see][for detailed discussion]{valenti+13}. \\
However, dating bulge stars is a very complicated task, challenged by the stellar crowding, the patchy and highly variable extinction, the uncertainties in the distance modulus, the distance spread due to the spatial depth of the bulge/bar along the line of sight, the metallicity dispersion and finally the contamination by foreground disk stars. The different contribution of all these factors prevents accurate location in terms of magnitude and color of the main sequence turnoff (MSTO) of the bulge population, so far among the most reliable age diagnostics \citep[see][ and \S5 by G. Bono]{renzini+88}.

Historically, the earliest age constraint by \citet{vandenbergh+74} in the Plaut field along the bulge minor axis at $b=-8^\circ$ ($\sim$1\,kpc) indicated a globular cluster (GC) like age. \citet{terndrup88} fit the photometry of other bulge fields at a range of latitudes with GC isochrones of varying metallicity, but because lacking a secure distance for the bulge he derived only a weak age constraint (11\--14\,Gyr). \citet{ortolani+95} solved the problem of contamination and distance uncertainties by comparing the bulge population with the NGC\,6528 and NGC\,6553 clusters. Forcing the bulge field and cluster luminosity function to match the HB clump luminosity level, it was possible to show for the first time that the relative ages of the bulge and metal\--rich cluster population could not differ by more than 5\%. \citet{feltzing+00} used HST\--based photometry of Baade's window and another low extinction field known as the Sgr\--I (i.e. at $l=1.25^\circ$ and $b=2.65^\circ$) to argue that while the density of the bulge MSTO stars increases for field closer to the centre, the foreground population does not change. They concluded that the bulk of the bulge population must therefore be old. The case for an {\it old bulge} has been further strengthened by later and more accurate photometric studies of different bulge fields, and by tackling the problem of contamination by foreground disk stars either kinematically by using proper motions, or statistically by considering control disk fields. Table\,\ref{tab:age} lists the location of the each observed field together with the adopted decontamination approach. 
\begin{table}[h]
\caption{Position of the bulge fields for which an age estimate of the stellar population has been provided through MSTO determination.}
\label{tab:age}       
\begin{tabular}{cccr}
\hline\noalign{\smallskip}
longitude & latitude & Disk decontamination &Reference \\
\noalign{\smallskip}\hline\noalign{\smallskip}
+1.13$^\circ$ & -3.77$^\circ$& proper motions & \citet{kuijken+02} \\
 +1.25$^\circ$ & -2.65$^\circ$&proper motions & \citet{kuijken+02} \\
  0$^\circ$ & -6$^\circ$ & statistical & \citet{zoccali+03} \\
  +1.25$^\circ$ & -2.65$^\circ$ &proper motions & \citet{clarkson08,clarkson+11} \\
  +0.25$^\circ$ & -2.15$^\circ$ &proper motions & \citet{brown+10} \\
  +1.26$^\circ$ & -2.65$^\circ$ &proper motions & \citet{brown+10} \\
  +1.06$^\circ$ & -3.85$^\circ$ &proper motions & \citet{brown+10} \\
  -6.75$^\circ$ & -3.81$^\circ$ &proper motions & \citet{brown+10} \\
 +10.3$^\circ$ & -4.2$^\circ$ & statistical & \citet{valenti+13} \\
  -6.8$^\circ$ & -4.7$^\circ$  & statistical & \citet{valenti+13} \\
\noalign{\smallskip}\hline\noalign{\smallskip}
\noalign{\smallskip}\hline
\end{tabular}
\end{table}

\begin{figure}
\begin{center}
\includegraphics[width=0.7\textwidth]{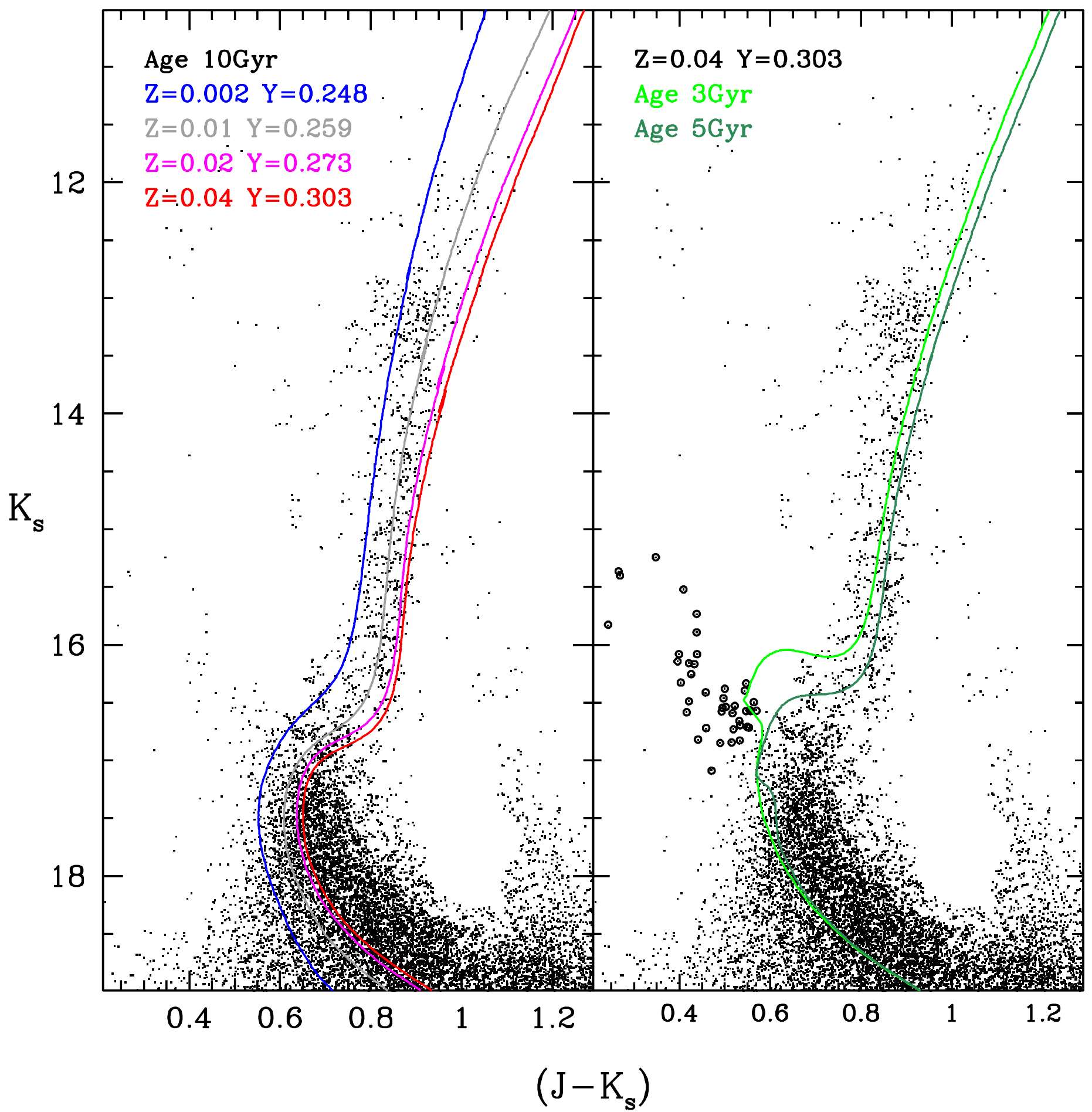}
 \caption{Disk\--decontaminated CMD of the field located at the far edge of the bar with over-plotted theoretical isochrones \citep{valcarce+12}. Isochrones ages, metallicity and helium abundances are indicated near the top\--left corner of each panel. Sub\--solar isochrones are $\alpha$\--enhanced, and solar\--scaled otherwise.The metallicity range covered by the bulk of the star in this field is $0.002\lesssim Z \gtrsim 0.060$, as derived from the observed photometric metallicity distribution. Open circles mark stars that are most likely blue stragglers. Figure adapted from \citet[][{\it Stellar ages through the corner of the boxy bulge}, their Figure\,10]{valenti+13}.}
 \label{fig:age}
 \end{center}
\end{figure}

As all previous studies, \citet{valenti+13} found that the bulk stellar population of the Milky Way bar edges is over $\sim$\,10\,Gyr old (see Figure\,\ref{fig:age}), with no obvious evidence of younger population. This age is indistinguishable from the one reported for more inner bulge fields, a few degrees from the Galactic centre or lying along the bulge minor axis.  
From the analysis of the MSTO in the HST\--based CMD kinematically decontaminated, \citet{clarkson+11} concluded that once the blue stragglers population is taken into account a significantly younger ($\lesssim$5\,Gyr) population in the bulge must be at most 3.4\%. 

However, there is a clear discrepancy between the ages inferred from the determination of the MSTO location in the observed CMDs and those derived by the microlensed project of Bensby and collaborators, which estimates single star age from its effective temperature and gravity (i.e. from isochrones in the $\rm T_{eff}$, log\,g plane) as obtained from high resolution spectra.
Indeed, based on a sample of 90 F and G dwarf, turnoff and subgiant stars in the bulge  (i.e. $|l|\lesssim6^\circ$ and $-6^\circ< b < 1^\circ$) observed during microlensing, \citet{bensby+17} found that about 35\% of the metal\--rich star ([Fe/H]$>0$) are younger than 8\,Gyr, whereas the vast majority of metal\--poor  ([Fe/H]$\lesssim-0.5$) are 10\,Gyr or older. In addition, from the derived age\--metallicity and age\--$\alpha$ elements distribution the authors concluded that the bulge must have experienced several significant star formation episodes, about 3, 6, 8 and 12 Gyr ago. 

As discussed by \citet{valenti+13}, each of the two approaches has its own {\it pros} and {\it cons}. The microlensing approach depends more heavily on model atmospheres that may introduce systematics especially in the metal rich regime, and it deals with small number statistics. At the same time, it has the advantage that the metallicity of each individual stars is very well constrained. Conversely,  by dealing with a statistically significant number of stars, the traditional CMD method should in principle be able to reveal the presence of young populations. However, in this case the metallicity of individual stars are unknown, therefore one does not know if, for instance, some of the stars above the MSTO of the $Z=0.060$ isochrones (see Figure\,\ref{fig:age}) are old and have lower metallicity, or whether they are metal\--rich stars younger than 10\,Gyr.  

The effect of the age metallicity degeneracy, specifically in terms of the color spread of the MSTO in the observed CMDs has been used by \citet{haywood+16} to argue in favor of the scenario suggested by the microlensing results. In particular,  \citet{haywood+16} compared the MSTO color spread observed in the CMD of \citet{clarkson+11} with that of synthetic CMDs, obtained by using two scenarios corresponding to different age\--metallicity relation (AMR). In {\it scenario I} a simulated CMD was obtained by using the AMR presented by \citet{bensby+13} (i.e. based on a total sample of 59 micorlensed drawf), whereas for {\it scenario II} an AMR that extends from [Fe/H]$=-1.35$\,dex at 13.5\,Gyr to [Fe/H]$=+0.5$\,dex at 10\,Gyr was adopted. When taking into account distance, reddening and metallicity effects, \citet{haywood+16} showed that the MSTO color spread of a {\it purely} old stellar population would be wider than what observed, which in turn appears to be consistent with the simulation obtained from the {\it scenario I}. Unfortunately, what the \citet{haywood+16} paper does not address is the fact that the simulation using the AMR of \citet{bensby+13} produced a CMD that not only
has a smaller MSTO color spread like the observed one but also show a remarkable number of stars 
just above (i.e. brighter) the MSTO, which are not matched by the observations (see their Figure 8). In 
this respect, the comparison between observations and simulations presented by \citet{zoccali+03} to 
infer the age of the bulge population would seem more appropriate because the synthetic CMD was 
obtained by using the observed luminosity function, and therefore the comparison was done such as 
to match not only the location and spread in color of the MSTO, but also the number of stars at the MSTO level.

\subsection{Summary and conclusions}
\label{sec:end}
Owing to the systematic and detailed study of RC star properties performed in the last decade by using wide area photometric surveys we have finally reached a good and complete comprehension of the 3D structure of the Milky Way bulge.
The bulge, as referred as the region in the inner $\sim$3\,kpc is a bar with an orientation with respect to the Sun\--Galactic centre line of sight of $\sim27^\circ$, and whose near side points in the first Galactic quadrant. 
The bar has a boxy/peanut/X\--shape structure in its outer regions, a characteristic morphology of bulges formed out the natural evolution of disk galaxies as the consequence of disk dynamical instabilities and vertical buckling of the bar. The observed split in the RC mean magnitude distribution in the outer regions is interpreted by the dynamical models as bar growing. 
In the innermost region ($|l,b|<2^\circ$), rather than a nuclear bar, there seems to be an axisymmetric high stellar density peak, which instead may be responsible for the observed change in the bar pivot angle. In addition, RC stars trace a thinner and longer structure with a semimajor axis of $\sim\,4.6$\,kpc, known as the long bar, which according to the latest study appears  to be the natural extension of the bulge main bar at higher longitude.\\
The bulge is the most massive stellar component of the galaxy, with a mass ($M_B=2 \-- 1.8\times10^{10}M_\odot$) close to 1/5 of the total stellar mass of the Milky Way, and about ten times larger than the mass of the halo. \\
The recent spectroscopic surveys (ARGOS, GIBS, Gaia-ESO) of RC stars, together with the ongoing that targets K and M giants ( i.e. APOGEE-North) provided a comprehensive and detailed view of the chemical content of the stellar population over an area that corresponds to more than 80\% of the entire bulge. The emerging picture is that the bulge MDF as traced by the RC is much more complex that previously thought, and it hosts two populations with different mean metallicity (i.e. metal\--poor and metal\--rich), spatial distribution and kinematics. The metal\--poor population as traced by RC, RRLs and T2C is more spherically concentrated, whereas the RC metal\--rich component traces the boxy/peanut bar. The observed properties of such metal\--poor population possibly older (i.e. spatial distribution and kinematics) do not necessarily implies the presence of a {\it classical bulge} (i.e. a merger\--driven structure dominated by gravitational collapse) embedded in the boxy bulge. 
Indeed, the recent N\--body simulations model of \citet{debattista+16} accounts for the presence of a metal\--poor population spherically concentrated, as well as for other observed trend of densities, kinematics and chemistries, without invoking the need for a {\it composite bulge} scenario (i.e. the coexistence of two structures, one merger\--driven and one boxy shaped formed out of disk and bar evolution). According to \citet{debattista+16}, the observed properties of the Milky Way bulge stellar populations are consistent with a bulge formed from a {\it continuum} of disk stellar populations kinematically separated by the bar.\\
Based on accurate abundances analysis of RC stars, the bulge show $\alpha$\--element enhancement typical of fast formation process. In particular, a possible interpretation of the observed relative trends of $\alpha$\--elements in the bulge, thin and thick disk is that the metal\--poor bulge population experienced a fast formation scenario similar to the thick disk, whereas the metal\--rich bulge population underwent a more extended (i.e. longer) star formation, on a timescale similar to that of the thin disk.\\
The innermost and still poorly unexplored regions (i.e. $|b|\lesssim1^\circ$) will be soon probed by new IR surveys planned for the near future (i.e. APOGEE-South, Multi-Object Optical and Near-infrared Spectrograph at VLT -- MOONS) hence allowing us, for the first time, to complete the puzzle with a clear understanding of the chemical properties of the bulge as a whole with unprecedented accuracy. \\
There is no doubt that the central regions of the Milky Way hosts an old stellar population. The strongest evidence being the presence of a prominent population of RRLs and T2C  found by OGLE and VVV \citep[][and Bhardwaj et al]{dekany+13,pietrukowicz+15,gran+16}, which are by far the largest photometric campaigns of variable stars.  Furthermore, an old age is also guaranteed by the existence of a bulge GCs system \citep[see e.g.][and reference therein]{valenti+10,bica+16}.
However, what still remains to be firmly assessed is the contribution of intermediate\--young (i.e. $\lesssim$5\,Gyr) stars to the global bulge stellar population. The AMR proposed by \citet{bensby+17} should be either confirm on much statistically robust sample, or by using a methodology for the reconstruction of the star formation history more sophisticated than the approaches adopted so far. In particular, the comparison between observations and simulations should be performed by using as many as possible features of the CMDs \citep[i.e.][]{gallart+05}.

In the coming years, the exquisite astrometry provided by the next Gaia data releases will most probably allow us to further refine the global picture of the bulge structure. Even though a large fraction of the bulge RC population is out of GAIA reach because of the crowding and high extinction, the information derived from RC stars in the low reddening regions can be used to obtain a very accurate distances map of the bulge outer regions. This can be used then as the {\it reference frame} upon which, through a differential analysis with the most obscured regions, we can build the entire bulge distances and structure maps.

Finally, further efforts should be put to characterize the chemical content of the RRLs and T2C, which among all tracers are those representing {\it univocally} and {\it purely} the oldest stars in the Bulge. Indeed, accurate MDF and elemental abundances from high\-- or medium\--resolution spectroscopy for these type of stars are still missing, or largely insufficient. 
The future LSST project will provide the position, magnitude and colors for thousands of variable stars, spanning a variety of ages. The spectroscopic follow up of a sizeable sample of variables would literately open new frontiers of our knowledge by allowing for the first time an accurate study of the metallicity trends as a function of the stellar ages. If such analysis would be extended also outside the bulge regions we could be in position to understand the interaction among different Galaxy structures, such for instance a clear view of the transition between disk and bulge.


\section{RR Lyrae variables in the Ultra-Faint satellites of the Milky Way}

In the $\Lambda$-Cold Dark Matter ($\Lambda$-CDM) scenario, large galaxies are the result of the assembling of smaller fragments, cold-dark matter dominated \citep[e.g.][]{Diemand2007, Lunnan2012}. The baryonic component of these fragments may eventually collapse, forming small galaxies. This idea is appealing when applied to the MW, since it echoes the early scenario envisioned by \cite{SZ1978}, in which the outer halo of the MW may have formed by a continuous infall of protogalactic fragments onto the Galaxy, for some time after that the collapse of its central part was completed. Indeed, first attempts to link the $\Lambda$-CDM cosmology with the Galactic environment foresaw the assembling of the Galactic halo starting from a number of satellites, and producing a number of fragments and streams, which are actually observed \citep[e.g.][]{McConnachie2012, Grillmair2016}. For decades, the survivors of such a process have been identified with the dwarf spheroidal (dSph) satellites of the MW, since they are old, metal-poor, gas poor and dark matter-dominated systems. However, it was soon realized that the observed number of observed dSph was one or two order of magnitude smaller than that expected from theory. This mismatch, dubbed the ''missing satellites problem" \citep{Klypin1999, Moore1999}, has been for several years a major problem in the comparison between theory and observations. A second problem, pointed out in the last few years, is that the circular velocities of the known dSph are too low, when compared to the expected values from their simulated sub-structures. In other words, the predicted densities of the massive subhaloes are too high, to host any of the bright dSphs. This mismatch, called the ''too big to fail problem" \citep{BoylanKolchin2012}, has heavy implications, since it means either: i) massive dark subhaloes exist as predicted, but they host faint ($L < 10^5 L_\odot$) satellites; ii) massive dark subhaloes does not exist as predicted, for instance they may be less concentrated than predicted.

As a matter of fact, in the last ten years a considerable number of new and faint MW satellites has been discovered \citep[e.g.][]{Belokurov2007, McConnachie2012}, most of them on the basis of the SDSS data and, more recently, thanks to the ongoing large surveys conducted with OMEGACAM@VST, DECAM@CTIO and Pan-STARRS \citep[e.g.][]{Koposov2015, Laevens2015}. These systems, called the ultra-faint dwarfs (UFDs), have integrated luminosities similar or even lower than those of the Galactic globular clusters, and are apparently dark matter dominated\citep[see][]{McConnachie2012}. 

The large number of systems currently available (dSphs + UFDs), allowed to trace a statistically significant analysis of their spatial distribution, leading to the discovery that they actually populate a relatively thin ring, perpendicular to the MW plane, and possibly rotationally supported \citep{Pawlowski2013}. Moreover, several of the recently discovered candidate MW satellites also seem to be clustered around the Magellanic Clouds, hinting that they may have fallen in as a group \citep[e.g.][]{Sales2015}, in line with the theoretical predictions \citep{Wetzel2016}. Similar aligned structures, showing a kinematic coherence, have been discovered around the Andromeda galaxy \citep{Ibata2013} and, outside the Local Group, around NGC 5557 \citep{Duc2014}.
Similar structures, but without a clear kinematic coherence, have been reported in the literature around NGC 1097, NGC 4216, NGC 4631 \citep[][, and references therein]{PK2014}, and possibly around the M81 and Cen A groups \citep{Muller2016, muller18}. The MW structure, dubbed the \textit{Vast POlar Structure} (VPOS), opens a wide scenario of cosmological problems, since at the present time it is not clear if it is made of primordial (dark matter-dominated) systems, or tidal (dark matter-free) galaxies. 

Interestingly, when the halo Galactic globular clusters are grouped in young halo (YH) and old halo (OH) on the basis of the variation of their HB morphology at constant $\rm [Fe/H]$, which is a rough approximation of the cluster age, they show up also a division by kinematics and spatial distribution \citep[e.g.][]{Zinn1993, Mackey2004, Lee2007}. In particular, YH clusters span a wide range in ages \citep[$\sim 5$ Gyr][]{Dotter2011} and are characterized by a hotter kinematics than the OH clusters. These occurrences suggest that YH clusters may be debris from accretion events. Finally, the discovery that YH clusters are part of the VPOS \citep{Pawlowski2013, Zinn2014}, strengthens the debris hypothesis. Moreover, it also suggests that a fraction of the accreted halo may have been originated in a number of moderately massive satellites that formed GCs, similar to Sagittarius, Fornax, or even the Magellanic Clouds \citep[but see][for new insights on the contribution of Fornax-like systems]{Fiorentino2016}. 

\subsubsection{The role of the RR Lyrae stars}
A fraction of the problem can be settled by carefully comparing the photometric and spectroscopic properties of the stellar populations of the halo of the MW and of its companions. Moreover, since their pulsational properties such as periods and amplitudes are a function of their structural and evolutive parameters, a detailed comparison of the pulsational properties of the RR Lyrae stars can add valuable information.

In particular, the \textit{ensemble} pulsational properties of the RR Lyrae stars can give important hints. Indeed, it is well known that cluster and field Galactic RR Lyrae stars are affected by the so-called Oosterhoff (Oo) dichotomy, where in the Oo~I group the fundamental mode variables show mean periods of $< P_{ab} > \sim 0.55$ days, while in the OO~II group they have $< P_{ab} > \sim 0.65$ days.
In fact, the bright MW companions have $< P_{ab} > \sim 0.6$ days and are generally classified as Oo-intermediate, which is difficult to reconcile with the dichotomy of the Galactic halo. On the other side, the RRLs hosted in the UFDs suggest an Oo~II classification \citep{DallOra2012}, consistent with an older population of the Galactic halo, possibly produced by an early dissipative collapse or merging \citep[e.g.][]{Miceli2008}.

\subsubsection{The ultra-faint dwarfs}
UFD galaxies are, at first glance, the low brightness tail of the dSph. From this point of view, there is no structural difference between the ''classical" dSphs and the low luminosity UFDs. However, a careful comparison of the central surface brightness as a function of the total luminosity, shows a ''knee" around $M_V \approx -8$ mag \citep[see][Figure~7]{McConnachie2012}. The galaxies brighter than $M_V \approx -8$ mag follow a linear trend, with the brightest galaxies having a higher central brightness, while galaxies weaker than this limit follow a horizontal distribution, with a constant central brightness no matter what is the total luminosity. In this work, we will therefore consider UFDs all the galaxies that follow such a horizontal distribution.

Stated in a different way, UFDs are characterized by low luminosities and projected densities. This means that it is difficult to recognize them as stellar overdensities in the field, and the problem becomes even more severe when one wants to detect their possible tidal tails. For these reasons, RRLs become a powerful tool to study the stellar populations of the UFDs and their spatial extent. Indeed, as suggested by \citet{Baker2015}, RRLs could be the \textit{only} method to unveil very faint satellites, with $M_V < 3.5$, especially at low Galactic latitudes when both extinction and field contamination can be important. All the UFDs searched for variability so far show at least one RRL. This is not surprising, since they are composed by (at least) old, metal-poor stellar populations, which are known to produce RRLs. The small statistics must not be misleading, since if one normalizes the observed number of RRLs by the integrated luminosity (i.e. a proxy of the baryonic mass), the fraction of RRLs is even higher than that observed in the bright dSphs. Indeed, adopting the specific frequency as parametrized by \citet{Mackey2003}

\begin{equation}
S_{RR} = N_{RR} \times 10^{0.4 (7.5 + M_V)}
\end{equation}

one finds that UFDs tend to have higher specific frequencies than dSphs, as shown in figure 3 of \cite{Baker2015}, here reproduced by kind permission. However, as suggested in \cite{Baker2015}, this could be due to incompleteness effects, being the census of the RRLs in the bright dSphs still not complete.

\begin{figure}
  \includegraphics[width=.8\linewidth]{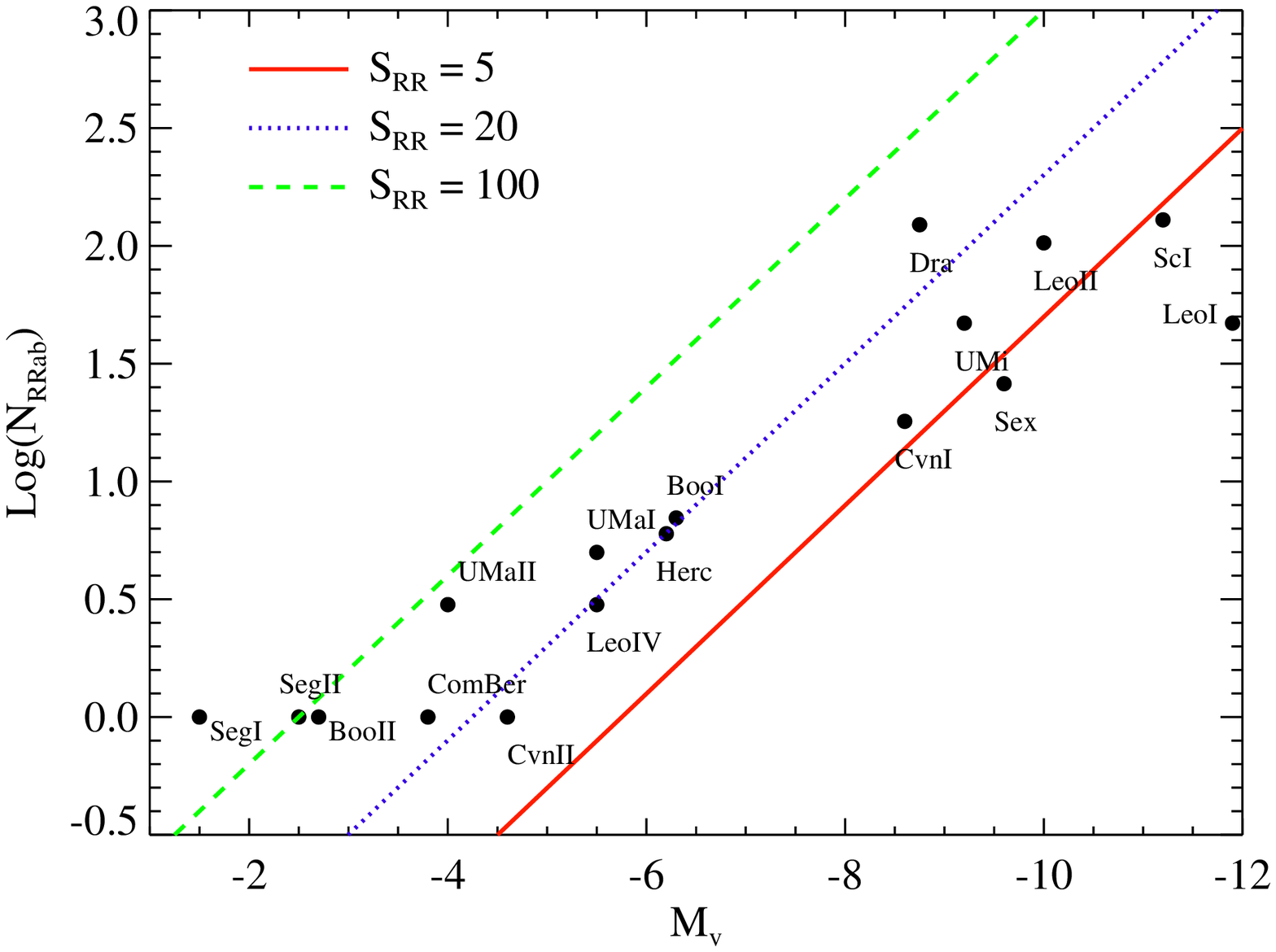}
\caption{Observed number of RRL stars as a function of the absolute magnitude. Overplotted are three specific frequencies (see text for details). The brightest galaxies have lower specific frequencies, possibly for incompletness. UFDs are characterized by specific frequencies of in the range 20-100. Figure by \cite{Baker2015}, here reproduced by kind permission.}
\label{fig:1}       
\end{figure}

\subsection{Our RRL sample}
Table 1 collects all the positional and pulsational parameters of the RRLs discovered in the UFDs so far. For each variable, we list the position, the period, the mean magnitudes and the luminosity amplitude in the $BVI$ bands (when available). In some cases, we merged the information on the same variable coming from different studies. We discuss these cases in the individual notes. Here, we point out that $B$-band photometry 
for the RRLs in Bootes~I comes from \cite{Siegel2006}, and the $VI$-bands from \cite{DallOra2006}. We also explicitly note that, when a star was listed in both studies, we adopt the coordinates listed in \cite{DallOra2006}. 
This table does not include the RRLs hosted in CVn~I, whose structural properties suggest a classification as a ``classical" dSph instead as a UFD, and those hosted in Leo~T, since it contains gas and a young stellar population, and in this sense is not a typical old, gas poor UFD. For these two galaxies, we refer the reader to the specific papers -- CVn~I, \cite{Kuehn2008}; Leo~T, \cite{Clementini2012}. The galaxies are ordered by RA, and the sources of the data are: Segue~II \citep{Boettcher2013}; UMa~II \citep{DallOra2012, Vivas2016}; UMa~I \citep{Garofalo2013}; Leo~IV \citep{Moretti2009}; Hyd~II \citep{Vivas2016}; Coma \citep{Musella2009}; CVn~II \citep{Greco2008}; Boo~II \citep{Sesar2014, Vivas2016}; Boo~III \citep{Sesar2014, Vivas2016}; Boo~I \citep{Siegel2006, DallOra2006}; Hercules \citep{Musella2012}.

\begin{table}
\small
\begin{tabular}{lcclcccccrl}
\hline\noalign{\smallskip}
ID & RA        & Dec       & Period & $<B>$ & $A_B $ & $<V>$ & $A_V $ & $<I>$ & $A_I $ & Type \\
   & (J2000.0) & (J2000.0) & (days) & (mag) & (mag)  & (mag) & (mag)  & (mag) & (mag)  & \\
\noalign{\smallskip}\hline\noalign{\smallskip}
V1\_SegueII & 02:19:00.06 & +20:06:35.2 & 0.748   & 18.62 & 0.62 & 18.25 & 0.51 &       &      & RR$_{ab}$ \\
V1\_UMaII   & 08:50:37.43 & +63:10:10.0 & 0.56512 & 18.61 & 1.27 & 18.24 & 1.01 &       &      & RR$_{ab}$ \\
V1\_UMaI    & 10:34:59.2\hspace{0.15cm}  & +51:57:07.3 & 0.56924 & 20.72 & 1.29 & 20.47 & 1.07 &       &      & RR$_{ab}$ \\
V2\_UMaI    & 10:35:05.5\hspace{0.15cm} & +51:55:39.8 & 0.584   & 20.85 & 1.09 & 20.37 & 0.78 &       &      & RR$_{ab}$ \\
V3\_UMaI    & 10:34:30.8\hspace{0.15cm} & +51:56:28.9 & 0.64315 & 20.73 & 0.99 & 20.40 & 0.75 &       &      & RR$_{ab}$ \\
V4\_UMaI    & 10:34:18.7\hspace{0.15cm}  & +51:58:29.2 & 0.74516 & 20.62 & 1.15 & 20.24 & 0.90 &       &      & RR$_{ab}$ \\
V5\_UMaI    & 10:35:37.5\hspace{0.15cm}  & +52:02:35.6 & 0.59967 & 20.87 & 1.38 & 20.49 & 0.96 &       &      & RR$_{ab}$ \\
V6\_UMaI    & 10:33:07.1\hspace{0.15cm}  & +51:50:05.1 & 0.39673 & 20.73 & 0.72 & 20.42 & 0.63 &       &      & RR$_{c}$  \\
V7\_UMaI    & 10:32:37.5\hspace{0.15cm}  & +51:49:55.7 & 0.40749 & 20.54 & 0.37 & 20.44 & 0.38 &       &      & RR$_{c}$  \\
V1\_LeoIV   & 11:32:59.2\hspace{0.15cm}  & -00:34:03.6 & 0.61895 & 21.82 & 0.99 & 21.47 & 0.73 &       &      & RR$_{ab}$ \\
V2\_LeoIV   & 11:32:55.8\hspace{0.15cm}  & -00:33:29.4 & 0.7096  & 21.86 & 0.76 & 21.46 & 0.64 &       &      & RR$_{ab}$ \\
V3\_LeoIV   & 11:33:36.6\hspace{0.15cm}  & -00:38:43.3 & 0.635   & 21.81 & 0.82 & 21.52 & 0.65 &       &      & RR$_{ab}$ \\
V1\_HydII   & 12:21:43.51 & -31:59:42.8 & 0.645   & 22.00 & 0.77 & 21.56 & 0.62 & 21.30 & 0.38 & RR$_{ab}$ \\
V1\_Coma    & 12:27:33.50 & +23:54:55.7 & 0.66971 &       &      & 18.44 & 0.78 & 17.74 & 0.53 & RR$_{ab}$ \\
V2\_Coma    & 12:26:50.89 & +23:56:00.6 & 0.31964 &       &      & 18.69 & 0.37 & 18.20 & 0.24 & RR$_{c}$  \\
V1\_CVnII   & 12:57:01.6\hspace{0.15cm}  & +34:19:33.4 & 0.358   & 21.74 & 0.83 & 21.49 & 0.68 &       &      & RR$_{c}$  \\
V2\_CVnII   & 12:57:11.8\hspace{0.15cm}  & +34:16:52.9 & 0.743   & 21.77 & 0.95 & 21.46 & 0.71 &       &      & RR$_{ab}$ \\
V1\_BooII   & 13:58:07.04 & +12:51:22.8 & 0.66349 &       &      & 18.23 & 0.71 &       &      & RR$_{ab}$ \\
V1\_BooIII  & 14:00:34.52 & +25:55:52.7 & 0.63328 &       &      & 18.56 & 1.07 &       &      & RR$_{ab}$ \\
V1\_BooI    & 13:59:29.36 & +14:10:43.6 & 0.303771& 19.78 & 0.628 &      &      &       &      & RR$_{ab}$ \\
V2\_BooI    & 13:59:51.34 & +14:39:06.0 & 0.3119  & 19.73 & 0.67 & 19.63 & 0.34 & 19.32 & $>$0.13 & RR$_{c}$  \\
V3\_BooI    & 14:00:26.86 & +14:35:33.1 & 0.3232  & 19.79 & 0.64 & 19.58 & 0.57 & 19.25 & $>$0.15 & RR$_{c}$  \\
V4\_BooI    & 14:00:08.90 & +14:34:24.1 & 0.3860  & 19.82 & 0.58 & 19.57 & 0.59 & 19.19 & 0.2  & RR$_{c}$  \\
V5\_BooI    & 14:00:21.56 & +14:37:28.8 & 0.6506  & 19.75 & 0.57 & 19.38 & 0.33 & 18.93 &      & RR$_{ab}$ \\
V6\_BooI    & 13:59:45.95 & +14:31:40.7 & 0.3919  & 19.86 & 0.61 & 19.58 & 0.53 & 19.13 & 0.22 & RR$_{c}$  \\
V7\_BooI    & 13:59:49.37 & +14:10:05.6 & 0.4011623&19.76 & 0.74 &       &      &       &      & RR$_{c}$  \\
V8\_BooI    & 13:59:59.69 & +14:27:34.0 & 0.4179  & 19.79 & 0.74 & 19.54 & 0.40 & 19.10 & 0.26 & RR$_{c}$  \\
V9\_BooI    & 13:59:47.28 & +14:27:56.3 & 0.5755  & 19.84 & 1.28 & 19.55 & 1.00 & 19.07 & $>$0.41& RR$_{ab}$ \\
V10\_BooI   & 14:00:25.75 & +14:33:08.4 & 0.628   & 19.78 & 1.32 & 19.47 & 1.09 & 18.98 & $>$0.24& RR$_{ab}$ \\
V11\_BooI   & 13:58:04.38 & +14:13:19.3 &0.6617310& 19.82 & 1.20 &       &      &       &      & RR$_{ab}$ \\
V12\_BooI   & 13:59:56.00 & +14:34:55.0 & 0.3948  & 19.90 & 0.54 & 19.59 & 0.53 & 19.13 & $>$0.20& RR$_{d}$ \\
V13\_BooI   & 13:59:06.36 & +14:19:00.1 &0.7061108& 19.77 & 0.82 &       &      &       &      & RR$_{ab}$ \\
V14\_BooI   & 13:59:25.75 & +14:23:45.3 & 0.7186  & 19.89 & 0.86 & 19.50 & 0.44 & 19.03 &      & RR$_{ab}$ \\
V15\_BooI   & 14:00:11.09 & +14:24:19.7 & 0.8456  & 19.83 & 0.48 & 19.45 & 0.48 & 18.91 & 0.28 & RR$_{ab}$ \\
V1\_Her     & 16:31:02.17 & +12:47:33.7 &0.639206 & 21.68 & 1.16 & 21.27 & 1.06 &       &      & RR$_{ab}$ \\
V3\_Her     & 16:30:54.93 & +12:47:04.2 & 0.39997 & 21.72 & 0.61 & 21.32 & 0.48 &       &      & RR$_{c}$  \\
V4\_Her     & 16:30:56.14 & +12:48:29.2 & 0.39576 & 21.59 & 0.69 & 21.23 & 0.58 &       &      & RR$_{c}$  \\
V5\_Her     & 16:30:52.28 & +12:49:12.0 & 0.40183 & 21.64 & 0.61 & 21.30 & 0.47 &       &      & RR$_{c}$  \\
V6\_Her     & 16:30:52.41 & +12:49:60.0 & 0.69981 & 21.76 & 1.19 & 21.35 & 0.90 &       &      & RR$_{ab}$ \\
V7\_Her     & 16:31:29.48 & +12:47:34.9 & 0.67799 & 21.65 & 1.03 & 21.22 & 0.81 &       &      & RR$_{ab}$ \\
V8\_Her     & 16:31:27.20 & +12:44:16.7 & 0.66234 & 21.69 & 1.09 & 21.25 & 0.90 &       &      & RR$_{ab}$ \\
V9\_Her     & 16:31:29.50 & +12:40:03.1 & 0.72939 & 21.60 & 1.00 & 21.18 & 0.80 &       &      & RR$_{ab}$ \\
V10\_Her    & 16:30:03.96 & +12:52:06.3 & 0.6616  & 21.69 & 1.32 & 21.28 & 1.17 &       &      & RR$_{ab}$ \\
\noalign{\smallskip}\hline
\end{tabular}
\caption{Positional and pulsational parameters of the RRLs in the UFD studied so far. See text for details.}
\end{table}

\subsubsection{Notes on individual variables}
\textbf{V1 UMa~II}: this RRL was discovered by \cite{DallOra2012}, where they proposed a period of $P = 0.6593$ days. Subsequently, on the basis of a more densely sampled light curve, \cite{Vivas2016} computed a new period of $P = 0.56512$ days. This period appears to be compatible with the photometry of this variable, available in our database. Therefore, we adopt the period proposed by \cite{Vivas2016}.
\\
\textbf{V1 Boo~II}: \cite{Sesar2014} proposed a period of $P = 0.63328$ days, while \cite{Vivas2016} computed a slightly different period of $P = 0.66349$ days. We keep the \cite{Vivas2016} estimate, together with their mean magnitude and amplitude.
\\
\textbf{V1 Boo~III}: \cite{Sesar2014} and \cite{Vivas2016} give practically the same period. However, here we present the mean magnitude and pulsational amplitude proposed by \cite{Vivas2016}.
\\
\textbf{V5 Boo~I}: this star was classified as a $RR_{c}$ type star by \cite{Siegel2006}, with a period of $P = 0.3863158$ days, and as a $RR_{ab}$ star by \cite{DallOra2006}, with $P = 0.6506$ days. We used the \cite{Siegel2006} period estimate to phase the data of this variable available in our database, but unfortunately we were not able to achieve a satisfactorily phased light curve. Therefore, we keep the \cite{DallOra2006} estimate.  It is worth noting that this variable is quite peculiar, since in the \cite{DallOra2006} photometry it appears redder and brighter than the HB, and it could be blended with a companion.
\\
\textbf{V12 Boo~I}: this RRL was classified as $RR_{ab}$ by \cite{Siegel2006}, with a period of $P = 0.6797488$ days, and by a double-pulsator $RR_{d}$ by \cite{DallOra2006}, with periods of $P_1 = 0.3948$ days and $P_0= 0.5296$ days. since the light curve shown by \cite{DallOra2006} convincingly shows a typical double-mode behavior (see their Fig. 2), here we will keep their classification.
\\
\textbf{V1 Hyd~II}: the photometry of this variable was presented by \cite{Vivas2016} in the $gri$ system. Here, for consistency we present its pulsational properties in the $BVi$ bands, where $gr$ magnitudes were transformed in $B$, $V$ magnitudes following \cite{Jester2005}.

\subsection{Discussion}
A glance at the data listed in Table 5 shows that, except for a few number of galaxies (namely UMa~I, Boo~I, Her), UFDs host a very small number of RRLs.  Of course, with such small statistics, if a distance estimate can be reliable (especially in presence of a color-magnitude diagram, to check the robustness of the measured mean magnitudes and colors), some \textit{caveats} must be recognized when using RRLs as population tracers. Indeed, since the Oo type is an \textit{ensemble} feature, this should be declared only when a substantial number of fundamental mode RRL is available, in order to properly put them on an amplitude-period diagram (known as Bailey's diagram). Nevertheless, a comparison with the classic Oo~I and Oo~II lines in the Bailey's diagram can give interesting insights.

\subsubsection{The Oosterhoff classification and the Galactic halo}

In Figure 16, we show the positions on the Bailey's diagram of the listed RRLs. For reference, we plot the loci of the Oo~I and Oo~II clusters, according to \cite{Zorotovic2010}. At first glance, the positions of almost all the RRLs of the UFDs are compatible with a Oo~II classification. The only apparent exception is UMa~I, which was classified as Oo~intermediate in \cite{Garofalo2013}. However, when we compare the positions of its variables with those of the Galactic globular clusters M3 (left panel) and M15 \citep[right panel, data made available by][]{Clement2001}, we favor a Oo~II classification also for this system. Indeed, from the left panel it appears that the distribution of the UMa~I $RR_c$ variables is in good agreement with those of the other UFDs and with those of the Oo~II cluster M15. Also, the mean period of the fundamental pulsators of UMa~I is $<P_{ab}> = 0.628 \pm 0.063$ days, which is in agreement with the values of other Oo~II systems, such as M15 ($<P_{ab}> = 0.643 \pm 0.063$ days), M92 ($<P_{ab}> = 0.631 \pm 0.048$ days) and M68 ($<P_{ab}> = 0.627 \pm 0.062$ days), where the listed values are computed on the basis of the compilation published by \cite{Clement2001}. However, it should also be noted that, when discarding the variable V4, significantly brighter than the others, the mean period of the fundamental pulsators lowers to $<P_{ab}> = 0.599 \pm 0.032$ days, as discussed in \cite{Garofalo2013}.
This suggest that, in general, when dealing with systems with a small number of RRLs, a correct Oo classification is a risky business, and we may consider not only to compare the positions of the $RR_{ab}$ stars with respect to the mean Oo~I and Oo~II loci, but to consider the whole plane instead, with the actual distributions of the RRLs belonging to some reference Oo~I and Oo~II clusters. 

Taken at face value, the Oo~II classification could suggest a major contribution of UFD-like objects in assembling the Galactic halo. However, as pointed out in \cite{Fiorentino2015}, a detailed comparison of the pulsational properties of the RRLs of the Galactic halo and of the set dSphs $+$ UFDs, shows that the latter lack the so-called high-amplitude, short-period (HASP) variables. \cite{Fiorentino2015} argue that the HASP region is filled only when RRLs are more metal-rich than $\rm [Fe/H] = -1.5$ dex. Thus, \textit{present-day} dSph- and UFD-like objects seem to have played a minor role, if any, in assembling the Galactic halo. 

\begin{figure}
\centering
\begin{minipage}{.7\textwidth}
  \centering
  \includegraphics[width=1\linewidth]{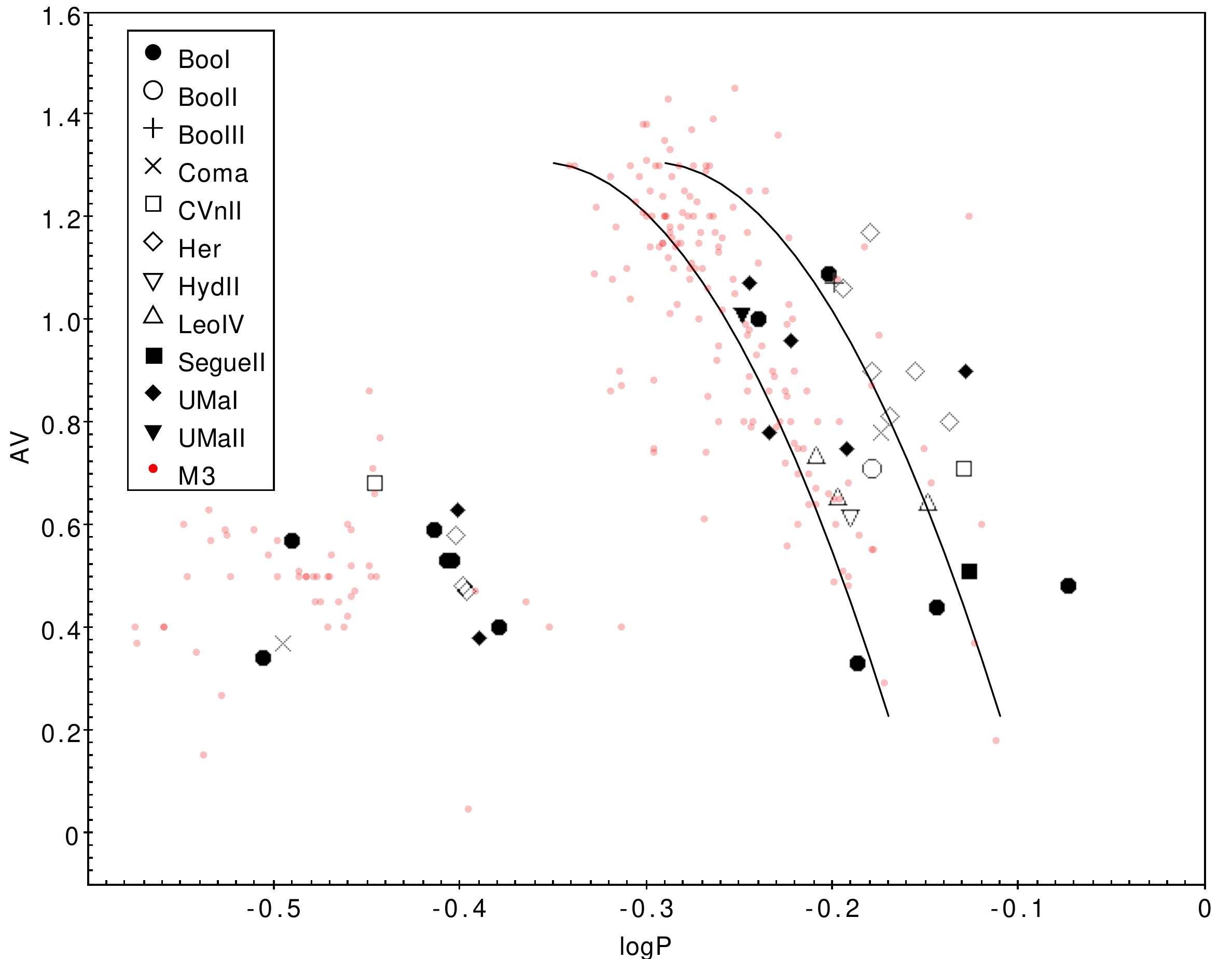}
\end{minipage}%
\begin{minipage}{.7\textwidth}
  \centering
  \includegraphics[width=1.\linewidth]{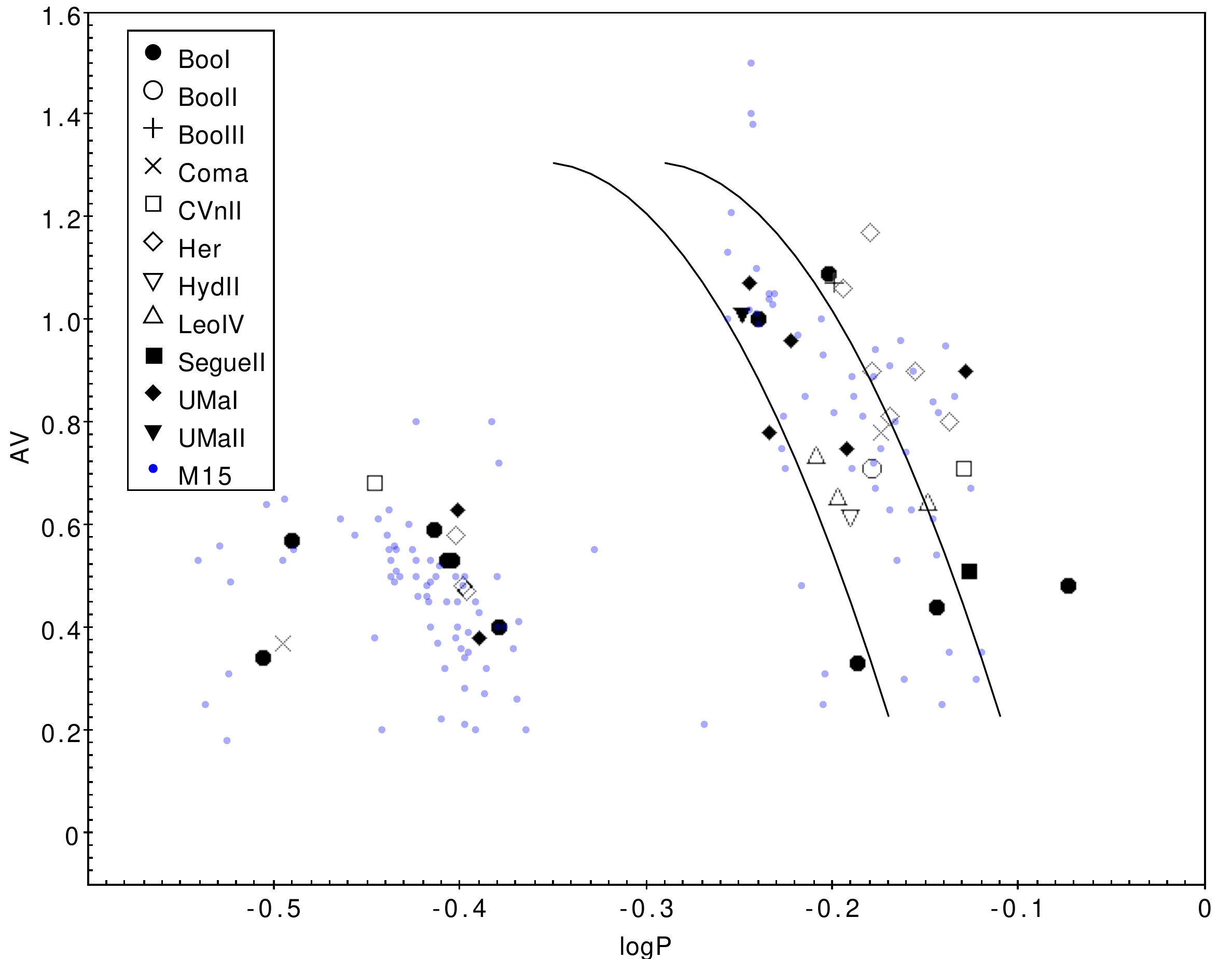}
\end{minipage}
\caption{Bailey's diagram of the RRL (black symbols of the UFD satellites of the Milky Way. Left panel shows the comparison with the Oo~I Galactic Globular cluster M3 (red filled circles) and with the Oo~I and Oo~II loci (dashed lines), according to \cite{Zorotovic2010}. Right panel shows the sample comparison, but with the Oo~II Galactic Globular cluster M15.} 
\end{figure}

\subsubsection{A homogeneous distance scale}
In Table 6, we propose a homogeneous distance scale for the MW UFDs, by using the same $M_V-[Fe/H]$ relation and the same reddening calibration from \citet{SF2011}. In particular, for the RRLs luminosity we adopt the absolute magnitude of $M_V = 0.54 \pm 0.03$ mag at $\rm [Fe/H] = -1.5$ dex \citep[based on a LMC distance of $18.52 \pm 0.09$ mag from][]{Clementini2003}, with a slope of $\Delta M_V [Fe/H] = 0.214 \pm 0.047$ mag dex$^{-1}$ \citep{Clementini2003}.
We do not make any attempt to homogenize the metallicity scale, where all the values have been taken from the collection listed in \cite{McConnachie2012}, except of Boo~III \citep{Carlin2009}, and Hyd~II \citep{Kirby2015}. The uncertainties of the distance are split in intrinsic error (the standard deviation of the mean, when at least two RRLs are available, or the typical photometric error when only one variable is present), plus a contribution due to the uncertainty in the $M_V-$$\rm[Fe/H]$ relation. As a matter of fact, there is also another source of uncertainty, that is the internal metallicity spread observed in several UFDs, but it is difficult to manage. The spread can be of the order of $\sim 0.6$ dex \citep[UMa~II,][]{Kirby2008}, which means an additional uncertainty up to $\sim 0.15$ mag.
In computing the distances of the individual galaxies, we dropped the variables V1 in Coma, V4 in UMa~I e V5 in Boo~I, since they are significantly brighter ($\sim 0.2$ mag) than the others and/or of the HB, and may be evolved variables not representative of the zero-age HB level.

The current radial distribution of both UFD and classical dwarf spheroidals (see Monelli contribution 
this book) seems to be quite similar. However, the uncertainties affecting the individual distances of 
UFDs are still too large. Individual distances based on the use of optical and/or near Infrared PL relations 
will be crucial to further constrain the similarities in radial distribution of gas poor stellar systems.

\begin{table}
\small
\begin{tabular}{cccc}
\hline\noalign{\smallskip}
Name & [Fe/H]   & $A_V$  & $(m-M)_0$  \\
     & (dex)    & (mag)  &  (mag)      \\
\noalign{\smallskip}\hline\noalign{\smallskip}
SegueII & $-2.00 \pm 0.25$ & 0.507 & $17.31 \pm 0.03 \pm 0.06 $ \\
UMaII   & $-2.47 \pm 0.06$ & 0.257 & $17.65 \pm 0.04 \pm 0.03 $ \\
UMaI    & $-2.18 \pm 0.04$ & 0.054 & $19.98 \pm 0.04 \pm 0.03 $ \\
LeoIV   & $-2.54 \pm 0.07$ & 0.069 & $21.10 \pm 0.03 \pm 0.04 $ \\
HydII   & $-2.02 \pm 0.08$ & 0.167 & $20.96 \pm 0.04 \pm 0.03 $ \\
Coma    & $-2.60 \pm 0.05$ & 0.046 & $18.09 \pm 0.03 \pm 0.03 $ \\
CVnII   & $-2.21 \pm 0.05$ & 0.027 & $21.06 \pm 0.02 \pm 0.03 $ \\
BooII   & $-1.79 \pm 0.05$ & 0.084 & $17.67 \pm 0.04 \pm 0.03 $ \\
BooIII  & $-2.1\hspace{0.15cm}  \pm 0.2~~$  & 0.058 & $18.09 \pm 0.04 \pm 0.05 $ \\
BooI    & $-2.55 \pm 0.11$ & 0.047 & $19.18 \pm 0.06 \pm 0.04 $ \\
Her     & $-2.41 \pm 0.04$ & 0.171 & $20.75 \pm 0.05 \pm 0.03 $ \\	 
\noalign{\smallskip}\hline
\end{tabular}
\caption{RR Lyrae-based distances to the UFD satellites of the Milky Way. Distances have been estimated according to the $M_V-$$\rm [Fe/H]$ relation provided by \cite{Clementini2003}. $A_V$ absorptions are based on the calibration by \cite{SF2011}.}

\end{table}


\section{On the absolute and relative ages of globular clusters}

The early estimates of the ages of globular clusters date back to more than 
half a century ago thanks to the pioneering papers from Alan Sandage 
\citep{sandage58} and 
Halton Arp \citep{arp62}
for the observational aspects and from Fred Hoyle \citep{hoyle59} 
and Martin Schwarzschild \citep{schwarzschild70} for theoretical analyses. 
The reader interested in a more detailed discussion concerning the dawn of 
cluster age determination is referred to the seminal presentations and 
discussions of the 
1957 Vatican Conference \citep{oconnell58}. 
Particularly enlightening 
was 
the empirical evidence brought forward by Walter Baade \citep{baade58b} 
concerning the age difference among the different stellar populations belonging 
to the Galactic components (Halo, Disk, Bulge).  
%
\subsection{Absolute cluster age estimates} \label{absolute}
%
In the following we will focus on the most reliable classical methods used
to estimate absolute ages of Galactic Globular Clusters (GGCs). The
first two rely on deep photometry of individual stars of a GGC down to
the Main Sequence Turn Off (MSTO) and the white dwarf cooling sequence
features, respectively. The second is observationally based on the detection of
radioactive heavy elements in individual stellar spectra 
in order to use 
direct cosmochronometry. 
We will highlight strengths and weaknesses of their application.

\subsubsection{The Main Sequence Turn Off}

The MSTO of a cluster is identified as the bluest point along 
its Main Sequence. This is the most important clock to date for both open and globular 
stellar systems.  The key advantages of this diagnostic are the following:

i) The anti-correlation between cluster age and brightness of MSTO stars is 
linear over a broad range of stellar ages \citep[see e.g.,][]{dicecco15b,valle13}. 


ii) Stars in this evolutionary phase are burning hydrogen in the core. This 
means that they evolve on a long
nuclear time scale, and therefore, the number of 
stars per unit magnitude tracing this evolutionary phase is quite large 
compared with evolved phases.

iii) Accurate apparent optical magnitudes of MSTO stars in 
GGCs are within the capability of 2-4~m class telescopes equipped with
CCD detectors
and can be easily measured.\\

\begin{figure*}
\includegraphics[width=1.14\textwidth]{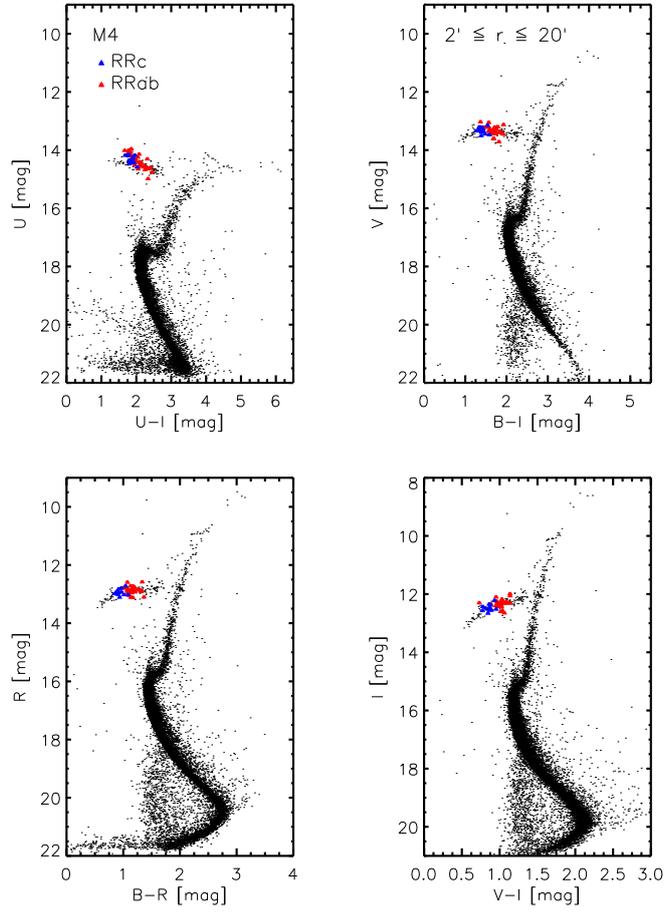}
\caption{
Optical Color-Magnitudes Diagrams (CMDs) based on UBVRI bands of the Galactic 
star cluster M4. Stars plotted in the CMDs were selected according to the 
position.  Blue and red triangles mark the position of first overtone (RRc) 
and fundamental (RRab) cluster RR Lyrae. Note the change in the slope of the 
Horizontal Branch (HB) when moving from the top left panel (U-I,U) to the 
bottom right panel (V-I,I). Bright red giant stars located across 
the tip of the red giant branch are missing due to saturation problems. The 
vertical plume of stars located at B-I$\sim$2.0-2.5, B-R$\sim$1.8-2.0  
and V-I$\sim$1.2-2.5 mag is caused by field star contamination.      
} 
\label{cmdopt}
\end{figure*}

\noindent The main cons of the MSTO are the following: 

i) The MSTO is prone to uncertainties on cluster distance and on cluster 
reddening. An uncertainty of 10\% in the error budget (reddening plus 
true distance modulus) of the MSTO, implies an uncertainty of about 1 Gyr in cluster 
age. The problem becomes even more severe if we are dealing with stellar 
systems either affected by large or by differential reddening. 

ii) The identification of the MSTO is not always trivial. In a broad range 
of stellar ages and chemical compositions, stars across the MSTO attain in 
optical bands similar colors and magnitudes.
In some traditional broad-band color-magnitude systems 
there is nearly a vertical distribution of MSTO stars \citep[e.g.,][]{salaris05}.
Fig.~\ref{cmdopt} shows the optical (UBVRI) CMDs of the Galactic 
globular M4 \citep{stetson14a,braga15}. The shape of the 
MSTO changes from ``cuspy" in the U-I,U CMD (top left panel) to 
``almost vertical" in the V-I,I CMD (bottom right panel).  

\begin{figure*}
\includegraphics[width=0.99\textwidth]{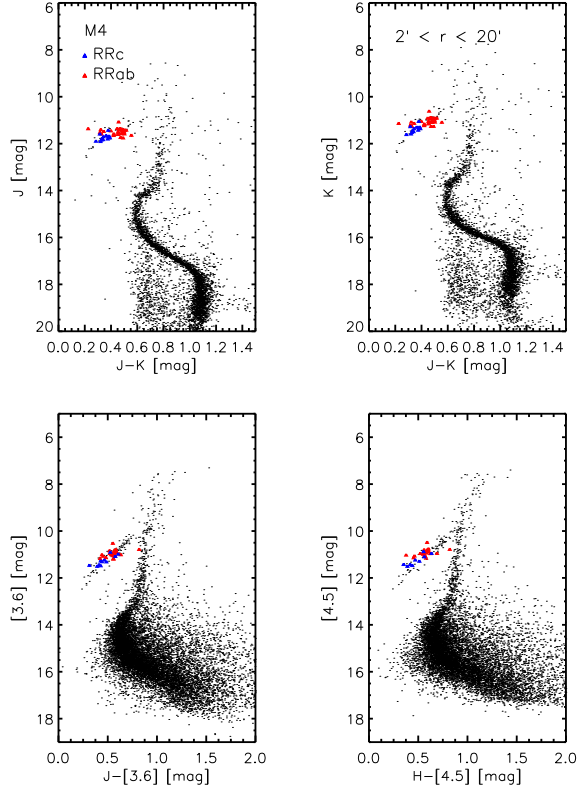}
\caption{ 
Same as Fig.~\ref{cmdopt}, but for NIR (JHK) and MIR ([3.6],[4.5]) bands. 
} 
\label{cmdnir}
\end{figure*}

Fortunately the variation in the color gradient of the 
region across the MSTO becomes more evident
 in the NIR and in the MIR regime. 
Data plotted in the top panel of Fig.~\ref{cmdnir} show that cluster 
stars display not only a well defined bending 
in the region across the MSTO, but also a sharp change in the slope of the 
lower main sequence across the Main Sequence Knee (MSK, see \S~5.4). 
A glance at the data plotted in the bottom 
panels shows that the bending across the MSTO is also in NIR/MIR CMDs. However, 
the photometric accuracy of the MIR bands do not allow us to clearly identify 
the MSK.  

\begin{figure*}
\includegraphics[width=0.99\textwidth]{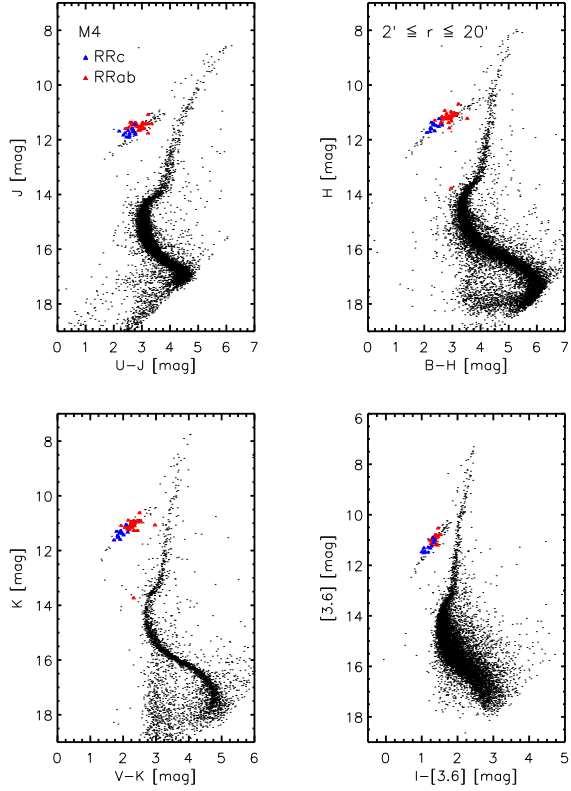}
\caption{ 
Same as Fig.~\ref{cmdopt}, but for optical and NIR/MIR bands. 
} 
\label{cmdopir}
\end{figure*}

The empirical scenario concerning the shape of both the MSTO and the MSK becomes 
even more interesting in dealing with optical/NIR/MIR CMDs, since they bring forward several advantages:  
a) They display a well defined bending across the MSTO and a sharp change 
in the MS slope across the MSK;  b) The broad range in central wavelengths 
among optical and NIR/MIR bands means also a strong sensitivity to the effective 
temperature. The consequence is that optical/NIR/MIR 
CMDs
display tight stellar sequences not only along the MS, but also in 
advanced evolutionary phases (RGB, HB, AGB).   

iii) Optical CMDs are affected in the region across the MSTO by field star 
contamination. This problem is less severe in optical/NIR CMDs, 
since the MSTO stars attain colors that are systematically bluer 
than typical field stars. The bottom left panel of Fig.~\ref{cmdopir}
shows that field stars are typically redder ($\rm V-K=2.8-3.5$) than MSTO 
stars ($\rm V-K\simeq$2.6).  However, field star contamination severely affects 
age and structural parameters of nearby stellar systems. 
Accurate and deep optical CMDs based on images collected with ACS on board of HST are less 
affected by the contamination of field stars. The pointing is typically located 
across the center of the cluster and in these regions cluster stars outnumber 
field stars. This rule of thumb does not apply to clusters either located or 
projected onto crowded Galactic regions such as the Bulge and Galactic thin disk
\citep{zoccali01b,ferraro09a,lagioia14} or nearby dwarf galaxies \citep{kalirai12}. 
A significant step forward in dealing with this problem was 
provided by HST photometry. 
The superb image quality of HST optical images provided the opportunity to 
measure the proper motion using images collected on a time interval of 
the order of ten years. This approach provided 
the opportunity to split not only field and cluster stars, but also to clearly 
identify stellar populations belonging to 
Sagittarius \citep{anderson02,king05,massari13,milone14}.     
The great news in this context is that similar results can also be obtained 
using NIR images as a second epoch collected with AO systems available at 8-10~m 
class telescopes \citep[e.g., NGC~6681,][]{massari16b}.

The results mentioned above are based on images that only cover 
a few arcminutes around the center of each cluster. 
The separation between cluster and field stars is 
much more challenging away from cluster centers,
 since field stars outnumber cluster stars in these regions.  
The reason why we are interested in tracing cluster stars in external 
cluster regions is twofold: 
a) There is mounting empirical evidence that stellar populations change 
(chemical composition, age) as a function of radial distance -- e.g.,
47 Tuc, \citep{kalirai12}; Omega Cen, \citep{calamida17};
b)  The estimate of structural parameters depends on the star counts in the 
outskirts of the clusters.  

These are the reasons why new photometric approaches for the separation 
between field and cluster stars are required. This approach does require 
photometric catalogs based on at least three photometric bands. To accomplish 
the goal we have used either multi-dimensional ridge lines as in dealing with 
$griz$ SDSS photometry of the metal-rich globular M71 \citep{dicecco15b} or 
the difference in the spectral energy distribution in dealing with the $ugri$
photometry of the complex globular $\omega$ Cen \citep{calamida17}. 
The latter approach was developed to deal with stellar systems characterized 
by multiple stellar populations \citep[see e.g.,][]{MartinezVazquez2016,MartinezVazquez2016b}.  

Initially, the ridge lines of the different sub-populations 
in $\omega$ Cen along the cluster evolutionary sequences (MS, SGB, RGB) 
were estimated. They are based on a 3D CMD (magnitude, color index, 
star counts) and the ridge lines trace the peaks of the stellar 
distribution.  
The horizontal branch (HB) stars are typically neglected, since they are
either bluer (hot and extreme HB) or they can be easily distinguished 
(RR Lyrae stars, red HB) from the field stars. 
These ridge lines were estimated in an annulus 
neglecting stars located in the innermost and in the outermost cluster 
regions and using several cuts in radial distance and in photometric 
accuracy. Once the ridge lines were estimated we performed a linear 
interpolation among them and generated a continuous multi-dimensional 
surface. Subsequently, the separation between field and cluster stars
was performed in two steps:
a) We estimated the total standard deviation
among the position of individual stars and the reference
surface;
b) We estimated the distance in magnitude and in color among 
the individual stars and the reference surface.

\begin{figure*}
\includegraphics[width=0.99\textwidth]{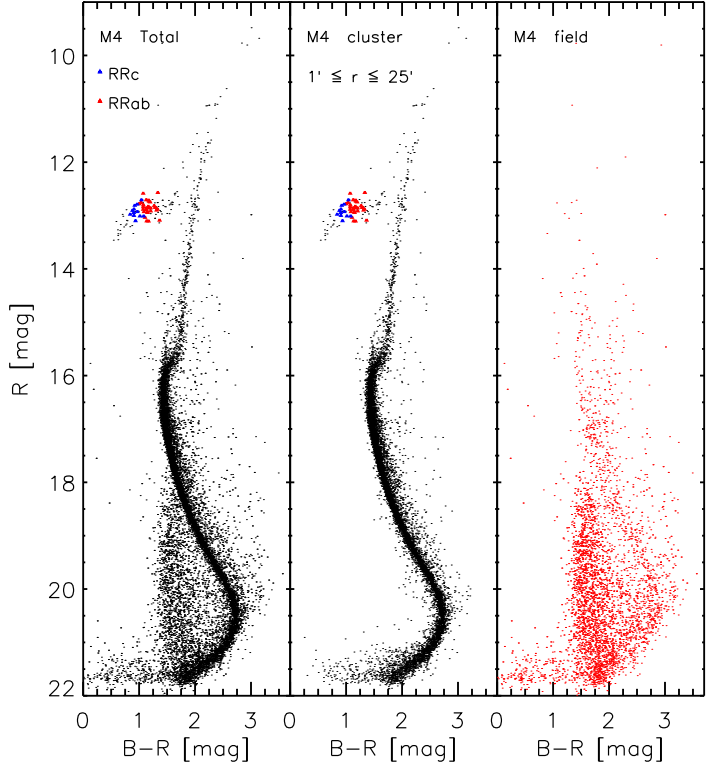}
\caption{ 
Left -- Optical CMD in B-R,R for stars covering a sky area of 35$\times$35 arcmin 
squared around M4. 
Middle -- Same as the left panel, but for candidate cluster stars. 
Right -- Same as the left panel, but for candidate field stars.  
} 
\label{cmdoptsepa}
\end{figure*}

The approach discussed above relies on a conservative assumption, 
i.e.  we do prefer to possibly lose some of the candidate cluster 
stars instead of including possible candidate field stars. Note that 
this assumption is fully supported by the fact that the age diagnostics 
we use for estimating cluster ages depend on the shape of MSTO and/or 
of the MSK. Stellar completeness mainly affects cluster ages based on 
cluster luminosity functions \citep{zoccali98}. 

Fig.~\ref{cmdoptsepa} shows the separation between field and cluster stars for the 
Galactic globular M4. This cluster is an acid test for the selection criteria 
based on photometry, since it is projected onto the Galactic Bulge and it is 
also affected by differential reddening. To overcome some of these problems 
we decided to follow the same approach we adopted for $\omega$ Cen, but the 
initial ridge lines were derived candidate field stars located outside the 
tidal radius of the cluster (Ferraro et al. 2018, in preparation). Data plotted 
in the left panel display stars located across the sky region covered by M4 in 
the B-R,R CMD. Field stars can be easily identified both along the MS and the RGB. 
The middle panel of the same figure shows the CMD, but for candidate cluster stars. 
The ``cleaning" appears quite good across the cluster sequences. Note that the 
sequence running parallel to the cluster MS is due either to binaries or 
to photometric blends. The plausibility of the criteria adopted to separate 
field and cluster stars are further supported by the CMD of candidate field 
stars plotted in the right panel. Once again there is evidence that a minor 
fraction of candidate field stars were misidentified, but the main 
peak of field stars even across the MSTO was properly identified.    
 
\begin{figure*}
\includegraphics[width=1.15\textwidth]{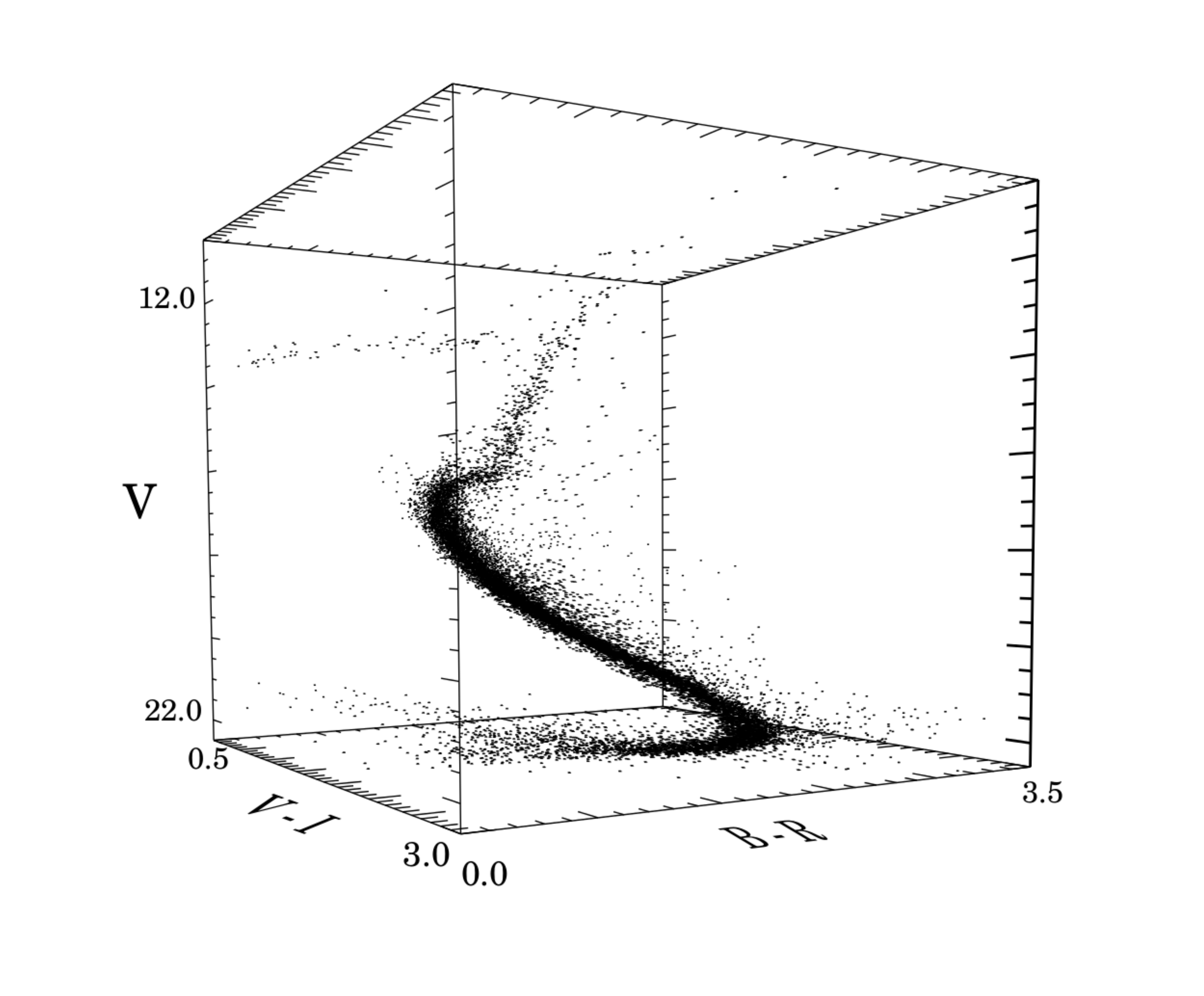}
\caption{Optical V, B-R, V-I color-color-magnitude diagram of candidate cluster 
stars for the globular M4 (see text for more details.}
\label{fig_vbrvi}  
\end{figure*}

To further support the photometric criteria adopted to separate field and 
cluster stars, Fig.~\ref{fig_vbrvi} shows a color-color-magnitude diagram of the 
selected candidate cluster stars. It is worth mentioning the smoothness of 
the cluster sequences when moving from the RGB to the MSTO and to the MSK.  

\subsubsection{The error budget} \label{error_budget}

The global error budget of absolute ages of globular clusters based on the 
MSTO includes theoretical, empirical and intrinsic uncertainties.\\

{\em Theoretical uncertainties} -- The precision of the clock adopted to date stellar 
systems depends on the precision of the input physics adopted to construct 
evolutionary models and, in turn, cluster isochrones. The main sources of 
uncertainties can be split into 
micro-physics
(nuclear reactions, opacity, equation of state, 
astrophysical screening factors) 
and macro-physics (mixing length, 
mass loss, atomic diffusion, rotation, radiative levitation). 
The uncertainties affecting the nuclear reactions, 
such as
the $pp$ and the $CNO$ cycle, have 
already been discussed by \cite{deglinnocenti04} and \cite{valle13}. 
They suggest that the uncertainties on age range from 3\% for the $^1H(p,\nu_e,e^+)^2H$ 
nuclear reaction to roughly 10\% for the $^{14}N(p,\gamma)^{15}O$ reaction.   
The same authors also suggest an uncertainty of the order of 5\% for the radiative 
opacities adopted for constructing MS and HB models. Moreover, they also suggest a 
similar uncertainty (5\%) for the conductive opacities affecting the energy transport in 
electron degenerate isothermal helium cores typical of RGB structures and, in turn, on 
the HB luminosity level \citep[e.g.,][]{marta08,chaboyer17}.

The impact of macro-physics, and in particular the treatment of the mixing in 
convective regions has been addressed 
comprehensively in the literature \citep[e.g.,][]{chiosi86,castellani99,maeder00,salaris08b,bertelli09,rosenfield17}). 
The treatment of atomic diffusion and radiative levitation 
has also been investigated in many studies
\citep{proffitt91,ciacio97,michaud04,vandenberg13}.

Finally, to transform theory into the observational plane, we also need predictions 
based on stellar atmosphere models \citep{gustafsson08}. We have to take account of 
uncertainties affecting bolometric corrections and  
color--temperature transformations. The impact that the quoted ingredients 
have on cluster isochrones has been discussed in detail in the
literature \citep[e.g.,][]{pietrinferni04,cassisi08,pietrinferni09,sbordone11,vandenberg13,salaris16b}. 

As a whole, the typical theoretical uncertainty in the adopted MSTO is of the 
order of 10\%. In this context it is worth stressing that theoretical 
uncertainties mainly affect the zero point of the absolute age determinations.
The relative age determinations are minimally affected by these 
uncertainties \citep[see e.g.,][]{Dotter2011,vandenberg13,chaboyer17}.

In this context, it is also worth mentioning that the helium to metal
enrichment ratio ($\Delta\,Y/\Delta\,Z$), a fundamental ingredient in
evolutionary prescriptions, is poorly known. This is used to derive
the current stellar helium content through the following relation: 
$Y = Y_P + \Delta\,Y/\Delta\,Z \times Z$, where $Y_P$ is the primordial 
helium content.  
However, empirical estimates range from $\Delta\,Y/\Delta\,Z$=3$\pm$2 \cite{pagel98}  
to $\Delta\,Y/\Delta\,Z$=5.3$\pm$1.4 \cite{gennaro10}. Moreover and even more importantly,
there are no solid reasons why this relation should be linear over the 
entire metallicity range and that the current local estimate is universal
\citep{peimbert10}.\\  

{\em Empirical errors} -- The main uncertainties in the absolute age estimate
of globulars are the cluster 
distances.  Indeed
an error of 10\% in true distance modulus 
($\Delta \mu_0 \sim$0.1 mag) implies an uncertainty 
of $\sim$1 Gyr in the absolute age. The error budget becomes even more severe 
once we also account for uncertainties in reddening corrections. An uncertainty 
of 2\% in color excess implies an uncertainty of 6\% in visual magnitude. The 
impact of this limitation becomes even more stringent if the cluster is 
also affected by differential reddening.  In this context, it is worth mentioning 
that more metal-rich globular clusters are mainly distributed across the Galactic 
Bulge, $i.e.,$ a region of the Galactic spheroid affected by large reddening and by 
differential reddening \citep{stetson14a}.

The ongoing effort in trying to quantify the systematics affecting the 
measurements of the MSTO 
of a large fraction of GGCs also faces
 the problem of the absolute photometric calibration. 
In handling this thorny problem, two independent approaches 
have provided the opportunity to limit and/or to overcome 
the uncertainties associated with the zero-points of different photometric 
systems: 

i) Cluster photometry collected either with 
HST/WFPC2 or HST/ACS has provided the unique opportunity 
to collect accurate and deep photometry in a single well-calibrated
photometric system
for a sizable sample of GGCs \citep[$\sim$60 out of 160, 
e.g.,][]{zoccali01,sarajedini07,marinfranch09,vandenberg13};

ii) The major effort to provide local 
standard stars in GGCs also has significantly improved the accuracy of 
photometry and saved
significant amount of telescope time \citep{stetson00}.   

The age diagnostic adopted to estimate the cluster age also depends on the iron content.
The massive use of multi-object fiber spectrographs paved the way for the definition of a firm 
metallicity scale including a significant fraction of 
GGCs \citep[e.g.,][]{carretta09}. This translates to a systematic decrease in the uncertainties 
on the iron and on $\alpha$-element abundance scale.\\

{\em Intrinsic uncertainties} -- Dating back to more than forty years ago, 
spectroscopic investigations brought forward a significant star-to-star 
variation in C and in N among cluster stars \citep[e.g.,][]{osborn71}. 
This evidence was 
complemented by discoveries of variations
 in Na, Al, and in O \citep[e.g.,][]{cohen78,pilachowski83,leep86} 
and by anti-correlations in CN--CH \citep[e.g.,][]{kraft94} 
as well as in O--Na and in Mg--Al \citep[e.g.,][]{suntzeff+91,gratton12}.

The light element abundance variation was
further strengthened by the occurrence of multiple
stellar populations in more massive clusters \citep{bedin04,piotto05,piotto07}.
However, detailed investigations  concerning the different stellar populations
indicate a difference in age that is, in canonical GGCs (i.e. the most massive 
globulars are not included), on average shorter than
1~Gyr  \citep{ventura01,cassisi08}. The intrinsic uncertainty does not seem to
be the main source in the error budget of GGCs absolute ages.

\subsubsection{The cluster distance scale} \label{empirical_absolute}

Several approaches have been suggested in the literature to overcome some of the 
thorny problems affecting the estimate of the absolute age of
GGCs. The time scale within which we can significantly reduce the theoretical 
uncertainties can be barely predicted. On the other hand, the empirical 
uncertainties are strongly correlated with technological developments either 
in the detectors or in the observing facilities or in both. The uncertainties 
affecting individual cluster distances would significantly benefit by the 
development of a {\em homogeneous cluster distance scale}.

Although Galactic globulars are 
at the cross-road of major
theoretical and empirical efforts, we still lack a distance scale
based on the same diagnostic and on homogeneous measurements. The
problem we are facing is mainly caused by the intrinsic limitations 
in the diagnostics used to determine the distance. 
The difficulties of these methods can be summarized briefly.

i) The tip of the Red Giant Branch (TRGB) can only be applied to two clusters 
($\omega$ Cen, 47 Tuc), due to limited stellar 
(Poisson) statistics when approaching the tip of the RGB.

ii) Main sequence fitting is hampered by the fact that the number of 
field dwarfs with accurate trigonometric parallaxes is small and covers 
a limited range in iron abundance \citep{chaboyer17,vandenberg14a}. 

iii) Another possible approach to constrain cluster
distances is to fit the white dwarf cooling sequence. 
However, this method can only be applied to 
a few clusters \citep{zoccali01,richer13,bono13}
and we are still facing some discrepancies between the distances based on 
this diagnostic and other distance indicators \citep{bono08b}. 
Furthermore, the number of nearby White Dwarfs (WDs) for which are available accurate 
trigonometric parallaxes is limited. 

iv) Cluster distances based on the kinematic approach appear also to 
be affected by systematics. This diagnostic has the potential to be a powerful 
geometrical method, since it is only based on the ratio between the standard 
deviation  of proper motions and the standard deviation of radial velocities. 
There is mounting evidence that the current kinematic distances are slightly 
larger than those based on other primary distance indicators \citep{bono08b}.  

v) During the last few years, accurate distances have been provided for a 
few globulars using eclipsing binary stars. 
This is also a very promising approach, since it is based on a geometrical 
method \citep{thompson01}.    
The main limitation of 
this distance indicator as well as methods iii) and iv)
is that they are challenging from the observational point of view. 
This means that they have only 
been applied to a very limited number of clusters.  

vi) The Horizontal Branch (HB) luminosity level is one of the most 
popular distance indicator  for GGCs. 
The typical anchor for the HB luminosity is the 
$M_V^{RR}$--metallicity relation at the mean color of the RR Lyrae instability 
strip. 
This color region
 was selected because it is quite flat in CMDs using the 
$V$-band magnitude. This distance indicator might be affected by two possible 
sources of systematic errors:\par
a) The HB morphology -- The metallicity is the most important parameter 
affecting HB morphology, indeed the HB is mainly blue in the metal-poor 
regime and becomes mainly red in the metal-rich regime. 
The parameter used to trace the change in the HB morphology is
$\tau_{HB}$=(B-R)/(B+V+R), i.e. 
the number ratio among stars either bluer
(B) or redder (R) than variable (V) RR Lyrae stars. 
Globulars approaching the two extrema ($\pm$1) typically lack of RR Lyrae 
stars. This means that it is quite difficult from an empirical point of view 
to anchor the HB luminosity level. The impact of this limitation becomes even 
more stringent for globulars more metal-rich than 47 Tuc, since their HBs 
typically display a stub of red HB stars. There are two exceptions, NGC~6441 
and NGC~6388, that are metal-rich but display a well populated blue and red 
HB, plus a good sample of RR Lyrae stars \citep{pritzl03};\par
b) Predicted HB luminosity level -- The current evolutionary prescriptions 
concerning the zero age HB (ZAHB) luminosity are typically 0.10--0.15 mag 
brighter than the observed ones. 
New interior conductive opacities \citep{cassisi07} 
contribute to alleviate the problem, but the discrepancy is still present.  

vii) Near-infrared and mid-infrared observations of cluster RR Lyrae stars appear 
to be very promising for providing a homogeneous cluster distance scale. 
The RR Lyrae stars do obey to well defined PL relations in these bands. 
These relations are very narrow \citep[e.g.,][]{braga15,Marconi2015,neeley17} 
and are not affected by off-ZAHB evolution, thus reducing the systematics 
discussed above. 
NIR photometry has also the advantage of being less prone to 
uncertainties introduced by either large and/or by differential reddening 
corrections. This is a typical problem among the more metal-rich Bulge globulars.  
Indeed, the uncertainties affecting NIR/MIR bands are on average one order 
of magnitude smaller than the optical bands. 
However, the number of globulars hosting a sizable sample of RR Lyrae stars 
is roughly half of the entire Galactic sample.
Finally, 
we still lack a detailed knowledge on how the reddening law changes when moving 
across the Bulge, along the Bar and into the nuclear Bulge 
\citep[e.g.,][]{indebetouw05,nishiyama06,nishiyama08,nishiyama09,Nataf_etal2013,schultheis14}.   

\subsubsection{The white dwarf cooling sequence} \label{wd}
The white dwarf cooling sequence depends on different physics
than the previously-discussed methods, and as such should be an excellent 
diagnostic to constrain possible systematics in absolute cluster age 
estimates based on the MSTO \citep{hansen04,salaris10}.
This means that deep and accurate photometry of nearby globulars can provide 
crucial constraints on the plausibility of the physical assumptions adopted 
to construct either main sequence or cooling sequence models. The key advantages
in using cluster WD cooling sequences is that they display a well defined 
blue turn off (WDBTO) in deep and accurate I,V-I CMDs. Theory and observations indicate 
that this change in the slope of the WD cooling sequence is due to interplay between 
an opacity mechanism called Collisional Induced Absorption (CIA) mainly from molecular 
hydrogen and/or to the cooling time of more massive WDs 
\citep{brocato99,hansen04b,richer06,moehler08,salaris10,omalley13}.     

\begin{figure}
\includegraphics[width=0.95\textwidth]{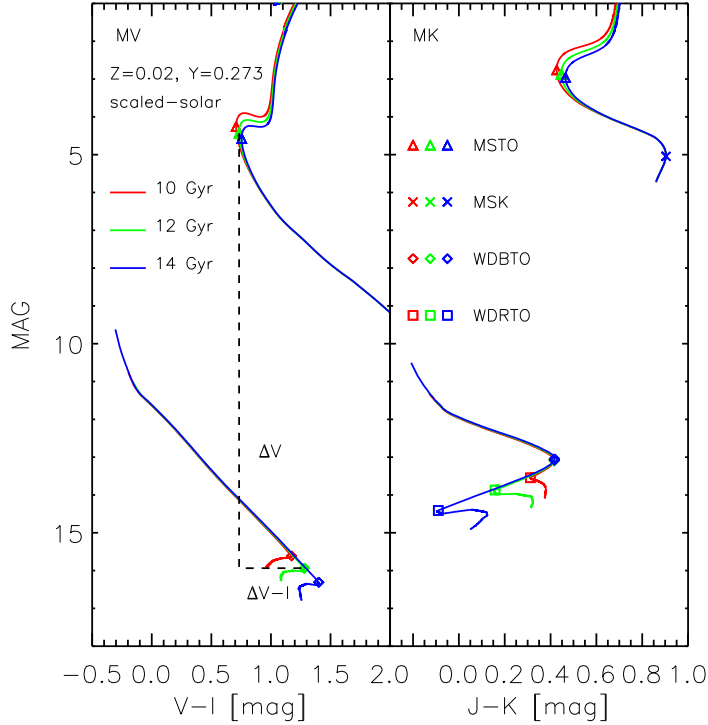}
\caption{Left -- Predicted optical (V-I,V) CMD for cluster isochrones 
at solar chemical composition and at scaled solar chemical mixture. The 
bright solid lines show isochrones of 10 (red), 12 (green) and 14 (blue) Gyr 
(BaSTI data base). The triangles mark the position of the Main Sequence Turn Off 
(MSTO). The 
faint solid lines show WD isochrones of 10 (red), 12 (green) and 14 
(blue) Gyr. The WD isochrones account for chemical separation 
\citep{salaris10}. The diamonds mark the position of the 
White Dwarf Blue Turn Off (WDBTO). 
The vertical and horizontal black dashed lines show the difference 
in magnitude and in color between the 12 Gyr MSTO and the WDBTO. 
Right -- Same as the left, but in the NIR (K,J-K) CMD. The triangles mark 
the position of the MSTO, the crosses mark the MS Knee \citep[MSK,][]{bono10a}, 
the diamonds mark the WDBTO, while the squares mark the 
White Dwarf Red Turn Off (WDRTO).
}
  \label{f1_ms_wd_iso}
\end{figure}

Recently, \cite{bono13}
performed a detailed theoretical investigation of cluster 
WD cooling sequences and found new diagnostics along the WD cooling 
sequences and in NIR Luminosity Functions (LFs).  
The interplay between CIA and the cooling time of progressively 
more massive WDs causes a red turn-off along the WD cooling sequences (WDRTO). This feature 
is strongly correlated with the cluster age, and indeed the faint peak in the K-band 
increases by 0.2--0.25 mag/Gyr in the range 10--14 Gyr. Moreover, they also suggested to 
use the difference in magnitude between the MSTO and the WDRTO, since this age diagnostic 
is independent of distance and reddening. 
These predictions appear very promising
in view of the unprecedented opportunity offered by JWST in the 
NIR/MIR regimes. 
This encouraging prospect also applies to ground-based extremely large 
telescopes equipped with state-of-the-art
multi-conjugated adaptive optics systems \citep{bono13}.

\subsubsection{Thorium Cosmochronometry} \label{tc}

\begin{figure}
\includegraphics[width=0.95\textwidth]{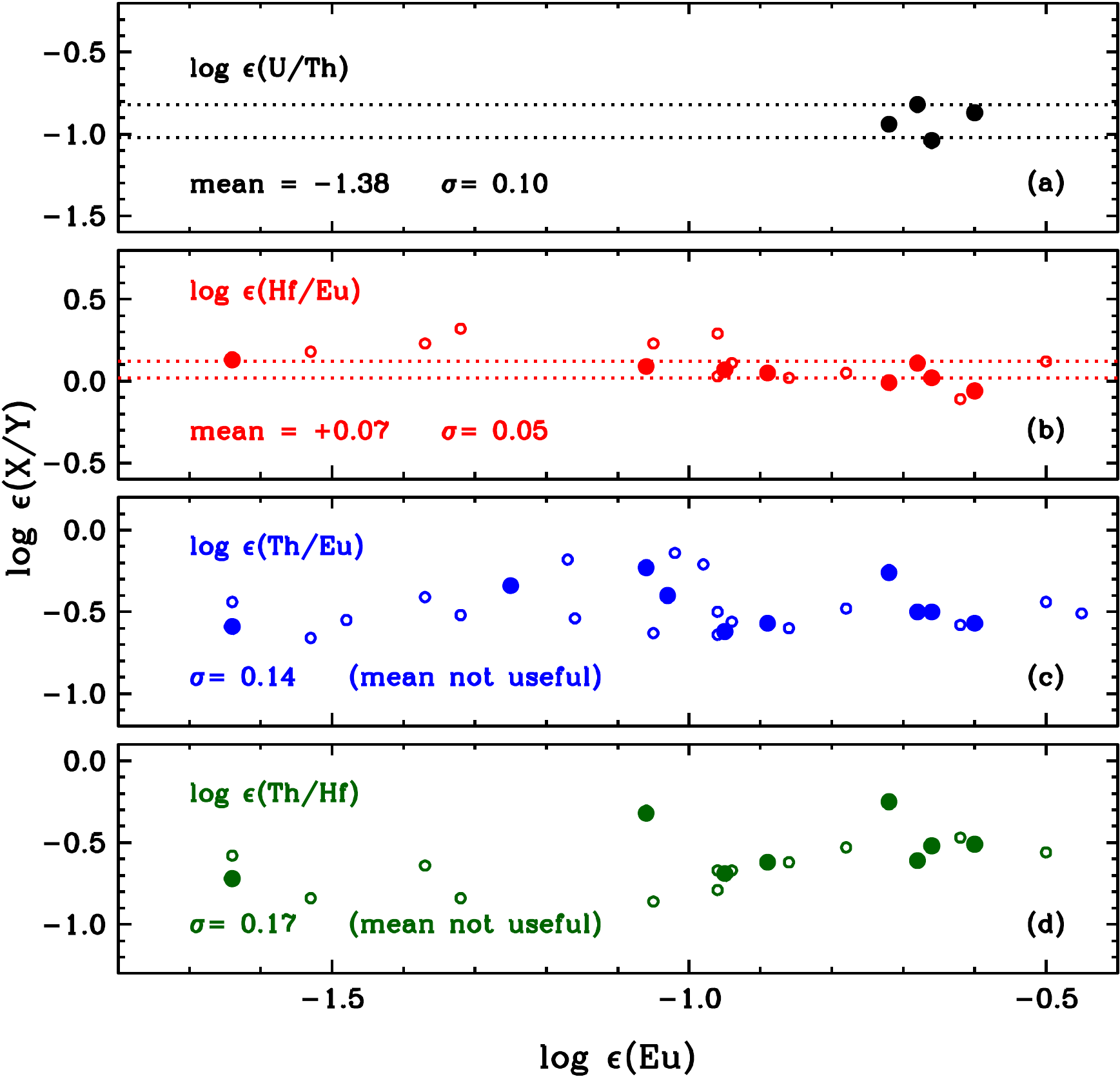}
  \caption{Abundance ratios of elements of interest to cosmochronometry
           plotted as functions of the Eu abundance.
           The large filled circles are from individual high-resolution
           spectroscopic studies \cite{westin00,hill02,cowan02,sneden03,hayek09,siqueira2013,mashonkina14,hill16},
           and the small open circles are from
           a single survey of many $r$-process-rich stars \cite{roederer09}.
           Means and standard deviations, computed only for the individual
           high-res studies, are written in each panel.
           Only in panels (a) and (b) are dotted lines drawn to indicate
           the $\sigma$ widths around the means; the scatters in panels
           (c) and (d) are too large to be useful for cosmochronometry.}
  \label{eucorrs4}
\end{figure}

Elements beyond the iron-group are overwhelmingly created in neutron-capture
fusion reactions by target heavy-element seed nuclei.
Their syntheses mostly are either slow ($\beta$-decay timescales fast
compared to neutron-capture timescales; called the $s$-process) or rapid
(neutron-capture timescales much faster than $\beta$-decay ones; the
$r$-process).
Very heavy radioactive elements thorium (Z~=~90) and uranium (Z~=~92) can
be created only via the $r$-process.
The heaviest stable element is bismuth (Z~=~83, with its sole natural
isotope $^{209}$Bi).
All isotopes of elements with Z~=~83$-$89 have very short half-lives, and
therefore cannot be created in the $s$-process.
Th and U decay on astrophysically interesting timescales: half-lives are
$1.4\times10^{10}$~yr for $^{232}$Th (its only naturally-occurring isotope),
$7.0\times10^{8}$~yr for $^{235}$U, and
$4.5\times10^{9}$~yr for $^{238}$U.
Therefore, derived abundances of Th and U with respect to stable
$r$-process-dominated neutron-capture elements in low-metallicity halo stars
have the potential to be translated into Galactic age estimates.

Attempts to use stellar Th abundances as chronometers began with 
\cite{butcher87}.
That study included only disk stars with metallicities [Fe/H]~$\geq$~$-$0.8.
As is the case for most Th abundance studies, \cite{butcher87} analyzed
just the 4019.2~\AA\ Th~II transition.
That line in high metallicity stars is at best a weak blending absorption
in this crowded spectral region.
Additionally, for a comparison stable neutron-capture element Nd was chosen
in the \cite{butcher87} study.
Unfortunately, Nd in the solar system (and probably in most disk stars) has a
$s$-process origin, accounting for $\simeq$58\% of its abundances;
the $r$-process fraction is 
only $\simeq$42\% (e.g. \cite{sneden08}).
With these limitations, \cite{butcher87} argued that the Th/Nd ratios were
roughly constant in their sample of disk stars, irrespective of assumed
stellar age.

Analyses of low-metallicity halo population $r$-process-rich stars have
yielded more easily-interpreted results.
The first such star, HD~115444, initially identified by \cite{griffin82},
has [Fe/H]~$\simeq$~$-$2.9, and [Eu/Fe]~$\simeq$ $+$0.9
(e.g. \cite{westin00,sneden09}).
Then CS~22892-052, a red giant from the \cite{beers92} low resolution
Galactic halo survey, was serendipitously discovered \citep{sneden94} to be
very $r$-process rich: [Fe/H]~$\simeq$ $-$3.0, [Eu/Fe]~$\simeq$ $+$1.7, and
an unambiguously strong 4019~\AA\ line, yielding [Th/Fe]~$\simeq$ $+$1.4
or [Th/Eu]~$\simeq$ $-$0.3 \citep[e.g.,][]{sneden03}.
The relatively depressed Th compared to Eu was taken to be a sign of
radioactive decay from an $assumed$ initial production ratio of
[Th/Eu]~$\equiv$~0.0.
Initial application of theoretical models suggested an ancient age for
the neutron-capture elements but with a large uncertainty:
11.5~$\leq$~$t$~$\leq$~18.8~Gyr \citep{cowan97}.

Detailed study of another $r$-process-rich star CS~31082-001
\citep{cayrel01,hill02}
revealed the first detection of U in a low metallicity star \citep{hill02}.
But the individual abundance ratios [Th/Eu] and [U/Eu] in this star turned
out to be too large to be sensibly interpreted as a straightforward
radioactive decay.
Instead, it was necessary to postulate an ``actinide boost'' with initial
production ratios [Th/Eu]~$>$~0 and [U/Eu]~$>$~0.
Fortunately, the ratio between neighboring elements Th and U should have
well-understood production ratios and, for example, \cite{schatz02} derived
an age from the U-Th abundance ratio of 15.5~$\pm$~3.2~Gyr, consistent with
but not constraining the age of the Galaxy determined from other indicators.

The problems and prospects of U and Th abundances are considered well in
\cite{hill02} and \cite{schatz02} and will not be repeated here.
The actinide boost problem effectively forces attention on detection of
both Th and U for meaningful radioactive cosmochronometry.
But the problem is the rarity of U detections even in low metallicity
stars with extreme $r$-process enhancements (e.g. [Eu/Fe]~$>$~$+$1).
Only a single U~II transition at 3539.5~\AA\ has been detected to date,
and it is at most a very weak bump among a clump of lines dominated
by two strong Fe I lines as well as weaker Nd~II and CN lines.  This is 
shown, for example, in Fig.~10 of \cite{hill02}, Fig.~9 of \cite{cowan02},
Fig.~2 of \cite{frebel07}.

If $r$-process production ratios [Th/Eu] cannot be predicted with certainty
given the large Periodic Table stretch between elements 63 and 90, can
another element closer to Th serve as the stable comparison element?
\cite{kratz07} suggested that Hf (Z~=~72) might be a good candidate, as their
computations showed that Hf is made in the $r$-process at similar neutron
densities to Th (see their Fig.~3).
We tested this idea by considering Eu, Hf, Th, and U abundances reported
for these kinds of stars in the literature.
In Fig.~\ref{eucorrs4} panel (a) we show that for radioactive elements
U and Th, their ``absolute'' number density ratios
log~$\epsilon$(U/Th)\footnote{
log~$\epsilon${A} $\equiv$ log$_{\rm 10}$(N$_{\rm A}$/N$_{\rm H}$) + 12.0 }
are essentially constant in the high-resolution studies published to date,
and \cite{hill16} report discovery of a fifth $r$-rich star that has a nearly
identical value of this ratio.
In panel (b) we show a similarly tight correlation between the stable
rare-earth elements Hf and Eu.
Star-to-star scatter increases markedly in ratios log~$\epsilon$(Th/Eu)
and log~$\epsilon$(Th/Hf).
Observations are clearly telling us that any actinide boost of the heaviest
$r$-process elements sets in beyond Z~=~72.
Abundance data on 3$^{rd}$ $r$-process peak elements Os, Ir, Pt (Z~=~76$-$78)
and Pb (Z~=~82); see individual papers cited above, \citep[e.g.,][]{cowan05,plez04}.
But these elements: (i) have detectable transitions only from their neutral
species, whereas all other neutron-capture elements with Z~$\geq$~56 are
represented only by ionized transitions, greatly increasing the derived
abundance uncertainties in the comparison; (ii) all of their detectable
transitions are in the near-UV or vacuum-UV.
Their abundances do correlate with those of other very heavy neutron-capture
elements, but do not add effective cosmochronometry information at present.

Moreover, predicted production ratios in the $r$-process can have significant
uncertainties, as very little experimental data exist on nuclei far from
the valley of $\beta$ stability.
The influence of these uncertainties have been discussed in several papers
published after Th and U detections were announced \citep[e.g.,][]{goriely99}.
A good summary of the nuclear issues is in \cite{niu09}, who considered
uncertainties in astrophysical $r$-process fusion conditions, nuclear mass
models, and $\beta$-decay rates.
They suggest that ``the influence from nuclear mass uncertainties on
Th/U chronometer can approach 2 Gyr.''; see their Fig.~5, which shows derived
ages of HE~1523-091 \citep{frebel07} and CS~31082-001 \citep{cayrel01,hill02}
with variations in all of the quantities that can influence the conversion
of derived observational U/Th abundance ratios into final age estimates.
Their computed mean ages are 11.8~$\pm$~3.7~Gyr for HD~1523-0901 and
13.5~$\pm$~2.9~Gyr for CS~31082-001, consistent with current age estimates
of the Galaxy, albeit with substantial error bars.

Cosmochronometry from radioactive elements U and Th is promising, but not
precise enough yet to set serious age constraints.
Since uncertainties abound in all phases of this exercise, one suspects
that a prerequisite for further progress is simply more detections of U
in very low metallicity $r$-process-rich stars, which should naturally have
strong Th transitions.
One looks forward to more discoveries of more such stars in field star
surveys in the near future.

This appears as a good viaticum for future developments, since age--dating 
methods based on MSTO and/or on the MSK can only be applied to ensemble of stars. 
The key advantage of the cosmochronometric method is that it can be applied 
to individual cluster and field stars and it is independent of both distance 
and reddening. The reader interested in a more detailed discussion concerning 
age-dating of field stars is referred to \cite{salaris16b}.

\subsection{Relative cluster age estimates} \label{relative}

Relative cluster ages are used to address different astrophysical problems,
e.g. they play a crucial role 
in investigating
the early formation of the 
Galactic components dominated by old stellar populations (Halo, Bulge). 
In particular, the spread in relative ages of Halo and Bulge stars is tightly 
correlated with the timescale of the collapse of the protogalactic cloud. 
Classical results based on accurate and homogeneous photometry of GGCs 
suggested that a significant fraction of Galactic globulars are, 
indeed, coeval \citep[e.g.,][]{buonanno98,rosenberg99}.  There are reasons, mainly 
kinematic, to believe that the ones that do not follow this trend are 
clusters that have been accreted. 

\begin{figure*}
\includegraphics[width=0.99\textwidth]{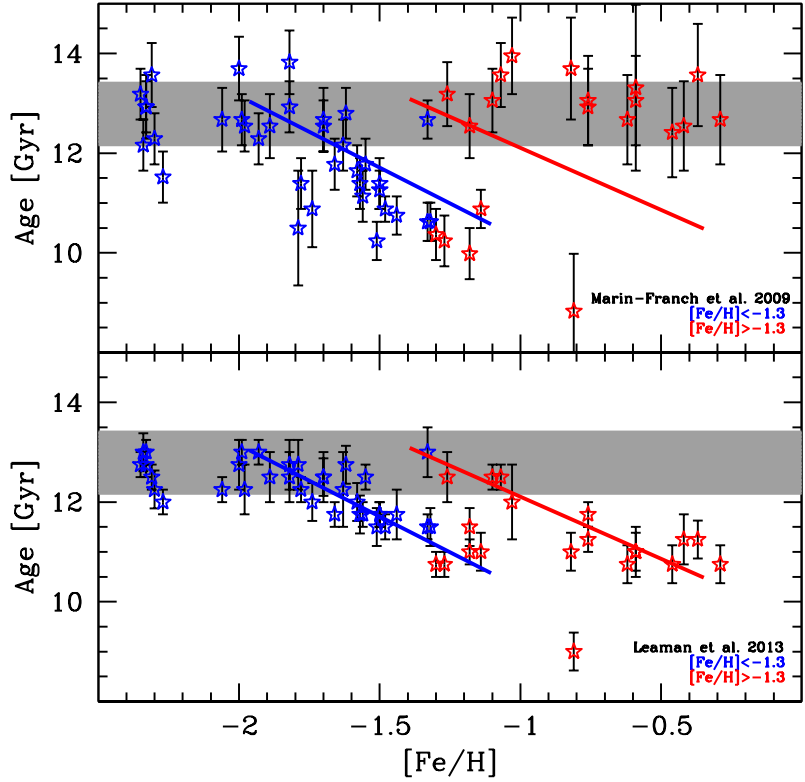}
\caption{Top: Relative age estimates of Galactic globular clusters 
according to \cite{marinfranch09}, these have been scaled to 12.8 Gyr,
i.e. a mean age coming assuming isochrones from \cite{dotter07b}. Clusters more metal-poor than [Fe/H]=-1.3 
were plotted as blue stars, while those that are more metal-rich as red 
stars. The blue and the red lines display the age--metallicity relations 
estimated by \cite{vandenberg13} and by \cite{leaman13}. The 
metallicities of the GCs are in the \cite{carretta09} metallicity 
scale. The grey area shows the $\pm1\sigma$ uncertainty in the age 
estimate.      
Bottom: Same as the top, but for the absolute age estimates by 
\cite{vandenberg13} and used in \cite{leaman13}. This sample includes
the GCs observed by \cite{marinfranch09} and plotted on the top
panel. However, we have plotted only GGCs in common between the two studies.
} 
\label{fig_gc}
\end{figure*}

Cluster ages based on a relative diagnostic (vertical, horizontal) 
are less prone to systematic uncertainties. The key idea in these methods 
is to estimate the difference in magnitude between the MSTO and a brighter 
point that is either independent or mildly dependent on the absolute age 
of the cluster. This means that they are independent of uncertainties 
on cluster distance and reddening. In dealing with large 
homogeneous photometric data sets, the relative ages are also less 
affected by uncertainties on the photometric zero-point. The pros are 
also on the theoretical side, since the clock is used, over the entire 
metallicity range, in relative and not in absolute sense. 

The relative cluster ages of GGCs is lively debated in the recent 
literature due to their impact on the early formation and evolution 
of the Galactic halo. 
The new results are based on deep and accurate photometry collected 
with ACS at HST. On one hand, it has been found by Marin-Franch and collaborators 
\citep{marinfranch09} that 
a sample of 64 GCs, covering a broad range in metallicity, in Galactocentric distance 
and in kinematic properties, show a bi-modal distribution. The former group is coeval
within the errors, i.e. its mean age is $\sim$ 12.8 Gyr with a small dispersion 
(5$\%$). The latter group is more metal-poor and follows an age--metallicity 
relation. This group might be associated to dwarf galaxies that have 
been accreted into the Halo as supported by their kinematic properties. 
The top panel of Fig.~\ref{fig_gc} summarizes the results obtained by \cite{marinfranch09}. 

On the other hand, based on independent ages estimated by \cite{vandenberg13}
and \citet{leaman13}, it was found that GGCs display a clear dichotomy, since at 
fixed cluster age there are two groups of GCs with roughly 0.4 dex of difference in
metallicity (see bottom panel of Fig.~\ref{fig_gc}). These two groups seem 
to follow two different, and well defined,  
age--metallicity relations. \cite{leaman13} also suggested that the metal-poor 
group were accreted, while the metal-rich ones are the truly globulars formed 
{\em in situ}.   
In this context it is worth mentioning that the globulars and the optical
photometry adopted by \cite{leaman13} significantly overlap with the 
sample adopted by \cite{marinfranch09}. We note here that data plotted
in Fig.~\ref{fig_gc} are only those in common between the two studies 
and a glance to the data indicates a significant difference in their 
age determination. In particular, Fig.~\ref{fig_gc} shows several interesting 
features worthy of discussion.

i) Absolute age estimates provided by \cite{vandenberg13} (VB) are, at 
fixed metal content, more precise than the relative age estimates provided by 
\cite{marinfranch09} (MF). The dispersion in age in the former sample is $\sim$17\% 
smaller than the latter one over the entire metallicity range.   

ii) The metal--poor globulars seem to show an age--metallicity relation independently 
of the approach adopted to estimate the absolute cluster ages. The
relation drawn through the VB's ages (blue solid line) is also a
reasonable `eye' fit to the MF's
age estimates. The quoted samples also show evidence of a flatting of the quoted 
age-metallicity relation when moving from metal-poor ([Fe/H]$\sim$-2) to more 
metal--poor GCs.  

iii) The more metal-rich GCs display a different trend in the two different 
absolute age estimates. The more metal-rich GCs in the MF sample are within 
the errors coeval, while in the VB estimates they show a slope that is quite 
similar to the slope of the more metal-poor GCs, but shifted by 0.4 dex in 
metallicity.   

The above findings clearly indicate that the main difference between the MF 
and the VB analysis is in the age estimates of metal-rich GCs.  
Note that to overcome possible deceptive uncertainties on the cluster iron 
abundance we adopted the homogeneous metallicity scale provided by 
\cite{carretta09}. The current uncertainties on cluster iron abundances 
are on average smaller than 0.1 dex \citep{gratton04a}.  
To unveil possible deceptive systematics either in age-dating globulars or in 
their metallicity scale the use of NIR diagnostics appears very promising. 
This applies not only to NIR photometry to use the MSK (see \S~5.4), but also 
to high-resolution NIR spectroscopy to use different sets of iron and 
$\alpha$-element lines.

\subsubsection{The empirical routes for relative age estimates} \label{empirical_relative}

The two most popular approaches adopted to estimate relative 
ages are the vertical and the horizontal photometric methods. 
The key advantages of these approaches are that they are
independent of uncertainties affecting the distance modulus and the cluster 
reddening. 

i) {\em Vertical Method} -- The vertical method relies on the difference in magnitude between the HB luminosity 
level and the MSTO. The HB luminosity level is typically chosen across the RR Lyrae 
instability strip. The former anchor is assumed to be independent of age, while the 
latter is age dependent. In applying this method there are a number of caveats. The 
main one is that it is mainly based on visual magnitudes, since at shorter
(U,B) and at longer (R,I) wavelengths the HB shows a well defined negative/positive 
slope when moving from red to blue HB stars. This means that the anchor to the middle 
of the RR Lyrae instability strip might be affected by systematic errors, due to the 
quoted problems with the HB morphology. Moreover, this approach is also hampered by 
the lack either of RR Lyrae stars or of accurate and homogeneous multi-band photometry 
for cluster RR Lyrae stars. It is worth mentioning that the cluster age is considered to be 
one of the main culprits in the variation of the HB morphology (second parameter problem) 
when moving from more metal-poor to more metal-rich GCs. In case the cluster age is 
confirmed to be the second parameter, then the ages based on the vertical method 
would be affected by systematic errors. In particular, a decrease in the age of the 
progenitor causes a decrease in the HB luminosity level, therefore smaller vertical 
parameters, and in turn systematically younger ages. This effect becomes more relevant 
for stellar structures at the transition between forming or not forming an electron 
degenerate helium core. This means the young tail in the age distribution of GCs 
(see Fig.~8).  Note that a systematic 
decrease in age typically means a redder HB morphology, thus suggesting once again 
that more metal-rich clusters are more prone to possible systematic age 
uncertainties.    

ii) {\em Horizontal Method} -- The horizontal method relies on the difference in color  between the MSTO and 
the base of the red giant branch. The empirical estimate of the latter reference 
point is not trivial in the V,I bands. Therefore, it was suggested to use the 
difference in color between the MSTO and the color of the RGB at a luminosity 
level that is 2.5 magnitudes brighter than the MSTO \citep{buonanno98,rosenberg99}. 
This approach has the same advantages of the vertical 
method, since it is independent of uncertainties on the cluster distance and 
reddening. Moreover, the reference point along the RGB is less affected by age 
effects and the RGB morphology is well defined in GGCs. However, the use of the 
color (horizontal method) instead of the magnitude (vertical method) together 
with the dependence of predicted effective temperatures at the MSTO, and in 
particular along the RGB, by the adopted mixing length parameter, are a source 
of further concern. The color-temperature relations predicted by atmosphere 
models are also less accurate when compared with the bolometric corrections. 
Note that this limitation becomes more severe when moving from the metal-poor 
to the metal--rich regime \citep{tognelli15}. Moreover, it is worth 
mentioning that the ratio between the duration of the sub-giant phase (hydrogen 
thick shell burning) and the MSTO age is not constant when moving from 
metal-poor to metal-rich stellar structures \citep{salaris97}. Furthermore, 
the slope of the RGB depends on the metal content, therefore, the difference in 
color with the MSTO steadily decreases when moving from more metal-poor to more 
metal-rich globulars. 

\subsection{A new approach for absolute age estimates} \label{newapproach}

During the last few years it has been suggested a new vertical method to 
estimate the absolute age of stellar systems \citep{bono10a}. It relies on 
the difference either in magnitude or in color between the MSTO and a well 
defined knee in the low-mass regime of the main sequence (MSK). The MSK has 
already been detected in several GCs -- $\omega$~Cen \citep{pulone98}; M4
\citep{pulone99,milone14,braga15,correnti16};  NGC~3201 \citep{bono10a}; 
47~Tuc \citep{lagioia14,correnti16}; NGC~2808 \citep{milone12,massari16a}; 
M71 \citep{dicecco15b}; NGC~6752 \citep{correnti16}), in intermediate-age 
open clusters \citep{sarajedini09b,an08,an09} and in the field of the Galactic 
bulge \citep{zoccali+00} by using either near-infrared (NIR) and/or optical--NIR 
photometry.

Empirical and theoretical evidence indicates that this feature is mainly
caused by the collisionally induced absorption of H$_2$ at NIR wavelengths 
\citep{saumon94} in the atmosphere of cool dwarfs. The shape of the bending 
in NIR and optical--NIR CMDs depends on the metal content, but the magnitude 
of the knee seems to be independent of cluster age and of metallicity
(magnitude in K-band of $\sim$3.5 mag).
This means that the difference in magnitude between the MSTO and the MSK
should be
a robust diagnostic to constrain the absolute cluster age.
The new diagnostic shares the same positive features than the classical vertical 
and horizontal methods. It is independent of uncertainties in 
cluster distance and reddening and minimally affected by uncertainties 
in photometric zero-point.  

The key advantages of the MSK when compared with the classical vertical and 
horizontal methods are the following: 

i) {\em The MSK is a faint but well defined feature} -- 
The reference magnitude for the MSK is fainter than the MSTO, this implies
larger photometric uncertainties for individual stars than when belonging to
brighter features (RGB, HB). However, the number of faint, low
mass stars ($\le$ 0.5 M$_{\odot}$) is intrinsically larger than that of more 
massive and brighter stars. Furthermore, these stars are in
their slow central hydrogen burning phase, thus they are more populous 
than in subsequent evolved phases. As a result, old, intermediate and 
young stellar systems always display a well sampled faint MS, whereas: 
a) the HB is less populous because it is populated by
stars in their central helium burning phase and,
even more, this feature shows up only in old stellar systems and
depends on a number of parameters, including the metallicity; 
b) the RGB is typical of stellar 
systems that are either old or of intermediate age, and again the difference in 
color with the MSTO depends on the metallicity.\par 
The main consequence 
is that the MSK can be easily identified in all the 
stellar systems, provided that photometric data sets are pushed to
reach very faint magnitudes,
while the HB and the RGB detection/morphology strongly depend on the 
evolutionary properties of the underlying stellar population.\\ 
 
ii) {\em The MSK is less affected by theoretical uncertainties} --
The MSK only relies on the physics of hydrogen burning phase, 
while the vertical and the horizontal methods are affected 
by uncertainties in extra-mixing (RGB, HB), in microphysics (electron 
degeneracy, conductive opacities, 3$\alpha$ and $^{12}C$($alpha$,$\gamma$)$^{16}C$
nuclear reaction rates) and in dealing with the transition from the core 
helium flash to the core helium burning \citep[ZAHB,][]{sweigart04}).\par   

Furthermore, the MSK together with the vertical method are independent of the
theoretical uncertainties plaguing the color--temperature transformations
required in the horizontal method. The same outcome applies to the 
dependence on the adopted mixing length parameter. According to theory,
MS stellar structures with a stellar mass of $M\sim$0.5-0.4~$M_\odot$ 
are minimally affected by uncertainties in the treatment of convection, 
since they are almost entirely convective and the convective transport 
is nearly adiabatic \citep{saumon08}.\\

iii) {\em The MSK has a lower global error budget} -- 
Recent empirical evidence indicates that the MSK provides absolute cluster 
ages that are a factor of two more precise when compared with the classical 
method of the MSTO \citep{bono10a,sarajedini09a,dicecco15b,monelli15,massari16a,correnti16}.
Moreover, preliminary findings suggest that the correlation between the MSK and the 
cluster age is linear over the age range typical of old open cluster 
(a few Gyrs) to GGCs. We still lack a detailed theoretical investigation to 
constrain the dependence of the MSK on the chemical composition (metals, 
helium) when moving from optical to NIR and optical/NIR CMDs.\\  

The main cons in dealing with the MSK is that the identification of the 
knee requires accurate and deep photometry in crowded stellar
fields. However, the advent of HST and of modern Adaptive Optics (AO) assisted NIR
instrumentation overcame this problem. The MSK is a feature
that can be detected and used in both optical and NIR bands, whereas the
classical vertical and horizontal methods are robust age diagnostics 
only in optical CMDs. Accurate and deep NIR CMDs show that HB stars are far
from being horizontal. They become systematically 
fainter when moving from cool to hot and extreme HB stars 
\citep[positive slope][]{delprincipe06,coppola11,milone13,stetson14a}. The same 
problem shows up in CMDs based on near UV and far UV bands 
\citep[negative slope,][]{ferraro12}. This means that
the identification of the HB luminosity level needs a further anchor in
color along the HB, difficult to be uniquely identified. On the other hand, the difference in 
color between the MSTO and the RGB, required by the 
horizontal method, is hampered by the fact that MSTO and RGB have almost 
the same color in NIR bands. This means that the difference in color is 
steadily decreasing \citep{coppola11,stetson14a} when moving from
optical to NIR. 

NIR photometry is going to be exploited even more in the near
future when sophisticated AO will allow us to reach the diffraction
limit of ground based extremely large telescopes \citep{diolaiti16} 
and from the space with JWST. The use of new observables also means the 
opportunity to constrain possible systematics in evolutionary diagnostics 
currently adopted. 

\subsubsection{The absolute age of M15} \label{msk_M15}

To provide a new and independent absolute age estimate of the GC M15 (NGC 7078), 
\citep{monelli15} have collected collected AO images of the GC M15 (NGC~7078) using the First
Light AO \citep[FLAO,][]{esposito12} operating at the Large Binocular
Telescope (LBT). We remember here that M15 is located at $\sim$10~Kpc and 
it is affected by moderate extinction (E(B-V)=0.08, \citep[][updated 2010 version]{harris96} 
thus with current AO assisted 10~m telescopes we do expect to detect the MSK. 
Interestingly, M15 is supposed to be one of the oldest and most metal poor 
([Fe/H$\sim$-2.4]) GCs of the Milky Way Halo.

\begin{figure*}
\includegraphics[width=1.05\textwidth]{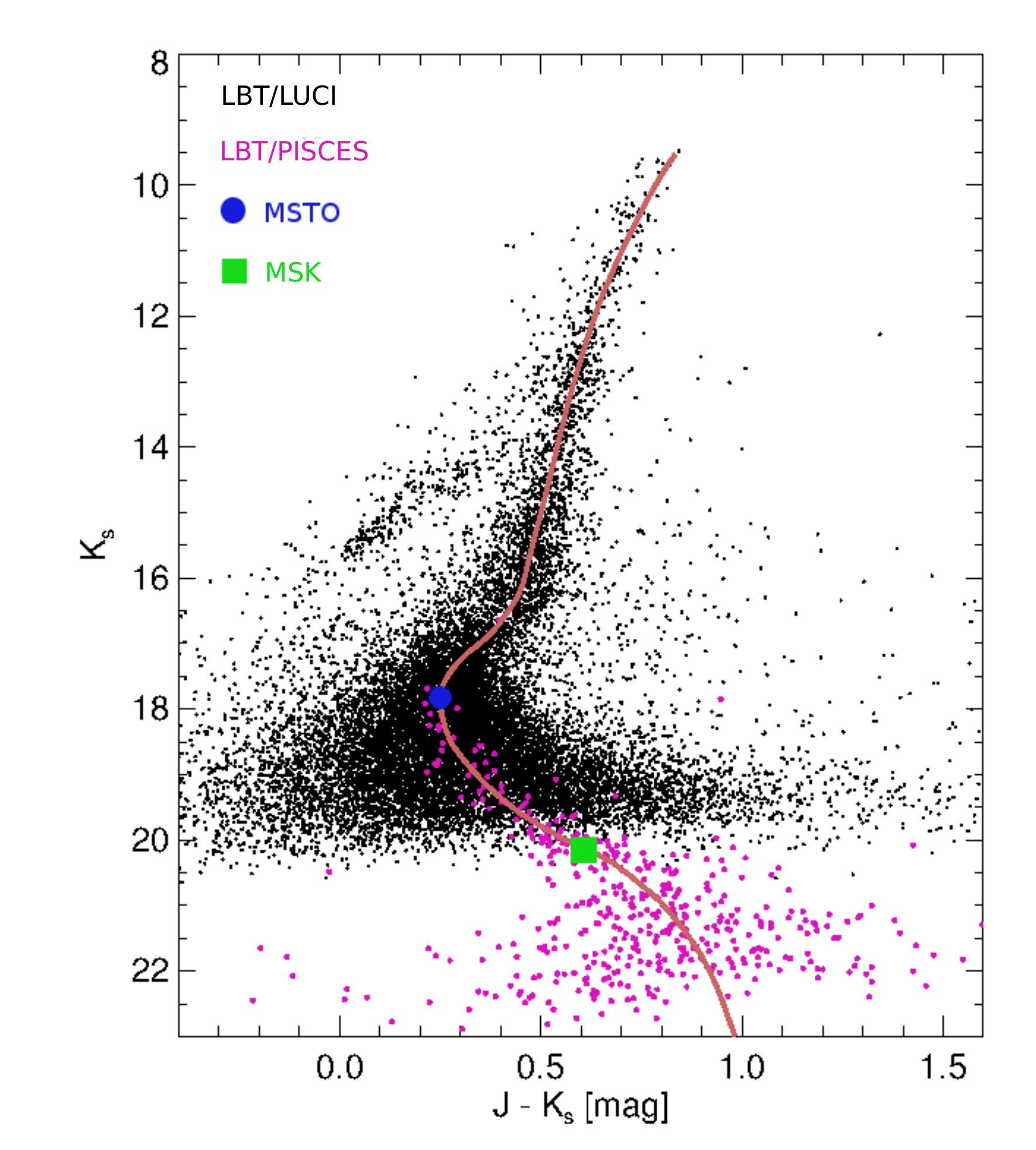}
\caption{Near-Infrared (K, J - K) CMD of the Galactic globular M15, based on 
images collected with LBT (black dots: LUCI1; red dots: PISCES). The orange solid 
line shows the cluster ridge line adopted to determine the position of both 
MSTO (blue circle) and of the MSK (green square).}
\label{fig_monelli}  
\end{figure*}

FLAO data have been taken with the NIR high resolution imager PISCES 
(pixel scale$=$0.0193"/pix) in J and K$_s$ bands. Due to the highly 
structured and asymmetric PSF shape, the data reduction was successfully 
performed with ROMAFOT suite of programs \citep[see details in][]{fiorentino14b, monelli15}. 
A natural guide star (NGS) of R=12.9 mag has 
been used to close the AO loop on a field located at $\sim$3' from the 
cluster center. This field was sufficiently uncrowded
to allow us to reach a very deep K$_s$ band magnitude of $\sim$22~mag, 
see Fig.~\ref{fig_monelli}. This limiting magnitude allowed us to measure 
the location of the MSK; see Table~3 in \cite{monelli15}. Note that the 
detection of faint stars in crowded stellar fields as the center of GCs  
is severely hampered by the large number of bright stars. 
However, as it is shown in Fig.~\ref{fig_monelli}, given the small PISCES field 
of view (FoV) $\sim$20" and the radial distance from the cluster center, 
we do not have sufficient sampling of the MSTO magnitude. We have used LUCI1 
data (FoV=4'$\times$4') to properly determine the location of the MSTO. 
These LUCI1 data have also been used to perform a proper calibration of 
PISCES data to the 2MASS photometric system. 

After measuring the difference MSTO-MSK, we are ready to compute the 
absolute age of M15. We compare this number with theoretical relations derived 
using a set of evolutionary isochrones provided by \cite{vandenberg14b} that 
relate the variation of MSTO-MSK with the absolute ages. Using only NIR, 
we derived an absolute age for M15 of 13.7$\pm$1.4 Gyr, which is compatible, 
but with a smaller uncertainty, to that obtained using the classical MSTO method 
in NIR (14.0$\pm$3.1 Gyr) or in purely optical HST bands (12.8$\pm$2.0 Gyr). 
This old age provides an upper limit to the age of the Universe and a lower 
limit to the Hubble constant H$_0$, since the former parameter is roughly 
the inverse of the latter one \citep{gratton03,monelli15}.

\subsection{Conclusions and final remarks} \label{conclusion}

There is mounting evidence of a difference between estimates of a Hubble constant 
based on direct measurements (Cepheids plus supernovae: \cite{riess16,freedman10},
Beaton this conference) and indirect methods (CMD, BAO, lensing, 
\cite{suyu13,bennett14,calabrese15}, Planck collaboration 2015). 
This critical issue has been addressed in several recent papers suggesting 
a difference that ranges from almost 2$\sigma$ \citep{efstathiou14} to more 
than 3$\sigma$ \citep{riess11a,riess16}. The quoted
uncertainties on the Hubble constant, once confirmed, can open the path to 
new physics concerning the number of relativistic species and/or the mass of 
neutrinos \cite{dvorkin14,wyman14,lukovic16}. Moreover, the quoted 
range in $H_0$ implies an  uncertainty on the age of the universe $t_0$ of 
the order of 2 Gyr. This uncertainty has a substantial impact not only on galaxy
formation and evolution, but also on the age of the most ancient stellar
systems, i.e. the globular clusters can play a crucial cosmological role. 

Based on the tantalizing evidence that stellar age is not an observable, the 
different stellar ``clocks" that can be applied to date globular clusters provide 
the unique opportunity to constrain the micro- and the macro-physics adopted 
to construct evolutionary models. This comparison becomes even more rewarding 
when comparing main sequence stellar structures with white dwarf models.  It might 
be possible that in the era of ``precision cosmology" we could use cosmological 
parameters to constrain the physics of stellar interiors. In the mean time, 
it is clear that calibrating clusters to which we can apply different age 
diagnostics (MSTO, MSK, white dwarf cooling sequence, cosmochronometry) 
become fundamental astrophysical and cosmological laboratories.  

In this investigation we reviewed the most popular methods to estimate both 
relative and absolute cluster ages. We focussed our attention on the error 
budget and critically discussed pros and cons of the different age 
diagnostics. In particular, we outlined the key advantages in using a new 
NIR age diagnostic --the main sequence knee-- and its application to M15. 
The main limitation being the faintness of this anchor. However, there 
are two new observing facilities that are going to play a fundamental 
role in the use of the MSK. 

i) {\em Multi-conjugated adaptive optics} -- The development of 
multi-conjugated adaptive optics systems at the 8-10~m class telescopes is a
real quantum jump. They provide NIR images that approach the diffraction 
limit of a field of view of the order of one arcmin. This means that they 
can provide accurate and deep NIR CMDs of the innermost crowded regions 
of GCs. Recent findings based on NIR images collected with GEMS at GEMINI
indicate that the mix between NGS and lasers allow us to reach the MSK in 
a sizable sample of Galactic globulars \cite{massari16a,turri16}. 
The empirical scenario becomes even more compelling if we take account for 
the next generation of NIR detectors AO assisted at VLT (ERIS). This instrument 
will simultaneously cover NIR bands and the L-band, and in turn, the unique 
opportunity to identify the MSK in the most crowded and most reddened regions 
of the Galactic Bulge.    

ii) {\em JWST} -- JWST is going to revolutionize the view of resolved stellar 
populations in the nearby Universe. The coupling between field of view and 
NIR/MIR bands would provide the unique opportunity to identify the MSK in 
a significant fraction of nearby dwarf galaxies. This means the opportunity 
to determine homogeneous ages for old stars in old stellar systems (dwarfs, 
globulars) to investigate whether they formed, as suggested by cosmological 
models, at the same epoch.    

The cosmic distance scale and the age-dating of nearby stellar systems have been 
for more than half a century the two fundamental pillars on which quantitative 
astrophysics build up. At the beginning of the new millennium they are waiting 
for massive solidification. The near future appears quite bright not only for 
the next Gaia data releases, but also for the near future ground-based 
(LSST, ELTs) and space observing facilities (JWST, WFIRST, EUCLID). The same
outcome applies for the wide spectroscopic surveys in optical and NIR regimes 
(DESI, PFS, 4MOST, MOONS-GAL, APOGEE, WEAVE).




\end{document}